\pdfoutput=1
\newcommand*{\ATLASLATEXPATH}{latex/}
\documentclass[cernpreprint,UKenglish,txfonts,texlive=2011]{\ATLASLATEXPATH atlasdoc}
\usepackage{\ATLASLATEXPATH atlaspackage}
\usepackage{\ATLASLATEXPATH atlasbiblatex}

\usepackage{\ATLASLATEXPATH atlascontribute}

\usepackage{\ATLASLATEXPATH atlasphysics}
\usepackage{amsmath}
\usepackage{mathrsfs}
\usepackage{longtable}

\graphicspath{{logos/}{figures/}}

\usepackage{paper-defs}

\AtlasTitle{Measurement of jet activity in top quark events using the $e\mu$ final state with two $b$-tagged jets in $pp$ collisions at $\sqrt{s}=8$\,\TeV\ with the ATLAS detector}

\author{The ATLAS Collaboration}

\AtlasRefCode{TOPQ-2015-04}

\PreprintIdNumber{CERN-EP-2016-122}

\AtlasJournal{JHEP}
\AtlasJournalRef{JHEP 09 (2016) 074}
\AtlasDOI{10.1007/JHEP09(2016)074}

\AtlasAbstract{%
Measurements of the jet activity in \ttbar\ events produced in proton--proton
collisions at \sxvt\ are presented, using 20.3\,\ifb\ of 
data collected by the ATLAS experiment at the Large Hadron Collider. The 
events were selected in the dilepton $e\mu$ decay channel with two identified
$b$-jets. The numbers  of additional 
jets for various jet transverse momentum (\pt) thresholds, and
the normalised differential cross-sections as a function of \pt\ for the
five highest-\pt additional jets, 
were measured in the jet pseudorapidity range $|\eta|<4.5$.
The gap fraction, the fraction of events which do not contain an additional jet
in a central rapidity region, was measured for several rapidity intervals
as a function of the minimum \pt\ of a single jet or the scalar sum of \pt\
of all additional jets. These fractions were also measured in different
intervals of the invariant mass of the \emubb\ system. All measurements
were corrected for detector effects, and found to be mostly well-described 
by predictions from
next-to-leading-order and leading-order \ttbar\ event generators with
appropriate parameter choices.
The results can be used to further optimise the parameters used in such generators.
}

\AtlasCoverSupportingNote{Differential cross-section analysis: }{https://cds.cern.ch/record/1983211}
\AtlasCoverSupportingNote{Gap fraction analysis:}{https://cds.cern.ch/record/1983214}
\AtlasCoverCommentsDeadline{Wednesday 18th May 2016}

 \AtlasCoverAnalysisTeam{Carolyn Bertsche, Jacquelyn Brosamer, Mark Cooke, Richard Hawkings, Ian Hinchliffe, Marjorie Shapiro, Michael Strauss, Mark Tibbetts}

\AtlasCoverEdBoardMember{Martine Fernandez-Bosman~(chair)}
\AtlasCoverEdBoardMember{Halina~Abramowicz}
\AtlasCoverEdBoardMember{Frederic~Deliot}
\AtlasCoverEdBoardMember{Jie~Yu}

\AtlasCoverEgroupEditors{atlas-topq-2015-04-editors@cern.ch}

\AtlasCoverEgroupEdBoard{atlas-topq-2015-04-editorial-board@cern.ch}

\hypersetup{pdftitle={ATLAS draft},pdfauthor={The ATLAS Collaboration}}

\begin{document}
\maketitle

\newpage

\section{Introduction}
\label{sec:intro}

The top quark plays a special role in the Standard Model and in some theories of
physics beyond the Standard Model. 
The large top quark mass and large \ttbar\ pair-production cross-section 
in $pp$ collisions
($242\pm 10$\,pb at \sxvt\ \cite{TOPQ-2013-04}) make top quark production
at the Large Hadron Collider (LHC)
a unique laboratory for studying the behaviour of QCD at the highest accessible 
energy scales. The decays of top quarks to charged leptons, neutrinos and 
$b$-quarks also make such events a primary source of background 
in many searches for new physics.
Therefore, the development of accurate modelling for events involving top quark
production forms an important part of the LHC physics programme.
Measurements of the activity of {\it additional jets} in \ttbar\ events,
{\em i.e.}\ jets not originating from the decay of the top quark and antiquark,
but arising from quark and gluon radiation produced in association
with the \ttbar\ system,
have been made by ATLAS~\cite{TOPQ-2011-21,TOPQ-2012-03} and 
CMS~\cite{CMS-TOP-12-018} using $pp$ data at \sxwt, and by 
CMS~\cite{CMS-TOP-12-041} at \sxvt. 
These data are typically presented 
as particle-level results in well-defined fiducial regions, corrected to 
remove detector efficiency and
resolution effects, and compared to the predictions of Monte Carlo (MC)
generators through tools such as the {\sc Rivet} framework \cite{rivet}. 
Such comparisons indicate that some state-of-the-art generators have 
difficulties in reproducing the data,
whilst for others  agreement with data can be improved with an appropriate
choice of generator parameter values or `tune', including those  controlling QCD factorisation 
and renormalisation scales, and matching to the parton shower 
\cite{ATL-PHYS-PUB-2014-005,ATL-PHYS-PUB-2015-002,ATL-PHYS-PUB-2015-007,ATL-PHYS-PUB-2015-011,ATL-PHYS-PUB-2016-004}.  

This paper presents two studies of the additional jet activity in \ttbar\  
events collected with the ATLAS detector in $pp$ collisions
at a centre-of-mass energy of \sxvt. 
Top quark pairs are selected in the same way in both measurements,
using the dilepton $e\mu$ final state 
with two jets identified (`tagged') as likely to contain $b$-hadrons.
Distributions of the properties of additional jets in these events are 
normalised to the cross-section (\sigemubb) for events passing this
initial selection, requiring
the electron, muon and two $b$-tagged jets  to have
transverse momentum $\pt>25$\,\GeV\ and pseudorapidity\footnote{ATLAS uses a right-handed 
coordinate system with its origin at
the nominal interaction point in the centre of the detector, and the $z$ axis
along the beam line. Pseudorapidity is defined in terms of the polar angle
$\theta$ as $\eta=-\ln\tan{\theta/2}$, and transverse momentum and energy
are defined relative to the beamline as $\pt=p\sin\theta$ and
$\et=E\sin\theta$. The azimuthal angle around the beam line is denoted by 
$\phi$, and distances in $(\eta,\phi)$ space by 
$\Delta R=\sqrt{(\Delta\eta)^2+(\Delta\phi)^2}$.
The rapidity is defined as $y=\frac{1}{2}\ln\left(\frac{E+p_z}{E-p_z}\right)$,
where $p_z$ is the $z$-component of the momentum and  $E$ is the energy 
of the relevant object.} $|\eta|<2.5$.

In the first study, the normalised particle-level cross-sections 
for additional jets with $|\eta|<4.5$ and $\pt>25$\,GeV\
are measured differentially in jet rank and \pt;
\begin{equation}
\frac{1}{\sigma}\frac{{\rm d}\sigma_{i}}{{\rm d}\pt} \equiv
\frac{1}{\sigemubb} \frac{{\rm d}\sigma^{\mathrm{jet}}_i } {{\rm d}\pt} \label{e:diffjet}
\ ,
\end{equation}
with  rank $i=1$ to 5, where $i=1$  denotes the leading (highest \pt) 
additional jet. These normalised differential cross-sections 
are then used to obtain the multiplicity distributions for additional jets
as a function of the minimum \pt\ threshold for such extra jets.

The additional-jet differential cross-section measurements are complemented by 
a second study measuring the jet `gap fraction', {\em i.e.} 
the fraction of events where no additional jet is present within a particular
interval of jet rapidity, denoted by \dely. The gap fraction is measured 
as a function of the jet \pt\ threshold, \qzero; 
\begin{equation}
\fqzero \equiv \frac{\sigma(\qzero)}{\sigemubb} \label{e:fgap}\ ,
\end{equation}
starting from a minimum \qzero\ of 25\,\GeV, where
$\sigma(\qzero)$ is the cross-section for events having no
additional jets with $\pt>\qzero$, within the rapidity interval \dely. 
Following the corresponding measurement at \sxwt\ \cite{TOPQ-2011-21},
four rapidity intervals \dely\ are defined: 
$|y|<0.8$, $0.8<|y|<1.5$, $1.5<|y|<2.1$ and the inclusive interval $|y|<2.1$. 
These intervals are more restrictive than for the normalised additional jet
cross-sections, which are measured over the wider angular range  $|\eta|<4.5$ 
corresponding to the full acceptance of the detector.

As well as \fqzero, the gap fraction is measured 
as a function of a threshold \qsum\ placed on the scalar sum of the \pt\ of all
 additional jets with $\pt>25$\,\GeV\ within the same rapidity intervals \dely:
\begin{equation}
\fqsum \equiv \frac{\sigma(\qsum)}{\sigemubb} \ .
\end{equation}
The gap fraction measured as a function of \qzero\ is sensitive to the leading
\pt\ emission accompanying the \ttbar\ system, whereas the gap fraction
based on \qsum\ is sensitive to all accompanying hard emissions.
Finally, the gap fractions \fqzero\ and \fqsum\ in the inclusive rapidity 
region $|y|<2.1$ are also measured
separately for four subsets of the invariant mass of the $e\mu\bbbar$ system 
\memubb,
which is related to the invariant mass of the produced \ttbar\ system and 
is on average higher if produced from quark--antiquark rather than 
gluon--gluon initial states.

This paper is organised as follows. 
Section~\ref{sec:detector} describes the ATLAS detector and the data sample used for these measurements.
Section~\ref{sec:simulation} provides information about the Monte Carlo 
simulated samples used to model signal and background processes, and 
to compare with the measured results.
The common object and event selection criteria are presented in Section~\ref{sec:selection}, and sources of systematic uncertainty are discussed in Section~\ref{sec:systematics}.
The measurement of the normalised jet differential cross-sections by rank 
and \pt\ is described in Section~\ref{sec:jets}
and the measurement of the gap fraction is presented in Section~\ref{sec:gapfrac}, in both cases including comparisons with   
the predictions of various \ttbar\ event generators.
Section~\ref{sec:conclusion} gives a summary and conclusions.

\FloatBarrier
\section{Detector and data sample}
\label{sec:detector}


The ATLAS detector \cite{PERF-2007-01} at the LHC covers almost the full solid 
angle around the collision point, and consists of an inner tracking detector
surrounded by a thin superconducting solenoid magnet producing a 2\,T axial
magnetic field, electromagnetic and hadronic calorimeters, and an external
muon spectrometer incorporating three large toroidal magnet systems.
The inner detector consists of a high-granularity silicon pixel detector and
a silicon microstrip tracker, together providing precision tracking in the
pseudorapidity range $|\eta|<2.5$,
complemented by a transition radiation
tracker providing tracking and electron identification information for
$|\eta|<2.0$. A lead/liquid-argon (LAr) electromagnetic calorimeter covers the
region $|\eta|<3.2$, and hadronic calorimetry is provided by 
steel/scintillator tile calorimeters for $|\eta|<1.7$ and copper/LAr
hadronic endcap calorimeters covering $1.5<|\eta|<3.2$. The calorimeter system
is completed by forward LAr calorimeters with copper and tungsten absorbers 
which extend the coverage to   $|\eta|=4.9$. The muon 
spectrometer consists of precision tracking chambers covering the region
$|\eta|<2.7$, and separate trigger chambers covering $|\eta|<2.4$. A 
three-level trigger system, using custom hardware followed by two
software-based levels, is used to reduce the event rate to about 400\,Hz 
for offline storage.

The analyses were performed on the 2012 ATLAS proton--proton collision data
sample, corresponding to an integrated luminosity of 20.3\,\ifb\ at 
$\sqrt{s}=8$\,\TeV\ after the application of detector status and 
data quality requirements. The integrated luminosity was measured using
the methodology described in Ref. \cite{DAPR-2011-01} applied to
beam separation scans performed in November 2012, and has a relative 
uncertainty of 2.8\,\%. Events were required to pass either a single-electron 
or single-muon trigger, with thresholds chosen such that the efficiency 
plateau is reached for leptons with $\pt>25$\,\GeV\ passing offline 
selections. Each triggered event also includes the signals from an average of
20~additional inelastic $pp$ collisions in the same bunch crossing
(referred to as pile-up).

\FloatBarrier
\section{Simulated event samples}
\label{sec:simulation}


Monte Carlo simulated event samples were used to evaluate signal efficiencies
and backgrounds, and to estimate and correct for resolution effects. The
samples were processed either through the full ATLAS detector simulation
\cite{SOFT-2010-01} based on GEANT4 \cite{Agostinelli:2002hh}, or through a 
faster simulation making use of parameterised showers in the calorimeters
\cite{ATL-PHYS-PUB-2010-013}. Additional simulated inelastic
$pp$ collisions, generated with {\sc Pythia8.1} \cite{Sjostrand:2007gs} using 
the MSTW2008 LO \cite{mstwnnlo} parton distribution functions (PDFs) and the 
A2 tune \cite{ATL-PHYS-PUB-2012-003},
were overlaid to simulate the effects of both in- and
out-of-time pile-up, from additional activity in the same and nearby 
bunch crossings. The resulting simulated events were processed using the
same reconstruction algorithms and analysis chains as the data.
The effects of pile-up were also studied with data recorded
from randomly selected bunch crossings (zero-bias data) as discussed 
in Section~\ref{sec:systematics}.

The baseline \ttbar\ full simulation sample was produced using the 
next-to-leading-order (NLO) QCD matrix-element generator {\sc Powheg-Box} 
v1.0 \cite{powheg,powheg2,powheg3}
using the CT10 PDFs \cite{cttenpdf} and interfaced
to {\sc Pythia6} (version 6.426) \cite{pythia6} with the CTEQ6L1 PDF set 
\cite{ctsix} 
and the Perugia 2011C (P2011C) tune \cite{perugia}
for the parton shower, fragmentation and
underlying event modelling. The renormalisation and factorisation scales
were set to the generator default value of $\sqrt{m_t^2+\pt^2}$, the
sum in quadrature of the top quark mass \mtop\ and transverse momentum
\pt, the latter evaluated for the underlying Born configuration 
before radiation.
The {\sc Powheg} parameter \hdmp, used in the damping function 
that limits the resummation of higher-order effects incorporated into the
Sudakov form factor, was set to infinity, corresponding to no damping.
The top quark mass was set to 172.5\,\GeV. The
total \ttbar\ production cross-section, used when comparing
predictions from simulation with 
data, was taken to be $253^{+13}_{-15}$\,pb, based on the 
next-to-next-to-leading-order (NNLO) calculation including the
resummation of next-to-next-to-leading logarithmic soft gluon terms
as described in Refs.~\cite{topxtheo1,topxtheo2,topxtheo3,topxtheo4,topxtheo5}
and implemented in the Top++ 2.0 program \cite{toppp}. The quoted 
uncertainties include PDF and $\alpha_{\rm s}$ uncertainties based on the 
PDF4LHC prescription \cite{pdflhc} applied to the 
MSTW2008 NNLO \cite{mstwnnlo,mstwnnlo2}, 
CT10 NNLO \cite{cttenpdf,cttennnlo} and NNPDF2.3 5f FFN \cite{nnpdfffn}
PDF sets, added in quadrature to the QCD scale uncertainty.

Alternative \ttbar\ simulation samples were used to evaluate systematic 
uncertainties, and were compared with the data measurements after unfolding
for detector effects.  Samples
were produced with {\sc Powheg} with $\hdmp=\infty$ interfaced to 
{\sc Herwig} (version 6.520) \cite{herwig1,herwig2} with the ATLAS AUET2 tune 
\cite{ATL-PHYS-PUB-2011-008} and {\sc Jimmy} (version 4.31) \cite{jimmy}
for underlying-event modelling. Samples with $\hdmp=\mtop$,
which softens the \ttbar\ \pt\ spectrum,
improving the agreement between data and simulation at 
 \sxwt~\cite{ATL-PHYS-PUB-2014-005}, were generated by combining {\sc Powheg} 
with either {\sc Pythia6} with the P2011C tune or {\sc Pythia8} (version 8.186)
with the A14 tune \cite{ATL-PHYS-PUB-2014-021}. Samples were also produced
with {\sc MC@NLO} (version 4.01) \cite{mcatnlo1,mcatnlo2} interfaced to 
{\sc Herwig} and {\sc Jimmy}, with the generator's default renormalisation
and factorisation scales of $\sqrt{m_t^2+(p_{{\rm T},t}^2+p_{{\rm T},\bar{t}}^2)/2}$ 
where $p_{{\rm T},t}$ and $p_{{\rm T},\bar{t}}$ are the transverse momenta of the
top quark and antiquark.
Several leading-order `multi-leg' generators were also
studied. The {\sc Alpgen} generator (version 2.13) \cite{alpgen} was used
with leading-order matrix elements for \ttbar\ production accompanied by
up to three additional light partons,
 and dedicated matrix elements for \ttbar\ plus 
\bbbar\ or \ccbar\ production, interfaced to {\sc Herwig} and {\sc Jimmy}.
An alternative sample was generated with {\sc Alpgen} interfaced to 
{\sc Pythia6} with the P2011C tune, 
including up to four additional light partons.
The MLM parton--jet matching scheme \cite{alpgen}
was applied to avoid double-counting of configurations generated by both
the parton shower and the matrix-element calculation. A further sample was 
generated using {\sc MadGraph 5} (version 1.5.11) \cite{madgraph} with up to
three additional partons and using MLM matching, 
interfaced to {\sc Pythia6} with the P2011C tune.
Finally, three pairs of samples with matching scale and parton
shower parameters tuned to 
explicitly vary the amount of additional radiation in \ttbar\ events were
used, generated using {\sc AcerMC} (version 3.8) \cite{acermc}, {\sc Alpgen} 
or {\sc MadGraph}, each interfaced to {\sc Pythia6} with
either the RadLo or RadHi P2011C tunes \cite{perugia}. The parameters of these
samples were tuned to span the variations in radiation compatible with
the ATLAS \ttbar\ gap fraction measurements at $\sqrt{s}=7$\,\TeV\
\cite{TOPQ-2011-21} as discussed in detail in Ref.~\cite{ATL-PHYS-PUB-2014-005}.

After the $e\mu\bbbar$ event selection, the expected non-\ttbar\ contribution is
dominated by $Wt$, the associated production of a $W$ boson and a single
top quark. This process is distinct from \ttbar\ production 
when considered at leading order.
But at NLO in QCD the two processes cannot be separated once the top quarks
decay to $Wb$: the resulting $WbW\bar{b}$ final state can appear
for example through both $gg\rightarrow\ttbar\rightarrow WbW\bar{b}$
and $gg\rightarrow Wt\bar{b}\rightarrow WbW\bar{b}$, and the two processes
interfere to an extent depending on the kinematics
of the final state. However, the currently available generators do 
not allow a full treatment of this interference; instead they consider
\ttbar\ and $Wt$ production as separate processes. Within this approximation,
the `diagram removal' and `diagram subtraction' schemes have been proposed
as alternatives for approximately handling the interference 
between the \ttbar\ and $Wt$ processes \cite{wtinter1,wtinter2}. 
For this paper, $Wt$ production was simulated as a process separate
from \ttbar, using {\sc Powheg\,+\,Pythia6}
with the CT10 PDFs and the P2011C tune. The diagram removal scheme was used
as the baseline and the diagram subtraction scheme was 
used to assess systematic uncertainties.
A cross-section of $22.4\pm 1.5$\,pb was assumed for $Wt$ production,
determined by using the approximate NNLO prediction described in 
Ref.~\cite{wttheo}.

Other backgrounds with two prompt leptons arise from diboson production 
($WW$, $WZ$ and $ZZ$) accompanied by $b$-tagged jets, modelled using 
{\sc Alpgen\,+\,Herwig\,+\,Jimmy} with CTEQ6L1 PDFs and with 
total cross-sections calculated
using MCFM \cite{dibmcfm}; and $Z\rightarrow\tau\tau(\rightarrow e\mu)$+jets, 
modelled using
{\sc Alpgen\,+\,Pythia6} with CTEQ6L1 PDFs, and including leading-order 
matrix elements for $Z\bbbar$ production. The normalisation of this background
was determined from data using $Z\rightarrow ee/\mu\mu$ 
events with two $b$-tagged jets as described in Ref.~\cite{TOPQ-2013-04}.
The remaining background originates from events with
one prompt and one misidentified lepton, {\em e.g.} a 
non-prompt lepton from the decay of a bottom or charm hadron, an electron
from a photon conversion, hadronic jet activity misidentified as an electron,
or a muon produced from an in-flight decay of a pion or kaon. Such 
events can arise from \ttbar\ production with one hadronically decaying $W$, 
modelled as for dileptonic \ttbar\ production 
with {\sc Powheg\,+\,Pythia6}; $W$+jets production, modelled as
for $Z$+jets; and $t$-channel single-top production, modelled using 
{\sc AcerMC\,+\,Pythia6} with CTEQ6L1 PDFs. Previous studies have shown 
that these simulation samples
provide a good model of the rate and kinematic distributions of 
$e\mu\bbbar$ events with one real and one misidentified lepton 
\cite{TOPQ-2013-04}. The expected contributions to
the additional-jet distributions from \ttbar\ production in association with a 
$W$, $Z$ or Higgs boson are below the percent level. Other backgrounds, 
including processes with two misidentified leptons, are negligible.

\FloatBarrier
\section{Object and event selection}
\label{sec:selection}
The two analyses use the same object and event selection as employed in the
ATLAS inclusive \ttbar\ cross-section analysis at $\sqrt{s}=8$\,\TeV\
\cite{TOPQ-2013-04}. Electrons were identified as described in
Ref.~\cite{PERF-2013-03}, required to have transverse energy $\et>25$\,GeV\
and pseudorapidity $|\eta|<2.47$, and to be isolated to reduce backgrounds
from non-prompt and misidentified electrons. Electron candidates within
the transition region between the barrel and endcap electromagnetic 
calorimeters, $1.37<|\eta|<1.52$, were removed. 
Muons were identified as described in Ref.~\cite{PERF-2011-01}, required
to have $\pt>25$\,GeV\ and $|\eta|<2.5$, and also required to be isolated.

Jets were reconstructed using the anti-$k_t$ algorithm 
\cite{Cacciari:2008gp,Cacciari:2005hq} 
with radius parameter $R=0.4$, starting from clusters of energy deposits in 
the calorimeters, calibrated using the local cluster 
weighting method \cite{PERF-2011-03}.
Jets were calibrated using an energy- and $\eta$-dependent
simulation-based scheme, with the effects of pile-up on the jet
energy measurement being reduced using the jet-area method described in
Ref.~\cite{ATLAS-CONF-2013-083}. After the application of in situ corrections 
based on data \cite{PERF-2012-01}, jets were required to satisfy
$\pt>25$\,\GeV\ and $|\eta|<4.5$. To suppress the contribution from low-$\pt$
jets originating from pile-up interactions, a jet vertex fraction (JVF)
requirement  was applied to jets with $\pt<50$\,\GeV\ and $|\eta|<2.4$ 
\cite{PERF-2014-03}. Such jets
were required to have at least 50\,\% of the scalar sum of the $\pt$
of tracks associated with the jet originating from tracks associated with the 
event primary vertex,  the latter being 
defined as the reconstructed vertex with the highest sum of associated 
track $\pt^2$.  Jets with no associated tracks were also selected.
To prevent double-counting
of electron energy deposits as jets, jets within $\Delta R=0.2$ of 
a reconstructed electron were removed.
Finally, to further suppress non-isolated leptons from 
heavy-flavour decays inside jets, electrons and muons within $\Delta R=0.4$ of
selected jets were also discarded.

Jets containing $b$-hadrons were tagged using the
MV1 algorithm, a multivariate discriminant making use of
track impact parameters and reconstructed secondary vertices \cite{PERF-2012-04}.
Jets were defined to be $b$-tagged if the MV1
discriminant value was larger than a threshold corresponding 
to a 70\,\% efficiency for tagging $b$-quark jets in \ttbar\ events, giving a 
rejection factor of about~140 against light-quark
and gluon jets, and about five against jets originating from charm quarks.

Events were required to have a reconstructed primary vertex
with at least five associated tracks. Events with any jets failing jet quality 
requirements \cite{PERF-2012-01}, or with any muons compatible 
with cosmic-ray interactions
or suffering substantial energy loss through  bremsstrahlung in the detector 
material, were removed. An event preselection was then applied, requiring 
exactly one electron and one muon selected as described above, with 
opposite-sign electric charges. At least  one of the leptons was required 
to be matched to an electron or muon object triggering
the event. Finally, selected events were required to have at least two 
$b$-tagged jets. The resulting \emubb\ event selection is similar to that of the
$\sqrt{s}=8$\,\TeV\ sample with two $b$-tagged
jets used in Ref.~\cite{TOPQ-2013-04}, except that events with 
three or more $b$-tagged jets are also accepted.\footnote{The event counts 
differ from those in Ref.~\cite{TOPQ-2013-04} as updated object calibrations
were used in this analysis, in particular for the jet energy scale.}
The numbers of preselected opposite-sign
$e\mu$ and selected \emubb\ events
are shown in Table~\ref{table:eventCounts}. The observed event count 
after requiring at least two $b$-tagged jets is in
good agreement with the prediction from the baseline simulation.

\begin{table}[tp]
\centering
\begin{tabular}{lrrrr}\hline
 & \multicolumn{1}{c}{$e\mu$} & \multicolumn{1}{c}{[\%]} & \multicolumn{1}{l}{$\geq 2$ $b$-jets} & \multicolumn{1}{c}{[\%]}   \\ \hline
Data & 70854 & & 12437 &  \\ 
Total simulation & 66200 & 100.0 & 12400 & 100.0  \\ \hline
$t\bar t$ & 40300 & 60.8 & 11900 & 96.3  \\
$Wt$ single top & 3840 & 5.8 & 360 & 2.9  \\
$Z(\rightarrow\tau\tau\rightarrow e\mu)$+jets & 12800 & 19.4 & 6 & 0.1  \\
Dibosons & 8030 & 12.3 & 2 & 0.0  \\ 
Misidentified leptons & 1200 & 1.8 & 96 & 0.8  \\ \hline
\end{tabular}
\caption{Selected numbers of events with an opposite-sign $e\mu$ pair, and with
 an opposite-sign $e\mu$ pair and at least two $b$-tagged jets in data, 
compared with the predictions from the baseline simulation, 
broken down into contributions
from \ttbar, $Wt$ and minor background processes. The predictions are 
normalised to the same integrated luminosity as the data.} 
\label{table:eventCounts}
\end{table}
Additional jets were defined as those other than the two $b$-tagged jets used 
to select the event. For the jet normalised differential cross-section 
measurements, in the 3\,\% of selected events with  three or more $b$-tagged 
jets, the jets with the two highest MV1
$b$-tagging weight values were taken to be the $b$-jets from
the top quark decays, and any other $b$-tagged
jets were considered as additional jets, along with all untagged jets. 
Distributions of the number of additional
jets are shown for various jet \pt\ thresholds in Figure~\ref{fig:recomult}. 
The \pt\ distributions for reconstructed additional jets are shown in 
Figure~\ref{fig:recopt}, with the estimated contribution from `unmatched jets' 
(defined in Section~\ref{ss:jetmatch} below) shown separately.
In both cases, the data are shown compared to the predictions from
simulation with the baseline {\sc Powheg\,+\,Pythia6} ($\hdmp=\infty$) 
\ttbar\ sample plus
backgrounds, and the predictions from alternative \ttbar\ simulation samples
generated with {\sc Powheg\,+\,Pythia6} and {\sc Powheg\,+\,Pythia8}
with $\hdmp=\mtop$, {\sc Powheg\,+\,Herwig} with $\hdmp=\infty$ and
{\sc MC@NLO\,+\,Herwig}.
The jet multiplicity distributions and \pt\ spectra in the simulation samples 
are generally in reasonable agreement with those from data, 
except for {\sc MC@NLO\,+\,Herwig}, which 
underestimates the number of events with three or more extra jets, and also
predicts significantly softer jet \pt\ spectra.

\begin{figure}[tp]

\centering
\hspace{-7mm}
\subfloat[][]{

\includegraphics[width=0.51\textwidth]{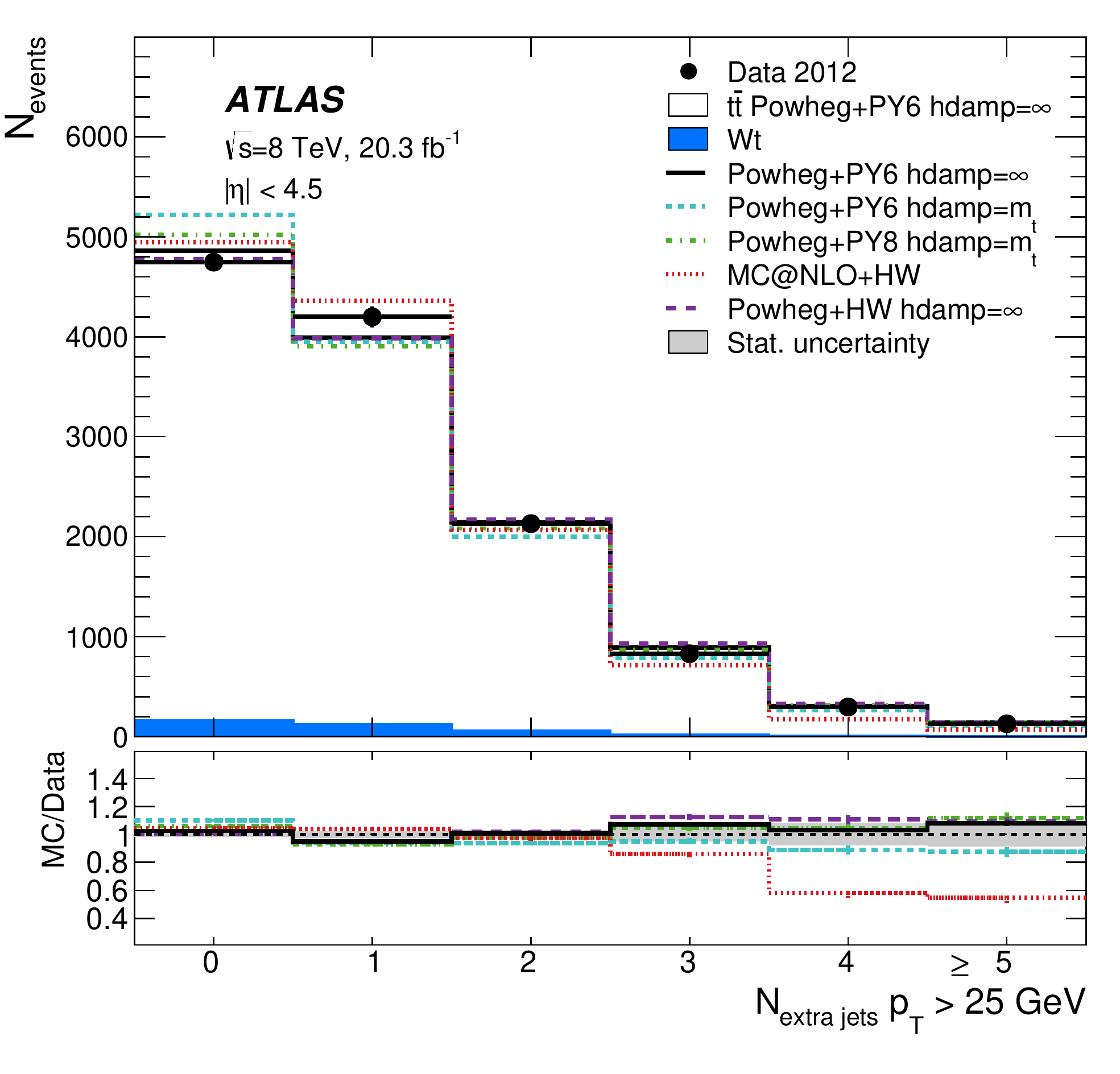}}
\subfloat[][]{

\includegraphics[width=0.51\textwidth]{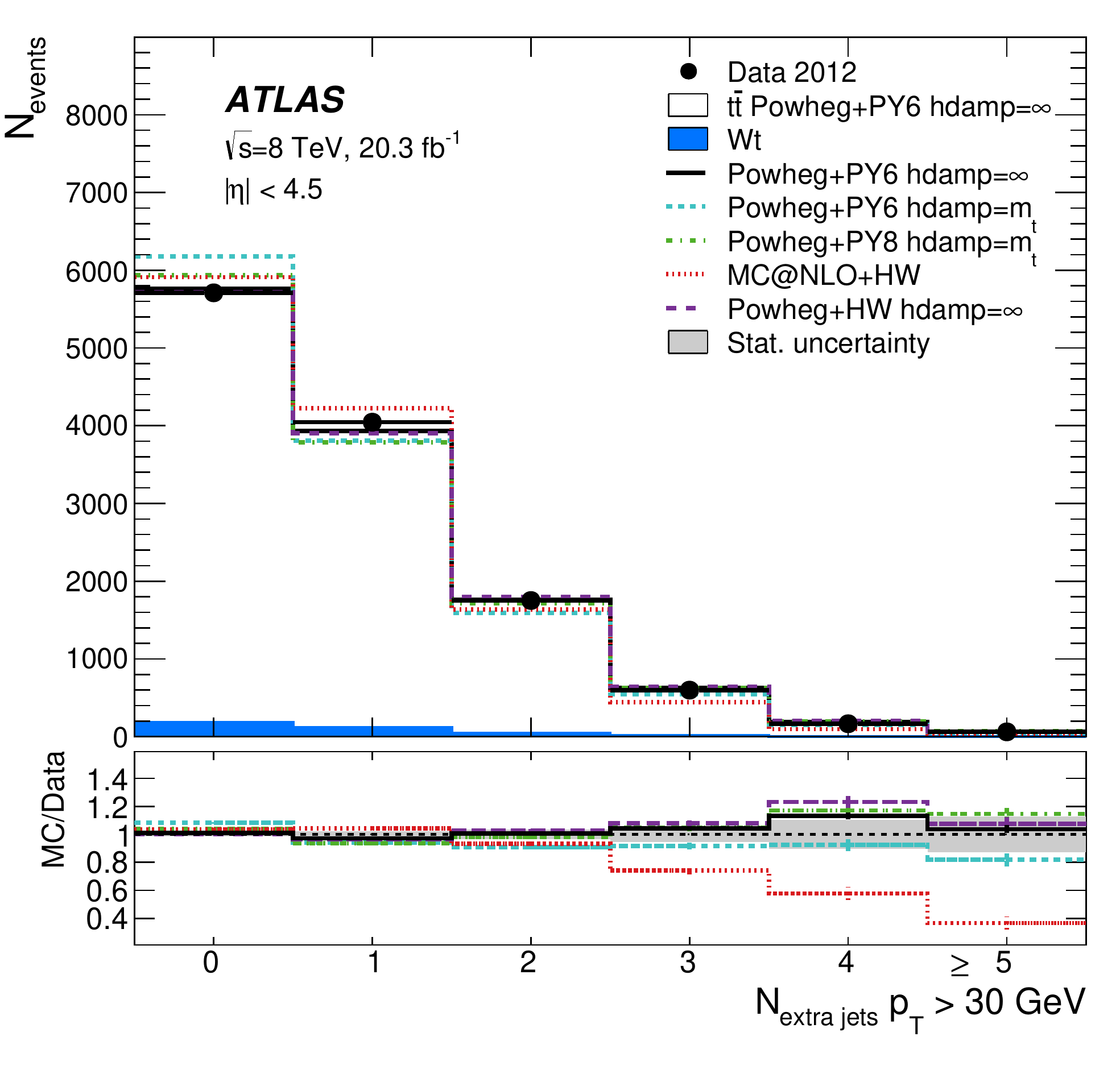}}
\\
\hspace{-7mm}
\subfloat[][]{

\includegraphics[width=0.51\textwidth]{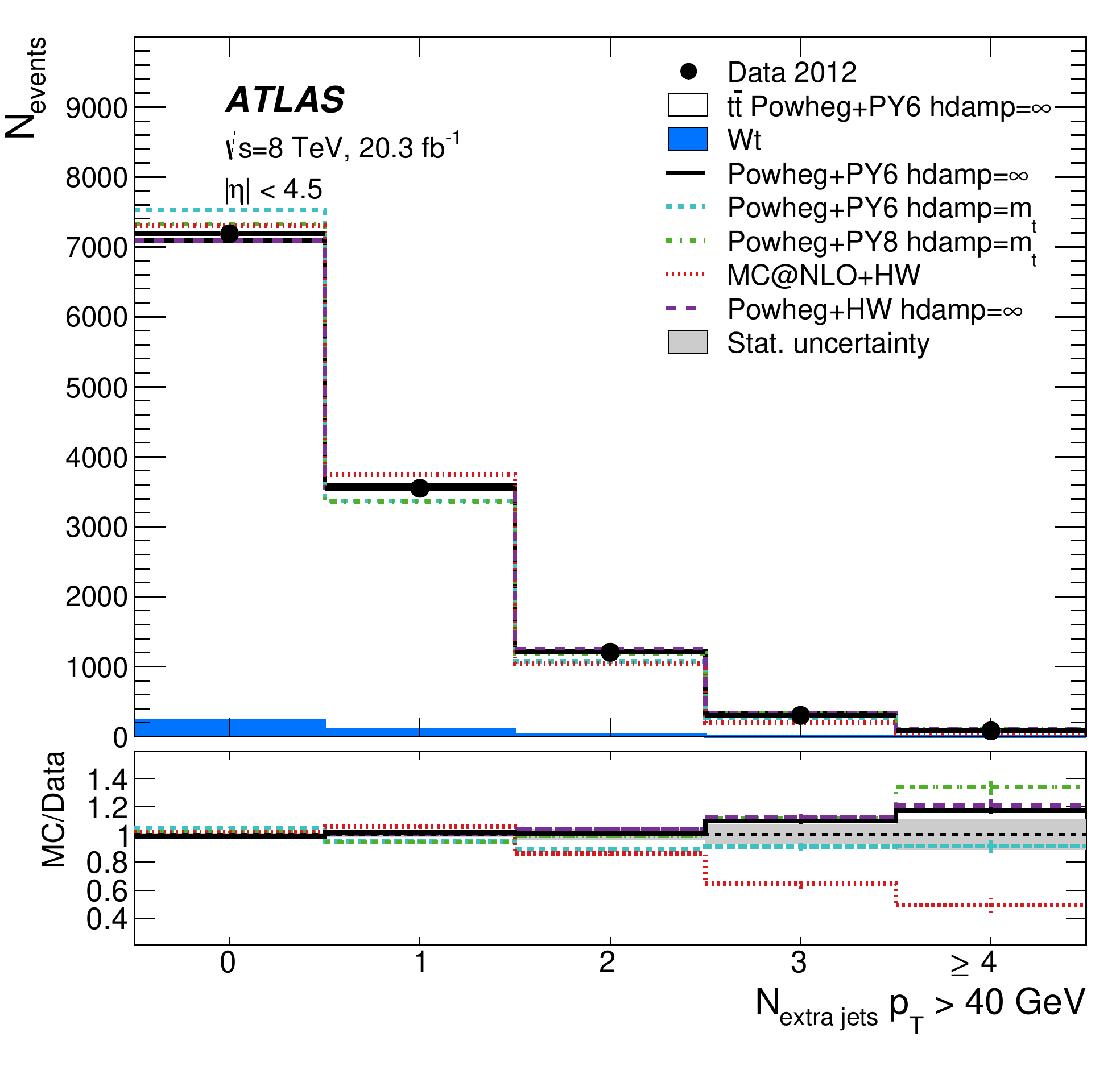}}
\subfloat[][]{

\includegraphics[width=0.51\textwidth]{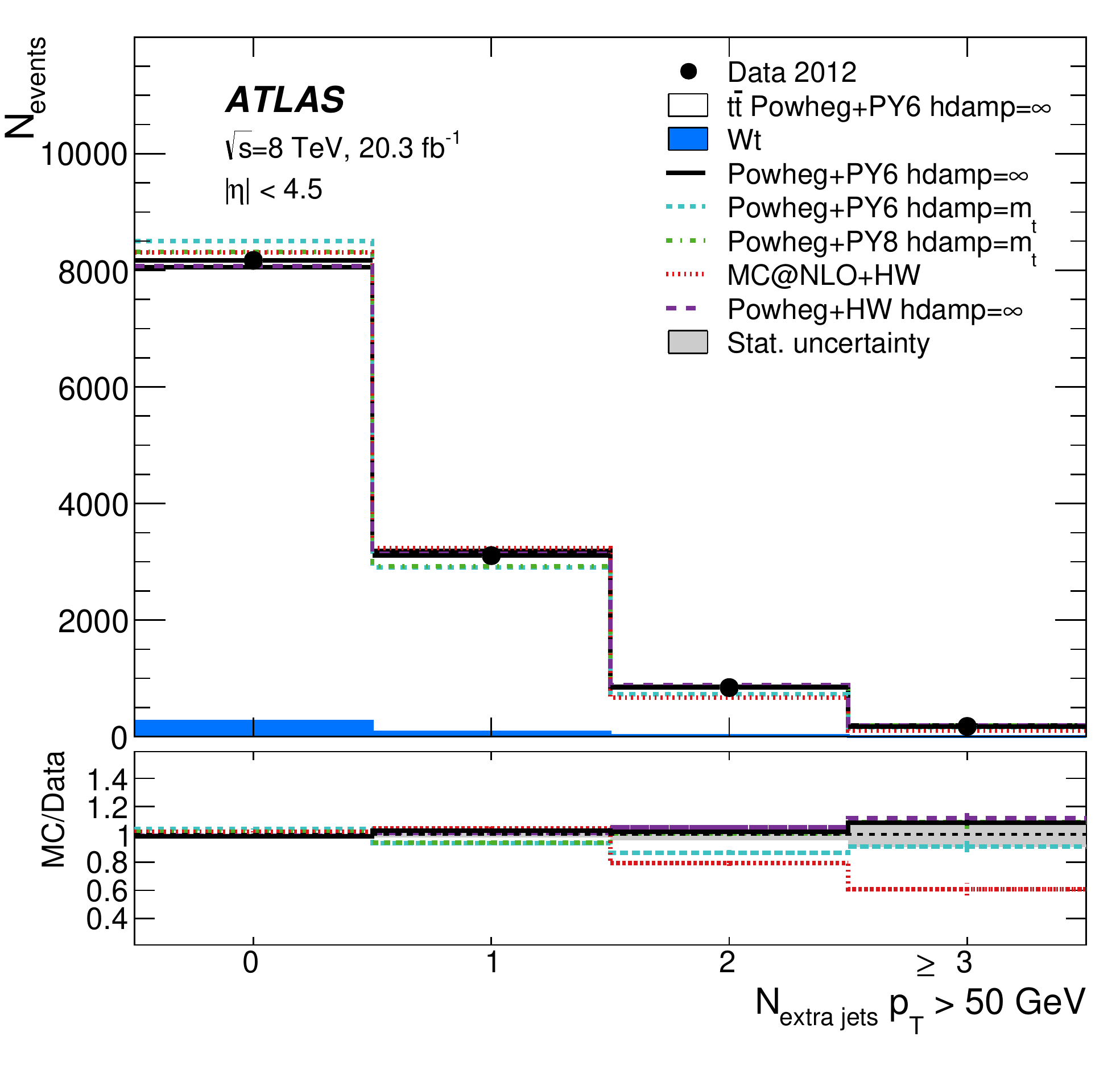}}
\caption{Distributions of the number of reconstructed extra jets with $|\eta|<4.5$ and \pt > (a) 25, (b) 30, (c) 40 and (d) 50 \GeV\ in selected \emubb\ events 
in data and in simulation, normalised to the same number of events as the data.
The simulation predictions for \ttbar\ and $Wt$ single-top production are shown 
separately, and the contributions from other backgrounds are negligible.
The ratios of different MC samples to data are shown with error bars corresponding to the simulation statistical 
uncertainty and a shaded band corresponding to the data 
statistical uncertainty. Systematic uncertainties are not shown.}
\label{fig:recomult}
\end{figure}

\begin{figure}[tp]
\centering
\hspace{-7mm}
\subfloat[][]{
\includegraphics[width=0.51\textwidth]{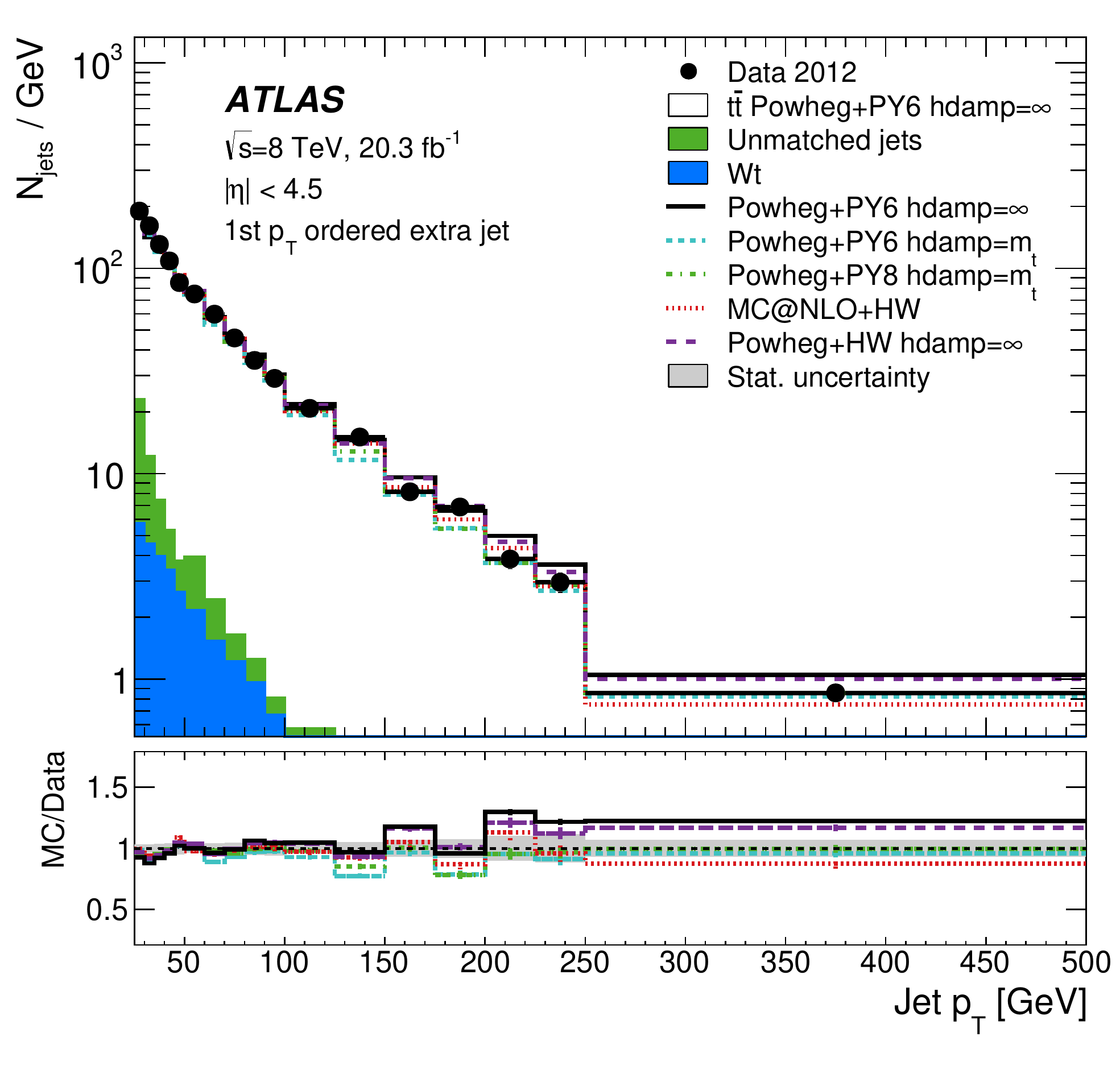}}
\subfloat[][]{
\includegraphics[width=0.51\textwidth]{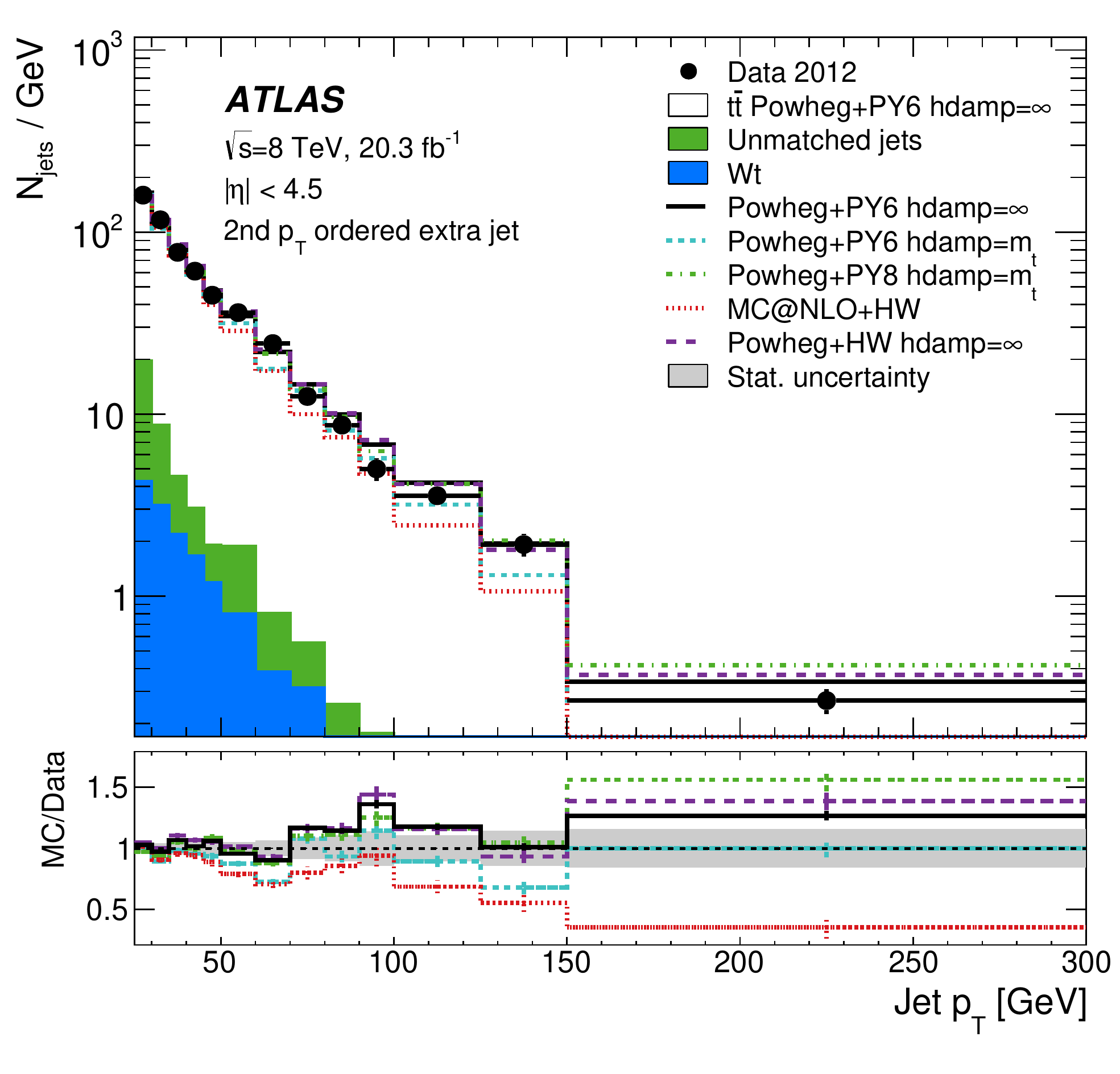}} \\
\hspace{-7mm}
\subfloat[][]{
\includegraphics[width=0.51\textwidth]{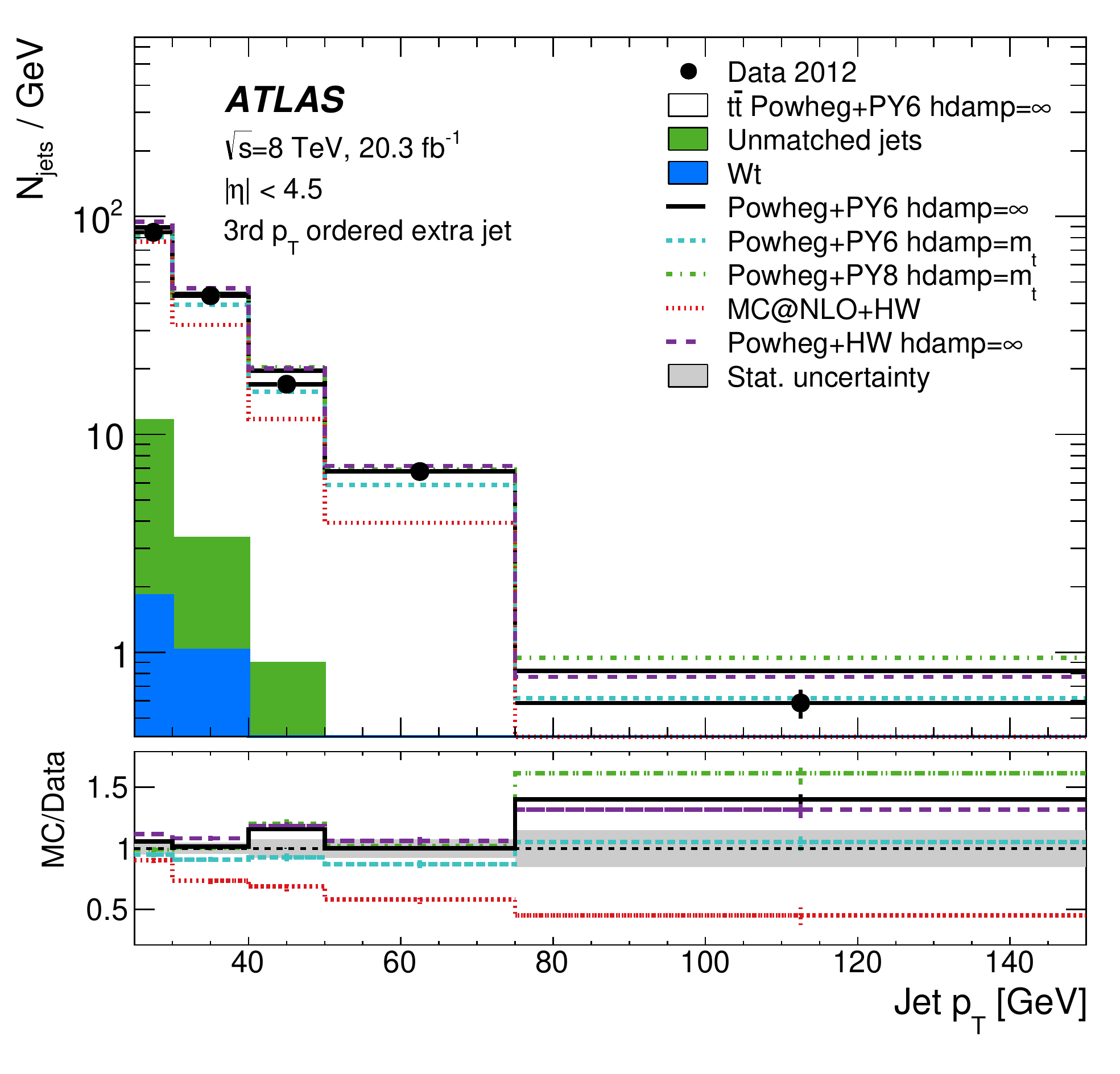}}
\subfloat[][]{
\includegraphics[width=0.51\textwidth]{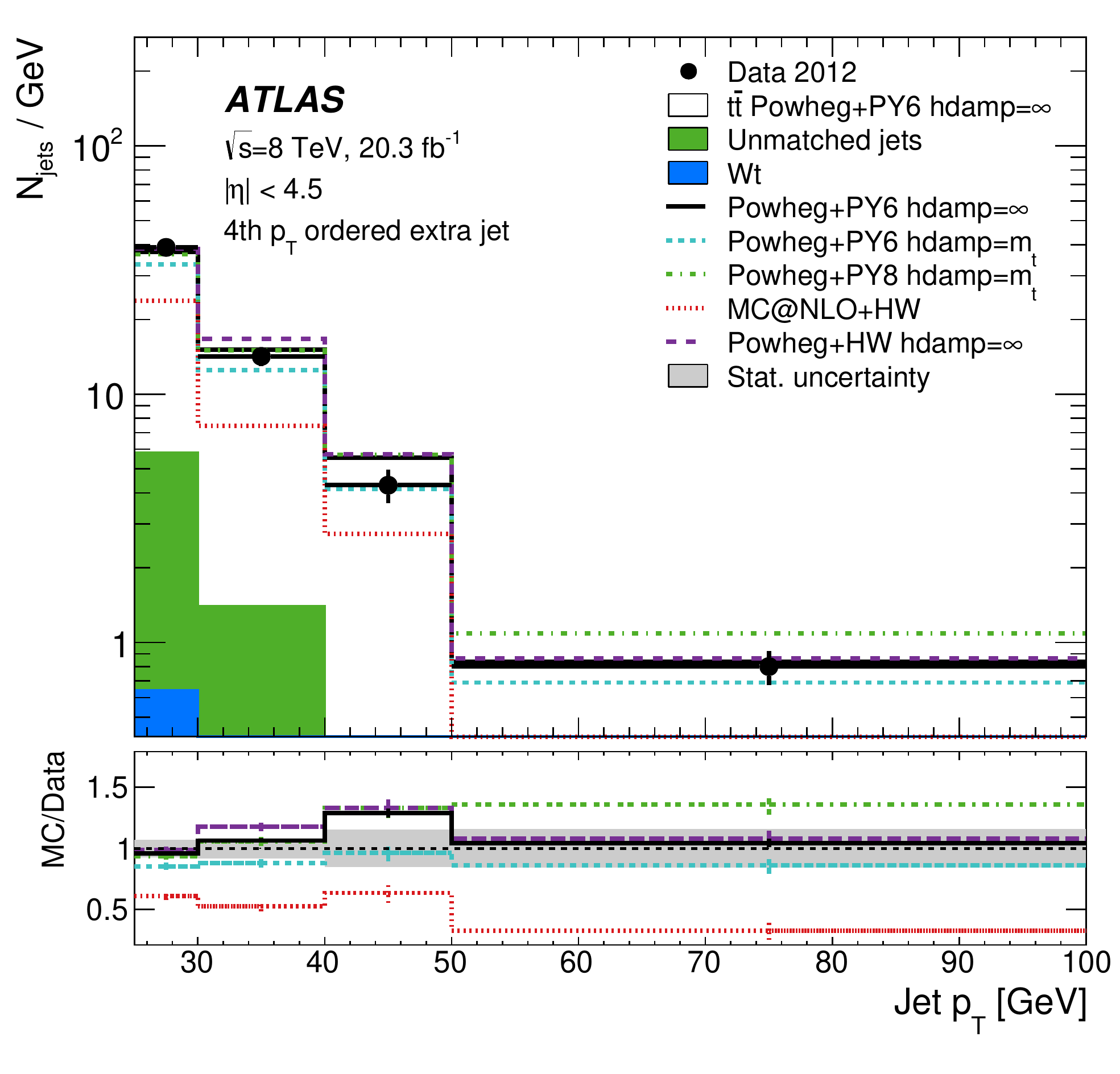}} \\
\caption{Distributions of reconstructed jet \pt\ for the (a) first to (d) fourth additional jet in selected \emubb\ events. The data are compared to simulation 
normalised to the same number of \emubb\ events as the data. Backgrounds from $Wt$ single-top and unmatched jets are estimated using the baseline {\sc Powheg\,+\,Pythia6} samples and shown separately. The contributions from
other backgrounds are negligible. 
The ratios of different MC samples to data are shown with error bars corresponding to the simulation statistical uncertainty and a shaded band corresponding to the data statistical uncertainty. Systematic uncertainties are not shown.}
\label{fig:recopt}
\end{figure}

The gap fraction measurements use the same basic $e\mu\bbbar$ event selection,
but restricting the additional jets to the central rapidity region,
$|y|<2.1$. If three or more jets were $b$-tagged, the two highest-\pt\ jets
were considered as the $b$-jets from the top quark decays, and the others as 
additional jets. This definition follows the \pt-ordered selection used at
particle level, and is different from that used in the 
differential cross-section analysis, as  
discussed in Sections~\ref{ss:partlevel} and~\ref{ss:jetmatch} below.
Distributions of the \pt\ and $|y|$ of the leading 
additional jet according to this definition are shown in 
Figure~\ref{fig:gfrecjet}. The predictions generally describe the data
well, and the trends seen are similar to those seen for
the leading jet over the full rapidity region in Figure~\ref{fig:recopt}(a).

\begin{figure}[tp]
 \centering
 \subfloat[][]{\includegraphics[width=.52\textwidth]{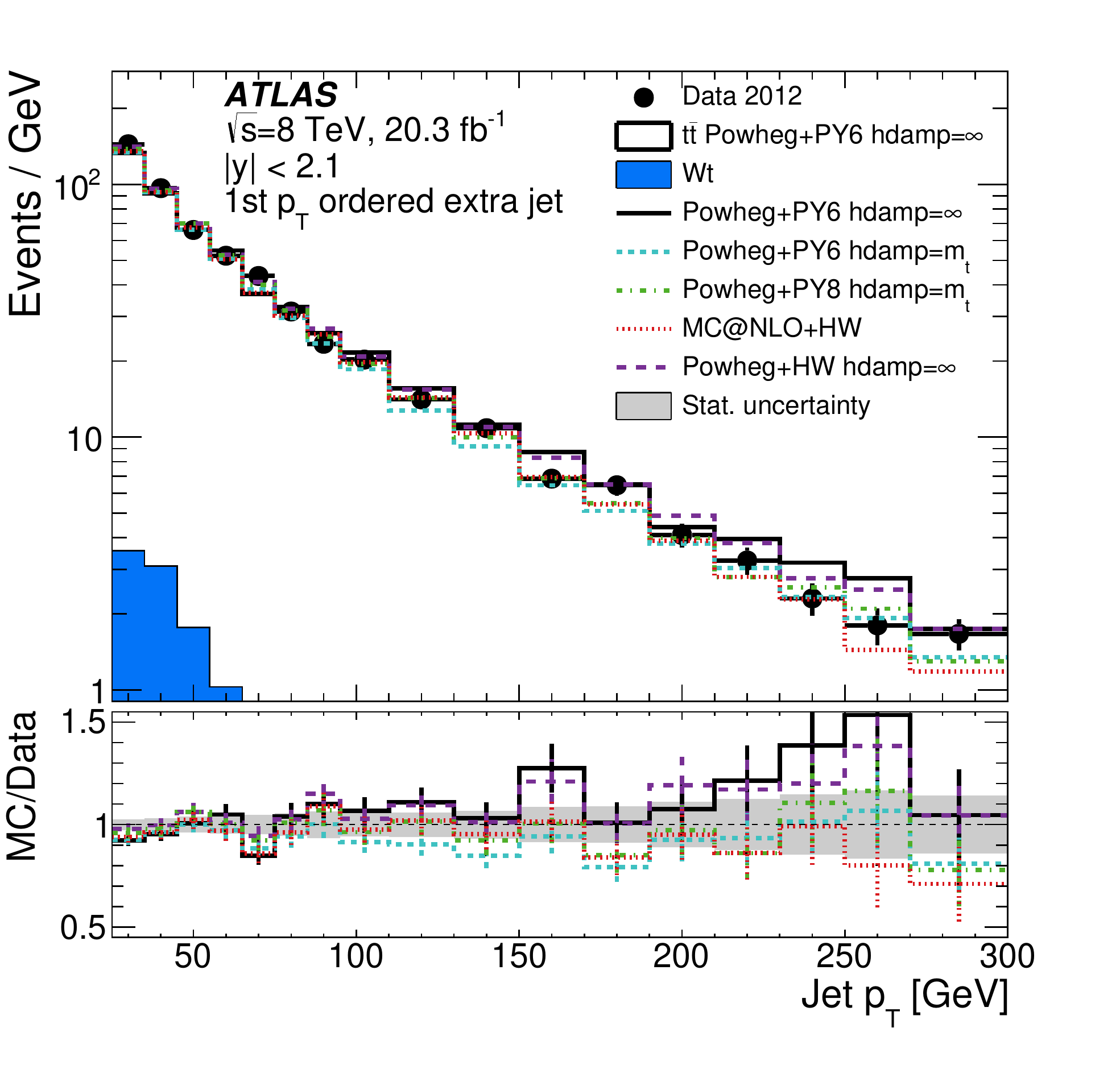}}
 \subfloat[][]{\includegraphics[width=.52\textwidth]{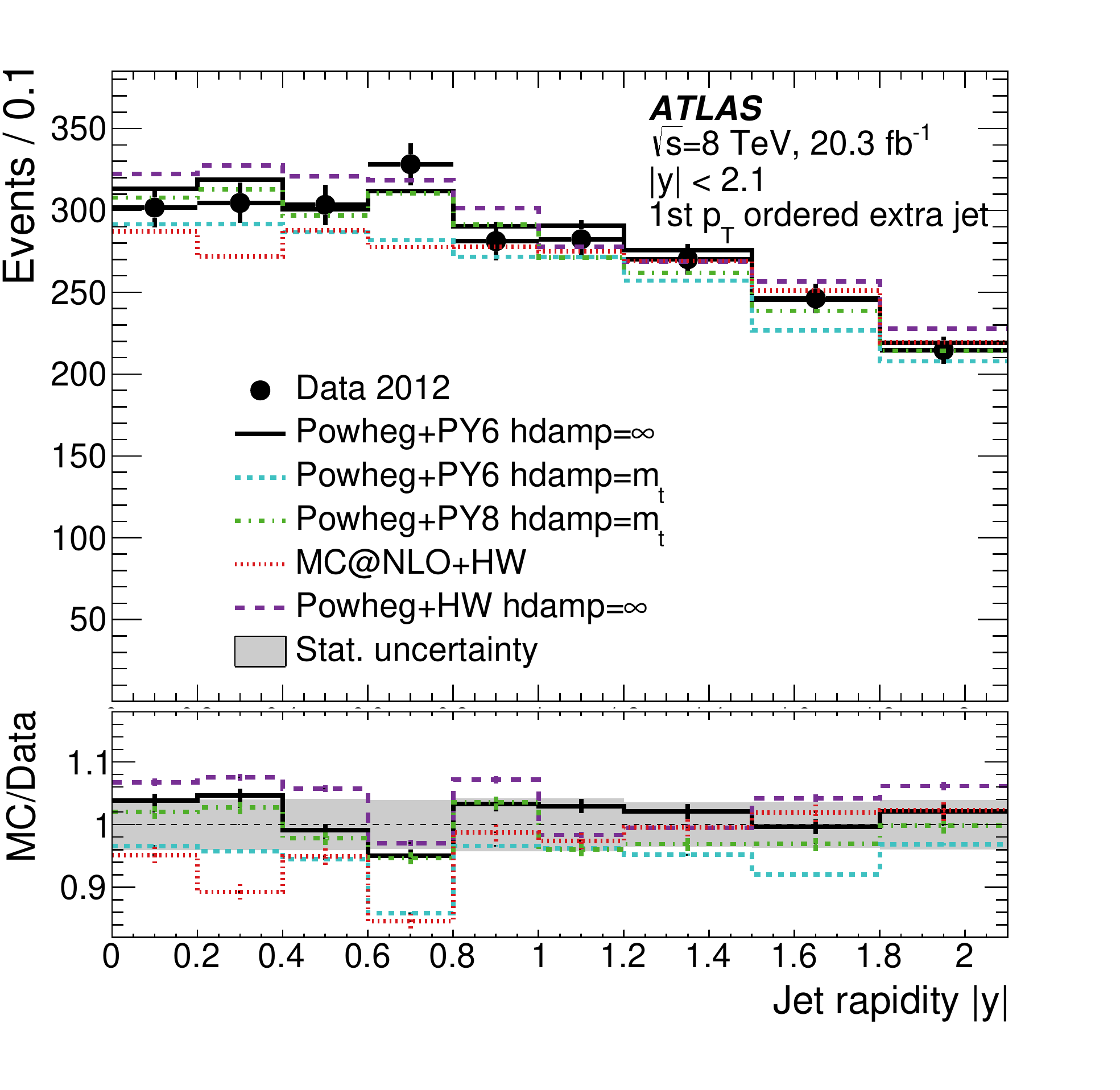}}
 \caption{Distributions of leading additional reconstructed jet (a) \pt\ and 
(b) $|y|$ in \emubb\ events as used in the gap fraction measurement. The data are shown compared to simulation predictions using several \ttbar\ generators, with
the $Wt$ background shown separately (not visible in (b)). 
Other backgrounds are negligible.
The ratios of different MC samples to data are shown with error bars corresponding to the simulation statistical uncertainty and a shaded band corresponding to the data statistical uncertainty. Systematic uncertainties are not shown.}
\label{fig:gfrecjet} 
\end{figure}

\subsection{Particle-level selection}\label{ss:partlevel}

To facilitate comparisons with theoretical predictions, the measured 
jet differential cross-sections and gap fractions were corrected to correspond
to the particle level in simulation, thus removing reconstruction 
efficiency and resolution effects.
At particle level, electrons and muons were defined as those originating from
$W$ decays, including via the leptonic decay of a $\tau$ lepton 
($W\rightarrow\tau\rightarrow e/\mu$). The electron and muon four-momenta were
defined after final-state radiation, and `dressed' 
by adding the four-momenta of all photons within a cone of size
$\Delta R=0.1$ around the lepton direction, excluding photons from
hadron decays or interactions with detector material.
Jets were reconstructed using the anti-$k_t$ algorithm
with radius parameter $R=0.4$ from all final-state particles with
mean lifetime greater than $3\times 10^{-11}$\,s, excluding dressed leptons
and neutrinos not originating from the decays of hadrons. Particles from the
underlying event were included, but those from overlaid pile-up collisions 
were not. Selected jets were required to have $\pt>25$\,\GeV\ and $|\eta|<4.5$,
and those within $\Delta R=0.2$ of a particle-level electron were removed.
Particle-level jets containing $b$-hadrons were identified
using a ghost-matching procedure \cite{ghostmatch}, where the four-momenta
of $b$-hadrons were scaled to a negligible magnitude and included in the set of
particles on which the jet clustering algorithm was run. Jets whose constituents
included $b$-hadrons after this procedure were labelled as $b$-jets.

The particle-level \emubb\ event selection was defined by requiring
one electron and one muon with $\pt>25$\,\GeV\ and $|\eta|<2.5$, each separated
from the nearest jet by $\Delta R>0.4$, and at least
two $b$-jets with $\pt>25$\,\GeV\ and $|\eta|<2.5$.  This closely 
matches the event selection used at reconstruction level.

\subsection{Jet matching}\label{ss:jetmatch}

For the definition of the gap fraction at particle level, if three or more
$b$-jets were found, the two highest-\pt\ jets were considered to be the 
$b$-jets from the top decays, and all other jets were considered to be
additional jets, whether labelled $b$-jets or not. In contrast,
the differential jet cross-section measurements require an explicit jet-by-jet
matching
of particle-level to reconstructed jets. This was achieved by first calculating
the $\Delta R$ between each particle-level jet passing a looser requirement 
of $\pt>10$\,\GeV\ and each
reconstructed $b$-tagged jet, considering the two with highest MV1 weight if 
more than two reconstructed jets were $b$-tagged. Ordering the $b$-tagged
jets by MV1 weight
was found to give a greater fraction of correct matches than the jet \pt\ 
ordering used for the gap fraction measurements, where no jet matching 
is needed.
If the closest reconstructed
$b$-tagged jet was within $\Delta R<0.4$, the particle-level and reconstructed
jets were considered matched. The procedure was then repeated with the remaining
particle-level and reconstructed jets, allowing each particle-level
and reconstructed jet to be matched only once. Reconstructed jets which remained
unassociated with particle-level jets after this procedure are referred to
as `unmatched' jets; these originate from single particle-level jets which 
are split in two at reconstruction level (only one of which is matched), and
from pile-up (since particles from pile-up collisions are not considered in the
particle-level jet clustering). The contributions from such unmatched jets
are shown separately in Figure~\ref{fig:recopt}.

\FloatBarrier
\section{Evaluation of systematic uncertainties}
\label{sec:systematics}

Monte Carlo simulation was used to determine selection
efficiencies, detector resolution effects and backgrounds. The 
corresponding systematic uncertainties were evaluated as discussed in
detail below, and propagated through the jet differential cross-section and
gap fraction measurements. 

\begin{description}

\item[\ttbar\ modelling:] Although the analyses measure the properties of
additional jets in \ttbar\ events, they are still slightly sensitive to the
modelling of such jets in simulation due to the finite jet energy resolution
and reconstruction efficiency, as well as the modelling of other \ttbar\
event properties related to the leptons and $b$-jets from the top quark decays.
The corresponding uncertainties were
assessed by comparing samples from the  different generator configurations 
described in Section~\ref{sec:simulation}. In
the differential cross-section measurement, which is sensitive
to the modelling of multiple additional jets, the 
uncertainty due to the choice of matrix-element generator was determined by
comparing the NLO generator {\sc Powheg} with the leading-order
multi-leg generator {\sc MadGraph}, both interfaced to {\sc Pythia6}. In the
gap fraction measurements, which are more sensitive to an accurate modelling
of the first additional jet, the corresponding uncertainty was assessed by 
comparing the NLO generators {\sc Powheg} and {\sc MC@NLO}, 
both interfaced to {\sc Herwig}. 
The choice of parton shower and hadronisation model was studied for both
analyses by comparing samples with {\sc Powheg} interfaced either to 
{\sc Pythia6} or to {\sc Herwig}. In all these cases, 
the full difference between the
predictions from the two compared samples was assigned as the corresponding
systematic uncertainty.
The uncertainty due to the modelling of additional radiation was calculated
as half the difference between the results using 
{\sc MadGraph\,+\,Pythia6} (differential cross-section) or  {\sc Alpgen\,+\,Pythia6}
(gap fraction) samples with tunes giving more or less parton shower radiation, 
spanning the results from the $\sqrt{s}=7$\,\TeV\ gap fraction measurement
\cite{TOPQ-2011-21}. These three systematic components were added in quadrature
to give the total \ttbar\ modelling uncertainty.

\item[Simulation statistical uncertainty:] In addition to the 
modelling uncertainties discussed above, the size of the \ttbar\
simulation samples was also taken into account.

\item[Parton distribution functions:] The uncertainties due to limited 
knowledge of the proton PDFs were evaluated by reweighting the 
{\sc MC@NLO\,+\,Herwig} simulated \ttbar\ sample based on the $x$ and 
$Q^2$ values of the partons participating in the hard scattering in each 
event. The samples were reweighted using the eigenvector variations of the 
CT10 \cite{cttenpdf},
MSTW2008 \cite{mstwnnlo} and NNPDF 2.3 \cite{nnpdfffn} NLO PDF sets.
The final uncertainty was calculated as half the envelope encompassing the
predictions from all three PDF sets along with their associated uncertainties,
following the PDF4LHC recommendations \cite{pdflhc}.

\item[Jet energy scale:] The uncertainty due to the jet energy scale (JES) 
was evaluated by varying it in simulation using a model with 23 separate 
orthogonal uncertainty components \cite{PERF-2012-01}. 
These components cover in situ measurement uncertainties,
the cross-calibration of different $\eta$ regions, and the dependence
on pile-up and the flavour of the jets.
The total jet energy scale uncertainty varies in the range
1--6\,\% with a dependence on both jet \pt\ and $|\eta|$.


\item[Jet energy resolution/efficiency:]  The jet energy resolution (JER)
was found to be
well-modelled in simulation \cite{PERF-2011-04}, and residual uncertainties
were assessed by applying additional smearing to the simulated jet energies.
The calorimeter jet reconstruction 
efficiency was measured in data using track-based jets, and found 
to be generally well-described by the simulation. Residual uncertainties
were assessed by discarding 2\,\% of jets with $\pt<30$\,\GeV; the 
uncertainties for higher-momentum jets are negligible. Both these uncertainties were symmetrised about the nominal value. The uncertainty due to the veto
on events failing jet quality requirements is negligible.

\item[Unmatched jets modelling:] The modelling of the component of unmatched
jets from pile-up collisions was checked by comparing the predictions
from simulated \ttbar\ events combined with either 
{\sc Powheg+Pythia8} pile-up simulation or `zero-bias' data. The latter were
recorded from randomly triggered bunch crossings throughout the data-taking 
period, and reweighted to match the instantaneous luminosity 
distribution in the simulated \ttbar\ sample. 
The estimated number of additional jets per event from pile-up is 
$0.017 \pm 0.002$ in
the central region used by the gap fraction measurements ($|y|<2.1$) and
$0.038 \pm 0.005$ over the full region used by the differential cross-section
measurements ($|\eta|<4.5$). The uncertainties
represent the full difference between the rate in zero-bias data and
simulation. The rate of unmatched jets in simulation was varied by these 
uncertainties in order to determine the effect on the results. 
In the differential cross-section measurements, the full rate of particle-level 
jets that were split in two at reconstruction level in the baseline simulation 
was taken as an additional uncertainty on the rate of unmatched jets.

\item[Jet vertex fraction:] In both measurements, the contribution of jets
from pile-up within $|\eta|<2.4$ was reduced by the JVF requirement 
described in Section~\ref{sec:selection}.
The uncertainties in the efficiency on non-pile-up jets of the JVF requirement  
were assessed by varying the cut value in simulation,
based on studies of $Z\rightarrow ee$ and $Z\rightarrow\mu\mu$ events
\cite{ATLAS-CONF-2013-083}. 

\item[Other detector uncertainties:] The modelling of the electron
and muon trigger and identification efficiencies, energy scales and resolutions 
were studied using $Z\rightarrow ee/\mu\mu$, $J/\psi\rightarrow ee/\mu\mu$ and
$W\rightarrow e\nu$ events in data and simulation, using the techniques
described in Refs. \cite{PERF-2013-03,PERF-2013-05,PERF-2014-05}. The 
uncertainties in the efficiencies for $b$-tagging $b$, $c$ and light-flavour
jets were assessed using studies of $b$-jets containing muons,
jets containing $D^*$ mesons, 
and inclusive jet events \cite{PERF-2012-04}.
The resulting
uncertainties in the measured normalised differential jet distributions 
and gap fractions are very small, since these uncertainties typically
affect the numerators and denominators in a similar way.

\item[Backgrounds:] As shown in Table~\ref{table:eventCounts}, the most
significant background comes from $Wt$ single-top events.
The uncertainty due to this background was assessed by conservatively
doubling and removing the estimated $Wt$ contribution, taking half the 
difference in the result between these extreme variations.
The sensitivity to the modelling of  $Wt$ single-top events
was also assessed by using a sample simulated with {\sc Powheg\,+\,Pythia6} 
using the diagram subtraction scheme \cite{wtinter1,wtinter2} instead of the
 baseline diagram removal scheme. The uncertainty due to $Z$+jets
and diboson background is negligible in comparison.
In the gap fraction measurements, the additional background uncertainty from
events with a misidentified lepton was also assessed by doubling and
removing it, a conservative range according to the studies of 
Ref.~\cite{TOPQ-2013-04}. In the jet differential cross-section measurements,
the misidentification of jets as leptons induces migration in the 
additional-jet rank distributions, and is corrected for as 
part of the unfolding procedure. The resulting effects on the unfolding
corrections are significantly smaller than the uncertainties from 
considering different \ttbar\ generators, and no additional uncertainty
was included.

\end{description}

Each independent uncertainty was evaluated according to the prescription above and then added in quadrature to obtain the total systematic uncertainty in the final measurements. Since both measurements are effectively ratios of 
cross-sections, normalised to the total number of selected $e\mu\bbbar$ events, 
many of the systematic 
uncertainties that typically contribute to a \ttbar\ cross-section measurement 
cancel, such as those in the integrated luminosity, lepton trigger and 
identification efficiencies, lepton momentum scales and resolution, 
and $b$-jet energy scale and tagging efficiency.
Instead, the significant systematic uncertainties are those
 that directly affect the measured additional-jet activity, {\em i.e.}\
systematic uncertainties in the jet energy scale and resolution, and the 
modelling of unmatched jets.


\FloatBarrier
\section{Measurement of jet multiplicities and \pt\ spectra}
\label{sec:jets}

The normalised differential cross-sections for additional jets, corrected to
the particle level, were measured as a function of jet multiplicity and \pt\ 
as defined in Equation~(\ref{e:diffjet}).
The fiducial requirements for event and object selection are defined in 
Section~\ref{ss:partlevel}, and include additional jets in the 
range $|\eta|<4.5$. As discussed in Section~\ref{sec:simulation}, the 
fiducial region receives contributions from both the \ttbar\ and $Wt$ processes.
Although the requirement for two $b$-tagged jets ensures that \ttbar\ is 
dominant,
once the $Wt$ process is considered at NLO, the two processes cannot 
in principle be cleanly separated. Therefore the results
are presented both with the $Wt$ contribution subtracted, to allow comparison
with the \ttbar\ generators discussed in Section~\ref{sec:simulation},
and for the combined $\ttbar+Wt$ final state, which may be compared with
future NLO calculations treating \ttbar\ and $Wt$ concurrently. In practice,
since the results are normalised to the number of selected $e\mu\bbbar$ events,
from \ttbar\ or $\ttbar+Wt$ as appropriate
in each case, and the predicted additional-jet distributions in 
simulated \ttbar\ and $Wt$ events are rather similar, the results from the two
definitions are very close.

\subsection{Correction to particle level}

The correction procedure transforms the measured spectra 
shown in Figure~\ref{fig:recopt}, after background subtraction, to the 
particle-level spectra for events that pass the fiducial requirements. 
The unfolding was performed using a one-dimensional distribution encoding both
the rank and \pt\ of each additional jet in each selected \emubb\ event, as
shown in Table~\ref{t:diffjet} and graphically in Figure~\ref{f:res}. 
The integral of the input (measured)
distribution is the number of measured jets in the \emubb\ sample and the 
integral of the output (unfolded) distribution is the number
of particle-level jets passing the fiducial requirements. 
This procedure involves several steps, as defined in the equation:
\begin{equation}
\sigmapti = \frac{1}{N_{\textrm{events}}}\frac{1}{\Delta^{k}}\; f^k \sum_j \left ({\bf M}^{-1}\right )_{\textrm{reco, }j}^{\textrm{unfolded, }k} g^j \left ({\mathscr N}^j_{\textrm {reco}}-
{\mathscr N}^j_{\textrm{bkgd}} \right ) \ .
\label{eqn:unffinal}
\end{equation}
Here, the bin indices $j$ and $k$ are functions of both jet \pt\ and rank, 
with $k$ corresponding to the appropriate \pt\ bin of the jet of rank $i$ 
at particle level under consideration.
The expression \sigmapti\ represents the measured differential 
cross-section, {\em i.e.} the final 
number of corrected jets per event in each bin divided by $\Delta^k$, 
the width of the \pt\ bin in units of \GeV. 
The number of events in data passing the \emubb\ selection requirements is 
represented by $N_{\textrm{events}}$.
The raw data event count reconstructed in bin $j$ is represented by 
${\mathscr N}^j_{\textrm {reco}}$.
The estimated additional-jet background, ${\mathscr N}^j_{\textrm{bkgd}}$, is subtracted from this raw distribution. 
The factor $g^j$ corrects for migration across the fiducial boundaries in
\pt\ and $\eta$
({\em e.g.} cases where the reconstructed jet has
$\pT>25$\,\GeV\ but the particle-level jet has $\pT<25$\,\GeV). 
The expression $\left ({{\bf M}}^{-1}\right )_{\textrm{reco, }j}^{\textrm{unfolded, }k}$ represents the application of an unfolding procedure mapping the number of jets reconstructed in bin $j$ to the number of jets in bin $k$ at particle
level in events which pass both the reconstruction- and particle-level 
selections.  The correction factor $f^k$
removes the bias in the unfolded additional-jet spectrum coming from
the reconstruction-level selection, as discussed further below.

\begin{figure}[htp]
\hspace{-8mm}
\subfloat[][]{
\includegraphics[width=0.55\textwidth]{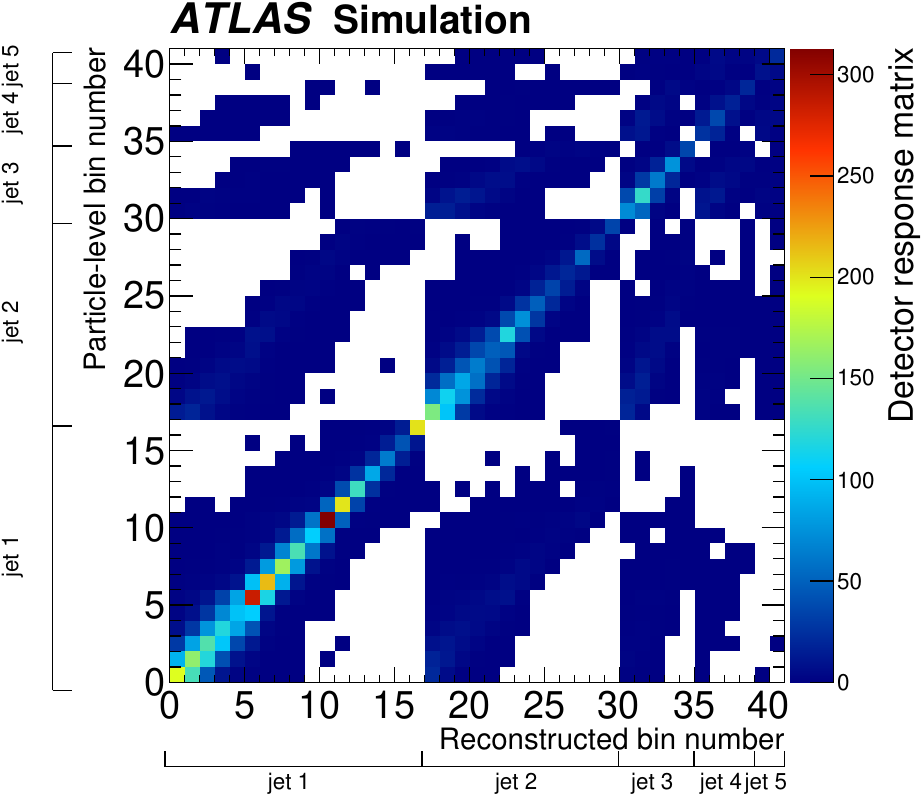}}
~
\subfloat[][]{
\includegraphics[width=0.497\textwidth]{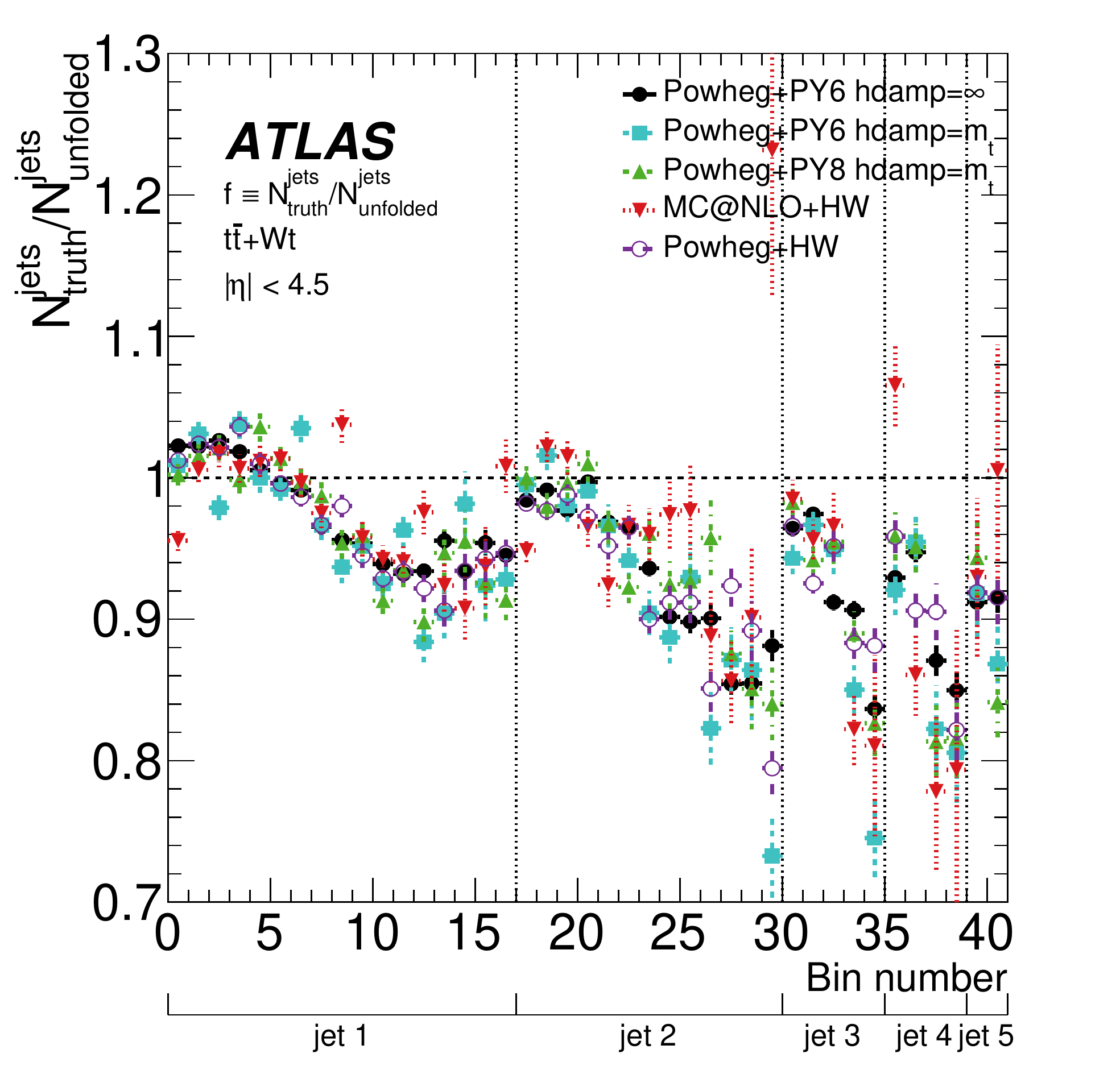}}
~
\caption{(a) Migration matrix between the particle-level and reconstructed number of additional jets in each bin, determined from the baseline $\ttbar+Wt$ 
simulation. Jets are binned according to both \pt\ value 
and rank; 
(b) bin-by-bin correction factor $f^{i}$ for the bias due to the
\emubb\ event selection, evaluated using both the baseline 
{\sc Powheg\,+\,Pythia6} sample and various alternatives.}
\label{f:res}
\end{figure}

The response matrix ${\bf M}_{\textrm{reco, }j}^{\textrm{unfolded, }k}$ encodes the
fractions of jets in particle-level bin $k$ which get reconstructed in bin
$j$, with both $k$ and $j$ being obtained from the corresponding jet \pt\ and
rank. The matrix is 
filled from simulated events that pass both the reconstructed and 
particle-level selection requirements.
Figure~\ref{f:res}(a) provides a graphical representation of ${\bf M}_{\textrm{reco, }j}^{\textrm{unfolded, }k}$. 
The matrix is largely diagonal, showing that jets are most likely to be 
reconstructed with the correct \pt\ and rank.
However, there are significant numbers of particle-level subleading jets 
reconstructed as leading jets and 
particle-level leading jets reconstructed as subleading jets, particularly
when several jets in the event have similar low \pt\ values.
This type of migration motivates the simultaneous binning in both rank and \pt. 

A Bayesian iterative unfolding method \cite{D'Agostini:1994zf}
implemented in the RooUnfold \cite{RooUnfold} software package was used. 
The response matrix ${\bf M}$ is not unitary because in mapping from
particle to reconstruction level, some events and objects are lost due to 
inefficiencies
and some are gained due to misreconstruction or migration of 
objects from outside the fiducial acceptance into the reconstructed
distribution.  This results in the response matrix being almost singular,
and it is therefore not possible to obtain stable unfolded results
by inverting the response matrix and applying it to the measured data.
Instead, an assumed particle-level distribution (the `prior')
was chosen, the response matrix  applied and the resulting trial reconstruction set was compared to the observed reconstruction set. A new prior was then 
constructed from the old prior
and the difference between the trial and the observed distributions.
The procedure was iterated until the result became stable.
For this analysis, two iterations were found to be sufficient, based 
on studies of the unfolding performance in simulated samples 
with reweighted jet \pt\ distributions and from different generators.

This unfolding procedure gives unbiased additional-jet distributions for
events passing both the particle-level and reconstruction-level event
selections. However, the reconstruction-level 
selection results in the unfolded distributions differing from
those obtained using the particle-level selection alone.
An additional  contribution to the bias results from events where one of the 
two reconstructed $b$-tagged jets is actually a mistagged light jet.
These biases were corrected using a bin-by-bin correction factor
$f^k={\mathscr N}^k_{\textrm{truth}}/{\mathscr N}^k_{\textrm{unfolded}}$, where 
${\mathscr N}^k_{\textrm{truth}}$ is the number of jets
in bin $k$ at particle level without the application of the reconstruction-level
event selection. The correction was applied after the unfolding, as shown in 
Equation~(\ref{eqn:unffinal}).
Figure~\ref{f:res}(b) shows the values of 
$f$ for both the baseline and some  alternative \ttbar\ generators. 
The corresponding systematic uncertainty was assessed as part of the
\ttbar\ modelling uncertainty as discussed in Section~\ref{sec:systematics}.

The procedure described above provides the absolute numbers of additional jets
in the number of events passing
the \emubb\ fiducial requirements ($N_{\textrm{events}}$).
This result was then normalised relative to $N_{\textrm{events}}$ to obtain the final distribution \sigmapti, which was finally integrated over jet \pt\ to 
obtain the jet multiplicity distributions.

\subsection{Determination of systematic uncertainties}
\label{ss:sysjets}

All systematic uncertainties were evaluated as full covariance matrices
including bin-to-bin correlations. The majority of uncertainties from 
Section~\ref{sec:systematics} are defined in terms of an RMS width,
with the assumption that the true distribution is Gaussian with a mean at
the nominal value. In these cases, the covariance matrix was calculated from 
pseudo-experiments drawn from this distribution. Each pseudo-experiment was
 constructed by choosing the size of the systematic uncertainty 
randomly according to a
Gaussian distribution, calculating the resulting effect at the 
reconstruction level and propagating it through the unfolding procedure.
The covariance was then given by
\begin{equation}
C_{ij} \equiv \frac{1}{N_{\textrm{pseudo}}} \sum_{x=1}^{N_{\textrm{pseudo}}} \left({\mathscr N}_x^i- \left \langle{\mathscr N}^i \right \rangle \right) \left({\mathscr N}_x^j- \left \langle{\mathscr N}^j \right \rangle \right) ,
\label{eqn:cov}
\end{equation}
where $N_{\textrm{pseudo}}$ is the number of pseudo-experiments (typically 1000), $\langle{\mathscr N}^i \rangle$ is the nominal number of jets in bin $i$, and  ${\mathscr N}_x^i$ is the number of jets in bin $i$ for pseudo-experiment $x$. Some systematic uncertainties were evaluated by comparing an alternative model 
to the baseline. In these cases, the covariance was approximated by
\begin{equation}
C_{ij} \equiv \delta_i \delta_j
\label{eqn:biascov} ,
\end{equation}
where $\delta_i$ is the bias in bin $i$. This bias was determined by analysing 
the alternative model using Equation~(\ref{eqn:unffinal}), with the response 
matrix and correction factors taken from the baseline.

The uncertainties calculated using Equation~(\ref{eqn:cov}) 
include all detector modelling effects ({\em e.g.}\ jet energy scale and 
resolution), PDFs, the
$Wt$ cross-section and statistical uncertainties associated with the simulated 
samples. Uncertainties evaluated using Equation~(\ref{eqn:biascov}) include
generator, radiation, parton shower and hadronisation contributions to the
\ttbar\ modelling uncertainty, and modelling of the unmatched jet background.
Figure~\ref{fig:sys} shows the fractional uncertainties in the corrected jet 
distributions. In most bins, the statistical uncertainty dominates, with the
largest systematic uncertainty coming from the jet energy scale. 

\begin{figure}[htp]
\centering
\subfloat[][]{
\includegraphics[width=0.5\textwidth]{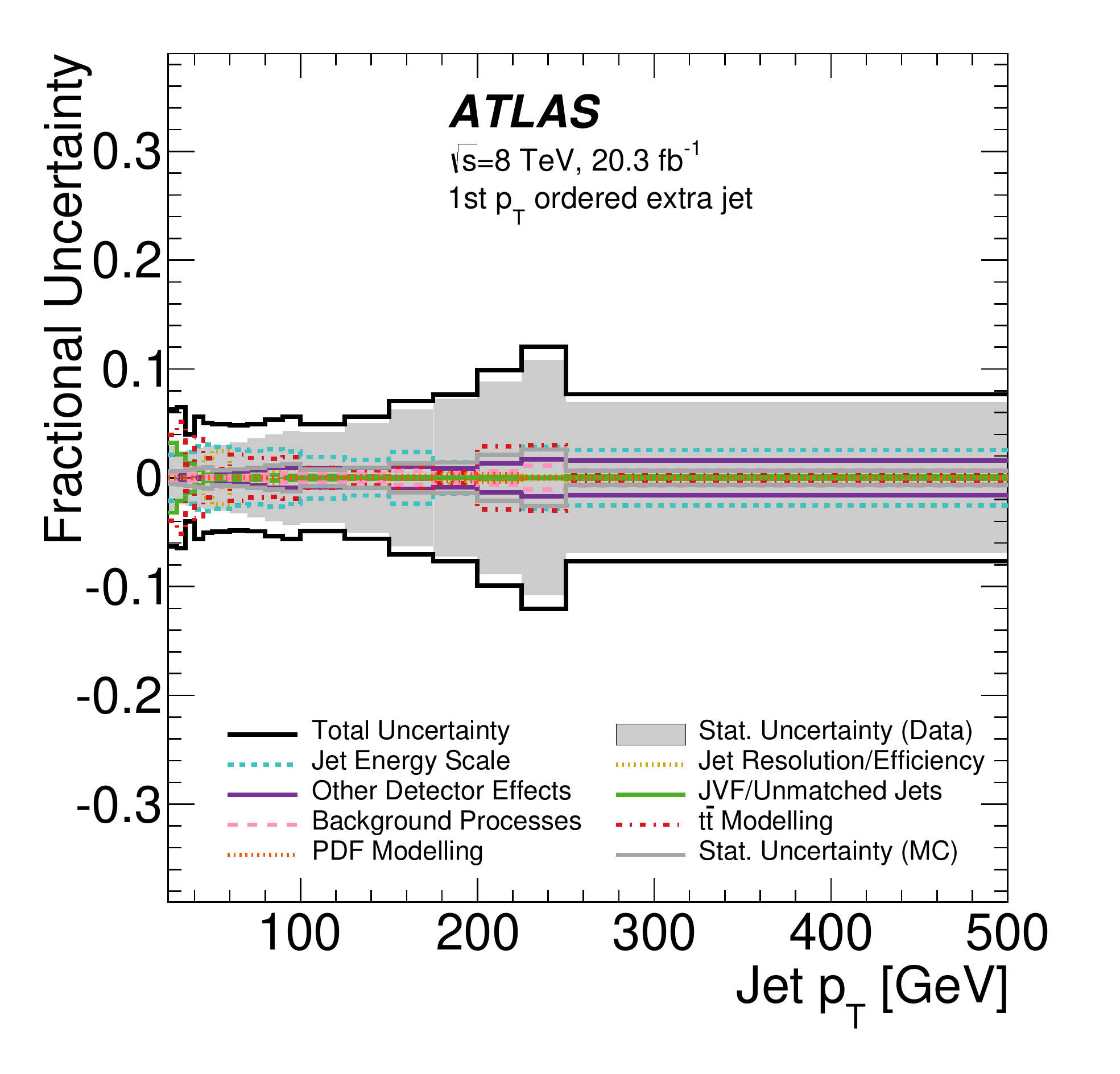}}
\subfloat[][]{
\includegraphics[width=0.5\textwidth]{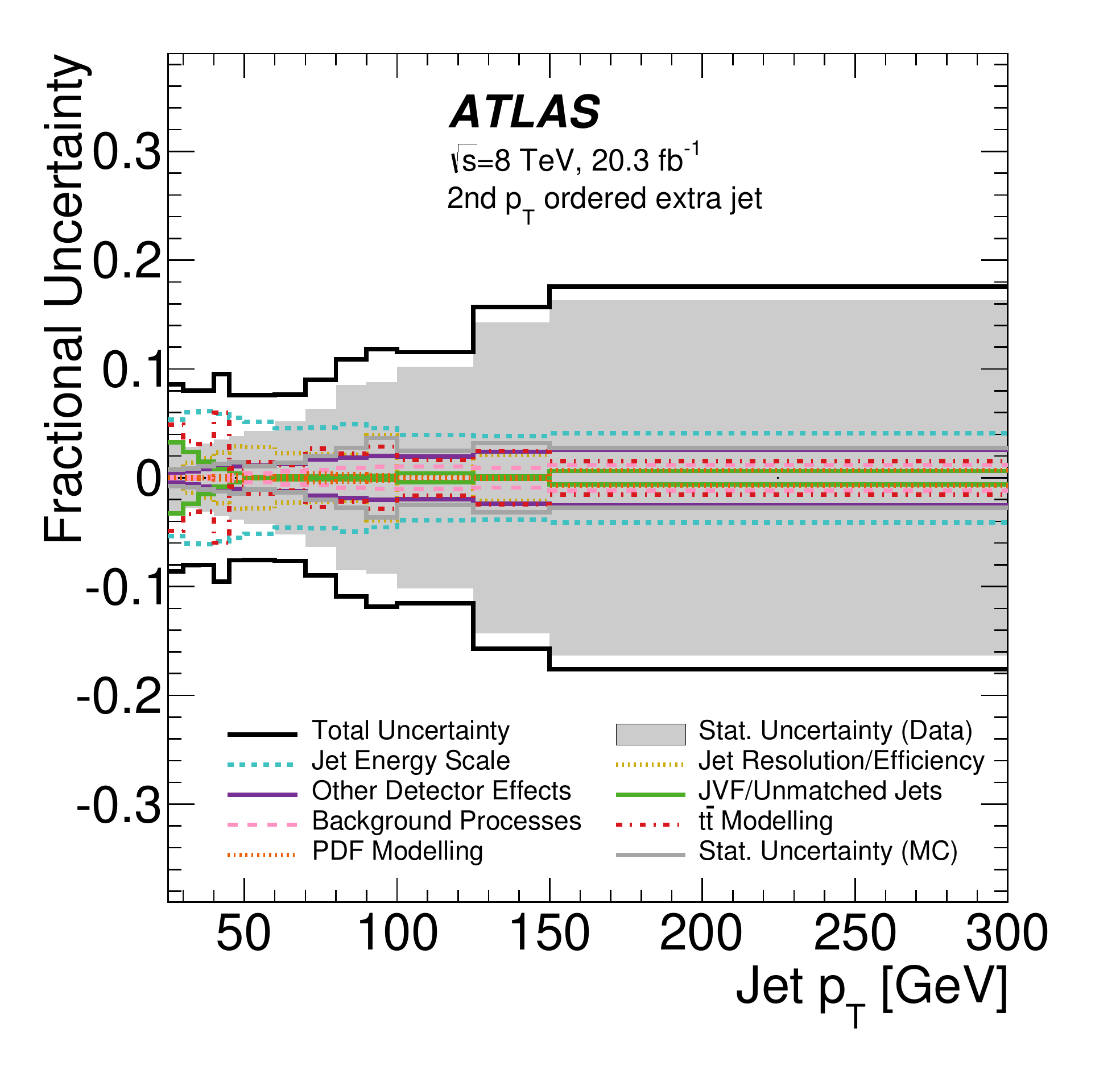}} \\
\subfloat[][]{
\includegraphics[width=0.5\textwidth]{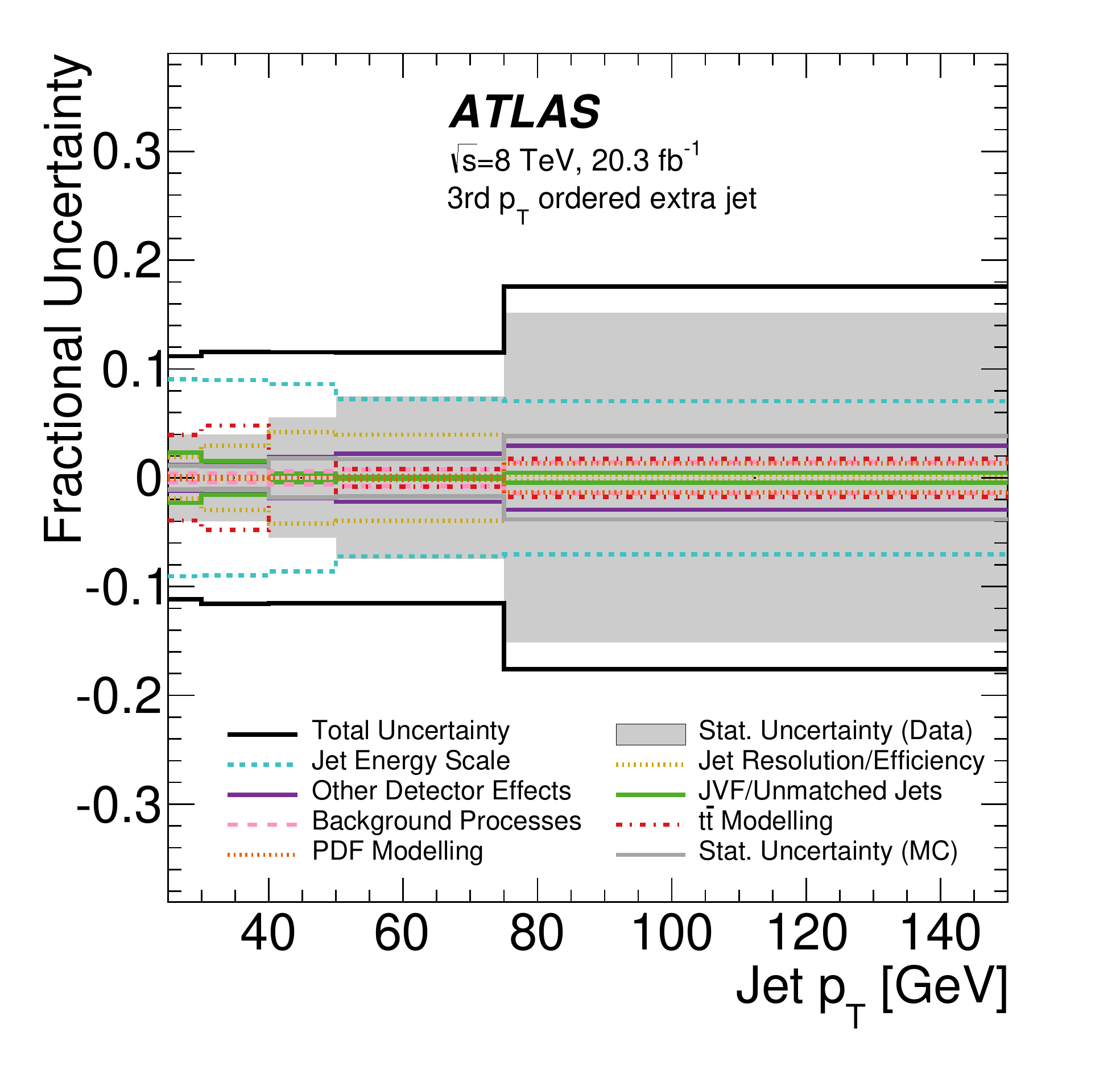}}
\subfloat[][]{
\includegraphics[width=0.5\textwidth]{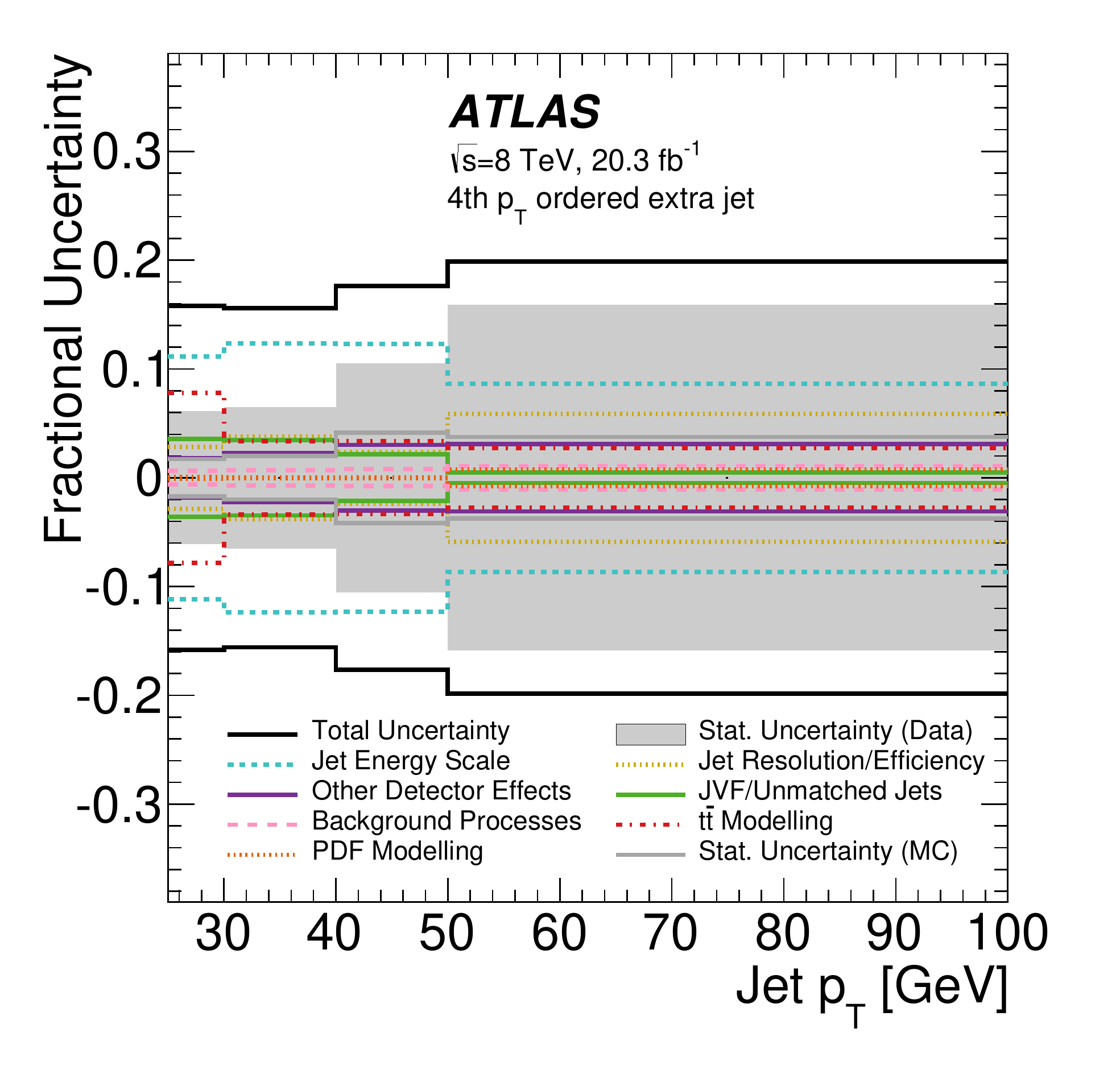}} 

\caption{Envelope of fractional uncertainties in the first (a) to the fourth (d)
additional-jet normalised differential cross-sections, as functions of the 
corresponding jet \pt. The total uncertainties are shown, together with the
separate contributions from the data statistical uncertainty and various 
categories of systematic uncertainty.}
\label{fig:sys}
\end{figure}

\subsection{Jet multiplicity and \pt\ spectra results}\label{ssec:jetres}

Figures~\ref{fig:unfmult1}--\ref{fig:unfmult2} show normalised distributions of the additional-jet 
multiplicity for different jet \pt\ thresholds, and compare the data to 
the NLO generator configurations
{\sc Powheg\,+\,Pythia6} with $\hdmp=\infty$ or \mtop,
{\sc Powheg\,+\,Pythia8}, {\sc MC@NLO\,+\,Herwig} and {\sc Powheg\,+\,Herwig}.
Figures~\ref{fig:unfpt1}--\ref{fig:unfpt2} show the normalised differential cross-sections 
\sigmapti\ for jets of rank $i$ from one to four. In both cases, the expected 
contributions from
$Wt$ events were subtracted from the event counts before normalising
the distributions, based on the baseline {\sc Powheg+Pythia6} $Wt$ simulation 
sample. The same data are presented numerically
in Table~\ref{t:diffjet}, both with and without subtraction of the $Wt$ 
contribution, and including two \pt\ bins for the fifth jet. The highest
\pt\ bin for each jet rank includes overflows, but the differential 
cross-sections are normalised using the bin widths  $\Delta$ derived from the
upper \pt\ bin limits listed in 
Table~\ref{t:diffjet} and shown in Figures~\ref{fig:unfpt1} 
and~\ref{fig:unfpt2}.

\begin{center}
\begin{longtable}{|c|c|r|r|r|r|}
\caption{\label{t:diffjet}Normalised particle-level differential jet cross-sections as a function of jet rank and \pt, both without ($\sigma^{\ttbar+Wt}$) and
with ($\sigma^{\ttbar}$) 
the $Wt$ contribution subtracted. The additional jets are required to have
$|\eta|<4.5$, corresponding to the full pseudorapidity range . The boundaries of each bin are given, together with the mean jet \pt\ in each bin. The last bin for every jet rank includes overflows, but the differential cross-section 
values are determined using the upper bin limit given for that bin.} \\ \hline
Bin & Rank & \pt\ range [\GeV] & \begin{tabular}[c]{@{}c@{}}Avg. \pt\ \\ $[$\GeV$]$ \end{tabular} & \begin{tabular}[c]{@{}c@{}}$\frac{1}{\sigma} \frac{{\rm d}\sigma_i}{{\rm d}\pt}(\ttbar+Wt) \pm$(stat.)$\pm$(syst.) \\ $[10^{-4} \gev^{-1}]$ \end{tabular} &  \begin{tabular}[c]{@{}c@{}}$\frac{1}{\sigma} \frac{{\rm d}\sigma_i}{{\rm d}\pt}(\ttbar)\pm$(stat.)$\pm$(syst.) \\ $[10^{-4} \gev^{-1}]$  \end{tabular}  \\
\hline 
\endfirsthead
\multicolumn{6}{l}%
 {\tablename\ \thetable\ --\textit{Continued from previous page}} \\ \hline
 Bin & Rank & \pt\ range [\GeV] & \begin{tabular}[c]{@{}c@{}}Avg. \pt\ \\ $[$\GeV$]$ \end{tabular} & \begin{tabular}[c]{@{}c@{}}$\frac{1}{\sigma} \frac{{\rm d}\sigma_i}{{\rm d}\pt}(\ttbar+Wt) \pm$(stat.)$\pm$(syst.) \\ $[10^{-4} \gev^{-1}]$ \end{tabular} &  \begin{tabular}[c]{@{}c@{}}$\frac{1}{\sigma} \frac{{\rm d}\sigma_i}{{\rm d}\pt}(\ttbar)\pm$(stat.)$\pm$(syst.) \\ $[10^{-4} \gev^{-1}]$  \end{tabular} \\
\hline 
\endhead
\hline \multicolumn{6}{l} {\textit{Continued on next page}} \\
\endfoot
\hline \hline
\endlastfoot
1 & 1 & 25--30 & 27.4  & $144.7 \pm 4.3\pm 8.0$ & $144.5 \pm 4.4\pm 8.2$  \\ 
2 & 1 & 30--35 & 32.4  & $122.7 \pm 3.0\pm 7.3$ & $122.8 \pm 3.1\pm 7.5$  \\ 
3 & 1 & 35--40 & 37.4  & $101.8 \pm 2.6\pm 3.1$ & $101.9 \pm 2.6\pm 3.2$  \\ 
4 & 1 & 40--45 & 42.5  & $84.0 \pm 2.3\pm 4.1$ & $84.0 \pm 2.4\pm 4.2$  \\ 
5 & 1 & 45--50 & 47.4  & $70.2 \pm 2.0\pm 2.9$ & $70.3 \pm 2.1\pm 3.0$  \\ 
6 & 1 & 50--60 & 54.8  & $58.0 \pm 1.7\pm 2.3$ & $58.1 \pm 1.8\pm 2.3$  \\ 
7 & 1 & 60--70 & 64.8  & $46.3 \pm 1.5\pm 1.6$ & $46.5 \pm 1.6\pm 1.7$  \\ 
8 & 1 & 70--80 & 74.8  & $35.3 \pm 1.3\pm 1.2$ & $35.4 \pm 1.3\pm 1.2$  \\ 
9 & 1 & 80--90 & 84.8  & $27.2 \pm 1.1\pm 1.0$ & $27.3 \pm 1.1\pm 1.0$  \\ 
10 & 1 & 90--100 & 94.8  & $21.9 \pm 0.9\pm 0.8$ & $22.0 \pm 1.0\pm 0.8$  \\ 
11 & 1 & 100--125 & 111.5  & $16.2 \pm 0.7\pm 0.4$ & $16.3 \pm 0.7\pm 0.5$  \\ 
12 & 1 & 125--150 & 136.7  & $11.18 \pm 0.56\pm 0.29$ & $11.26 \pm 0.58\pm 0.30$  \\ 
13 & 1 & 150--175 & 161.8  & $6.53 \pm 0.41\pm 0.22$ & $6.56 \pm 0.42\pm 0.22$  \\ 
14 & 1 & 175--200 & 186.7  & $5.24 \pm 0.38\pm 0.13$ & $5.29 \pm 0.39\pm 0.14$  \\ 
15 & 1 & 200--225 & 211.9  & $3.02 \pm 0.27\pm 0.14$ & $3.04 \pm 0.28\pm 0.14$  \\ 
16 & 1 & 225--250 & 236.8  & $2.17 \pm 0.23\pm 0.12$ & $2.18 \pm 0.24\pm 0.12$  \\ 
17 & 1 & 250--500+ & 344.4  & $0.66 \pm 0.05\pm 0.02$ & $0.67 \pm 0.05\pm 0.02$  \\ 
\hline 
18 & 2 & 25--30 & 27.4  & $110.6 \pm 3.5\pm 8.8$ & $110.6 \pm 3.6\pm 9.1$  \\ 
19 & 2 & 30--35 & 32.4  & $80.3 \pm 2.3\pm 6.0$ & $80.4 \pm 2.3\pm 6.2$  \\ 
20 & 2 & 35--40 & 37.4  & $59.2 \pm 1.9\pm 4.4$ & $59.5 \pm 1.9\pm 4.5$  \\ 
21 & 2 & 40--45 & 42.4  & $44.8 \pm 1.6\pm 4.0$ & $44.9 \pm 1.6\pm 4.1$  \\ 
22 & 2 & 45--50 & 47.4  & $35.4 \pm 1.4\pm 2.4$ & $35.5 \pm 1.4\pm 2.4$  \\ 
23 & 2 & 50--60 & 54.6  & $26.6 \pm 1.1\pm 1.7$ & $26.8 \pm 1.2\pm 1.8$  \\ 
24 & 2 & 60--70 & 64.6  & $17.1 \pm 0.9\pm 1.0$ & $17.3 \pm 0.9\pm 1.0$  \\ 
25 & 2 & 70--80 & 74.6  & $9.8 \pm 0.6\pm 0.7$ & $9.9 \pm 0.6\pm 0.7$  \\ 
26 & 2 & 80--90 & 84.7  & $5.88 \pm 0.50\pm 0.43$ & $5.92 \pm 0.51\pm 0.45$  \\ 
27 & 2 & 90--100 & 94.7  & $3.81 \pm 0.34\pm 0.33$ & $3.84 \pm 0.34\pm 0.34$  \\ 
28 & 2 & 100--125 & 110.9  & $2.43 \pm 0.25\pm 0.15$ & $2.44 \pm 0.25\pm 0.15$  \\ 
29 & 2 & 125--150 & 136.0  & $1.30 \pm 0.19\pm 0.09$ & $1.32 \pm 0.19\pm 0.10$  \\ 
30 & 2 & 150--300+ & 194.2  & $0.20 \pm 0.03\pm 0.01$ & $0.20 \pm 0.03\pm 0.01$  \\ 
\hline 
31 & 3 & 25--30 & 27.3  & $56.7 \pm 2.3\pm 6.0$ & $56.9 \pm 2.3\pm 6.2$  \\ 
32 & 3 & 30--40 & 34.3  & $29.6 \pm 1.2\pm 3.3$ & $29.8 \pm 1.2\pm 3.4$  \\ 
33 & 3 & 40--50 & 44.4  & $12.7 \pm 0.7\pm 1.4$ & $12.8 \pm 0.7\pm 1.4$  \\ 
34 & 3 & 50--75 & 59.3  & $4.68 \pm 0.35\pm 0.45$ & $4.74 \pm 0.36\pm 0.47$  \\ 
35 & 3 & 75--150+ & 97.9  & $0.40 \pm 0.06\pm 0.04$ & $0.41 \pm 0.06\pm 0.04$  \\ 
\hline 
36 & 4 & 25--30 & 27.3  & $23.5 \pm 1.4\pm 3.6$ & $23.7 \pm 1.5\pm 3.7$  \\ 
37 & 4 & 30--40 & 34.1  & $9.4 \pm 0.6\pm 1.4$ & $9.5 \pm 0.6\pm 1.4$  \\ 
38 & 4 & 40--50 & 44.2  & $3.07 \pm 0.32\pm 0.50$ & $3.10 \pm 0.33\pm 0.51$  \\ 
39 & 4 & 50--100+ & 64.1  & $0.55 \pm 0.09\pm 0.08$ & $0.55 \pm 0.09\pm 0.08$  \\ 
\hline 
40 & 5 & 25--30 & 27.2  & $7.3 \pm 0.9\pm 1.6$ & $7.4 \pm 0.9\pm 1.6$  \\ 
41 & 5 & 30--50+ & 38.8  & $1.95 \pm 0.29\pm 0.40$ & $1.97 \pm 0.30\pm 0.41$  \\ 
\hline 
\hline 
\end{longtable}
\end{center}

\begin{figure}[htp]
\centering
\subfloat[][]{
\hspace{-8mm}
\includegraphics[width=0.53\textwidth]{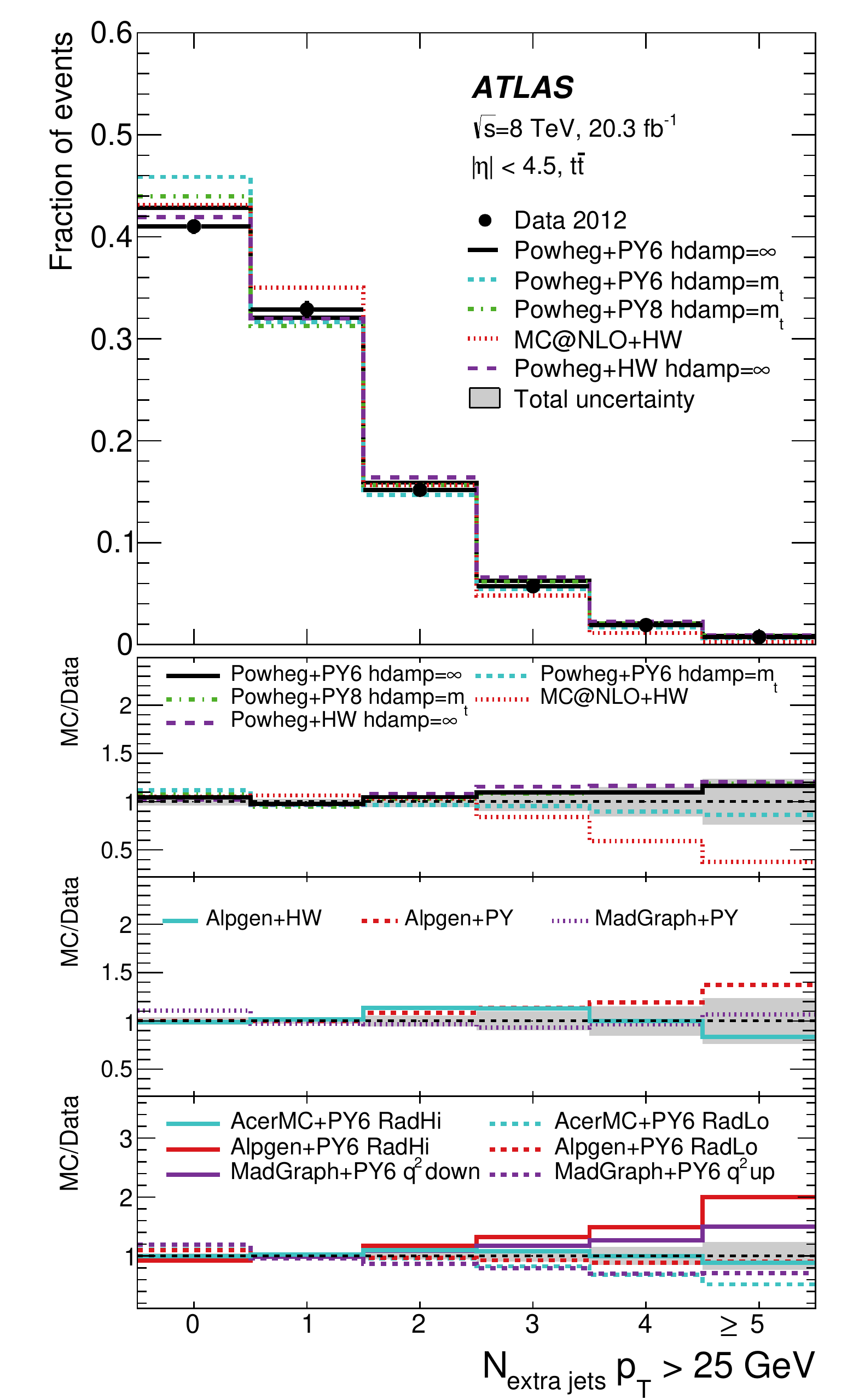}}
\subfloat[][]{
\includegraphics[width=0.53\textwidth]{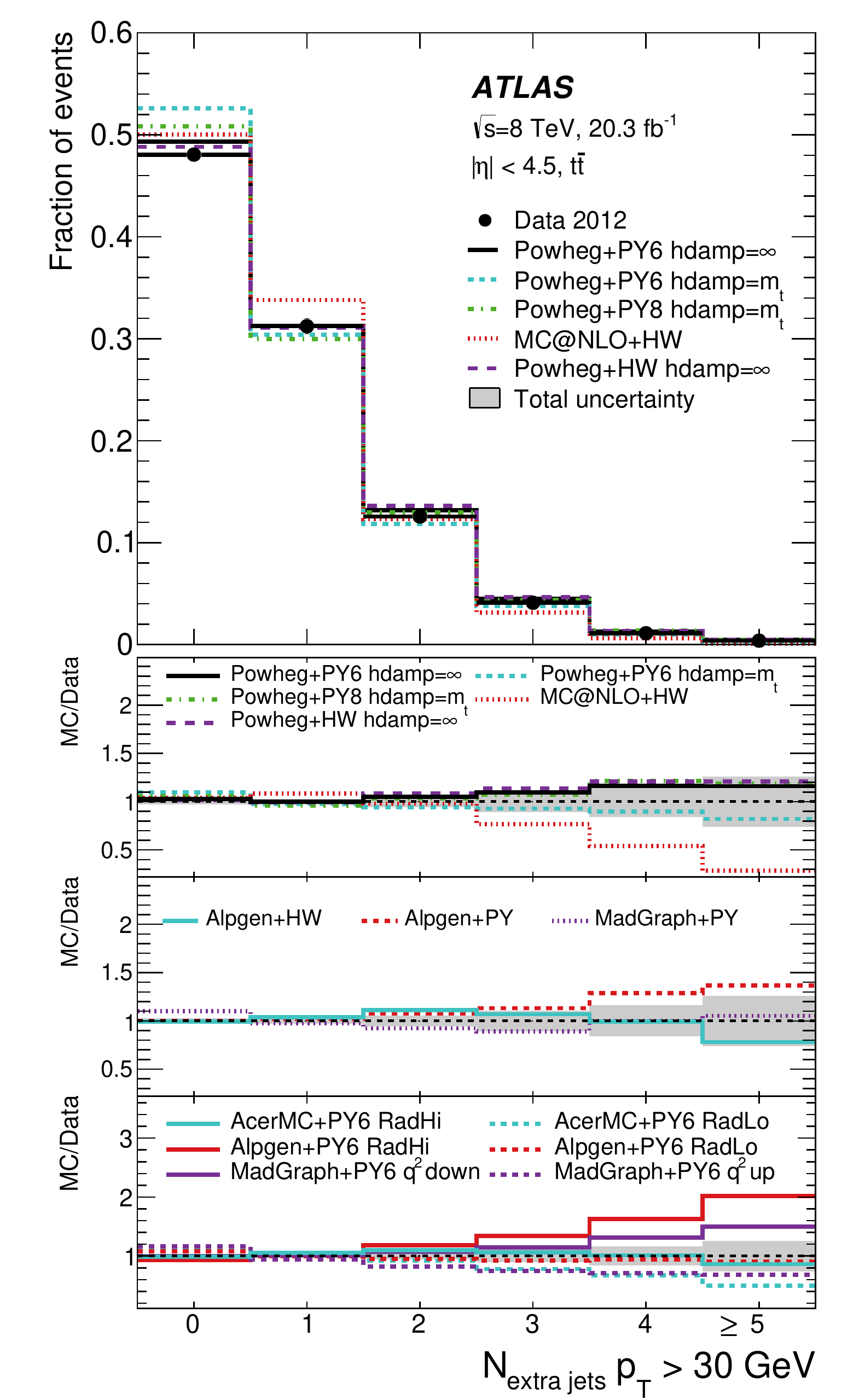}} \\
\caption{Unfolded normalised distributions of particle-level additional-jet 
multiplicity with \pt > (a) 25\,\GeV\ and (b) 30\,\GeV\ in selected
\emubb\ events. The data are 
shown as points with error bars indicating the statistical uncertainty,
and are compared to simulation from several NLO \ttbar\ generator 
configurations. The $Wt$ contribution taken from
{\sc Powheg\,+\,Pythia6} is subtracted from the data.
 The lower plots show the ratios of the different simulation predictions to
data, with the shaded bands including both the statistical and systematic 
uncertainties of the data.}
\label{fig:unfmult1}
\end{figure}
\begin{figure}[htp]
\centering
\subfloat[][]{
\hspace{-8mm}
\includegraphics[width=0.53\textwidth]{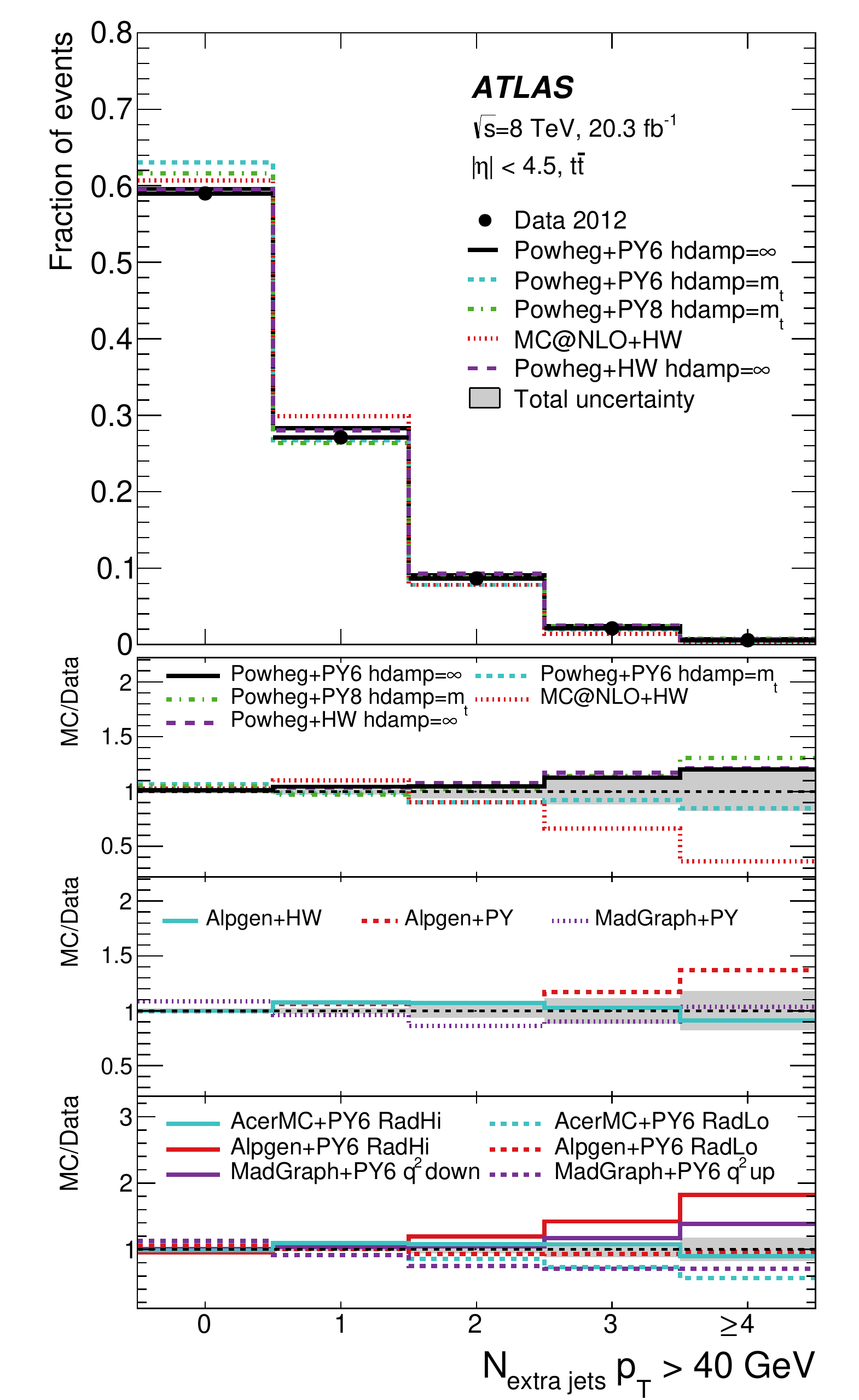}}
\subfloat[][]{
\includegraphics[width=0.53\textwidth]{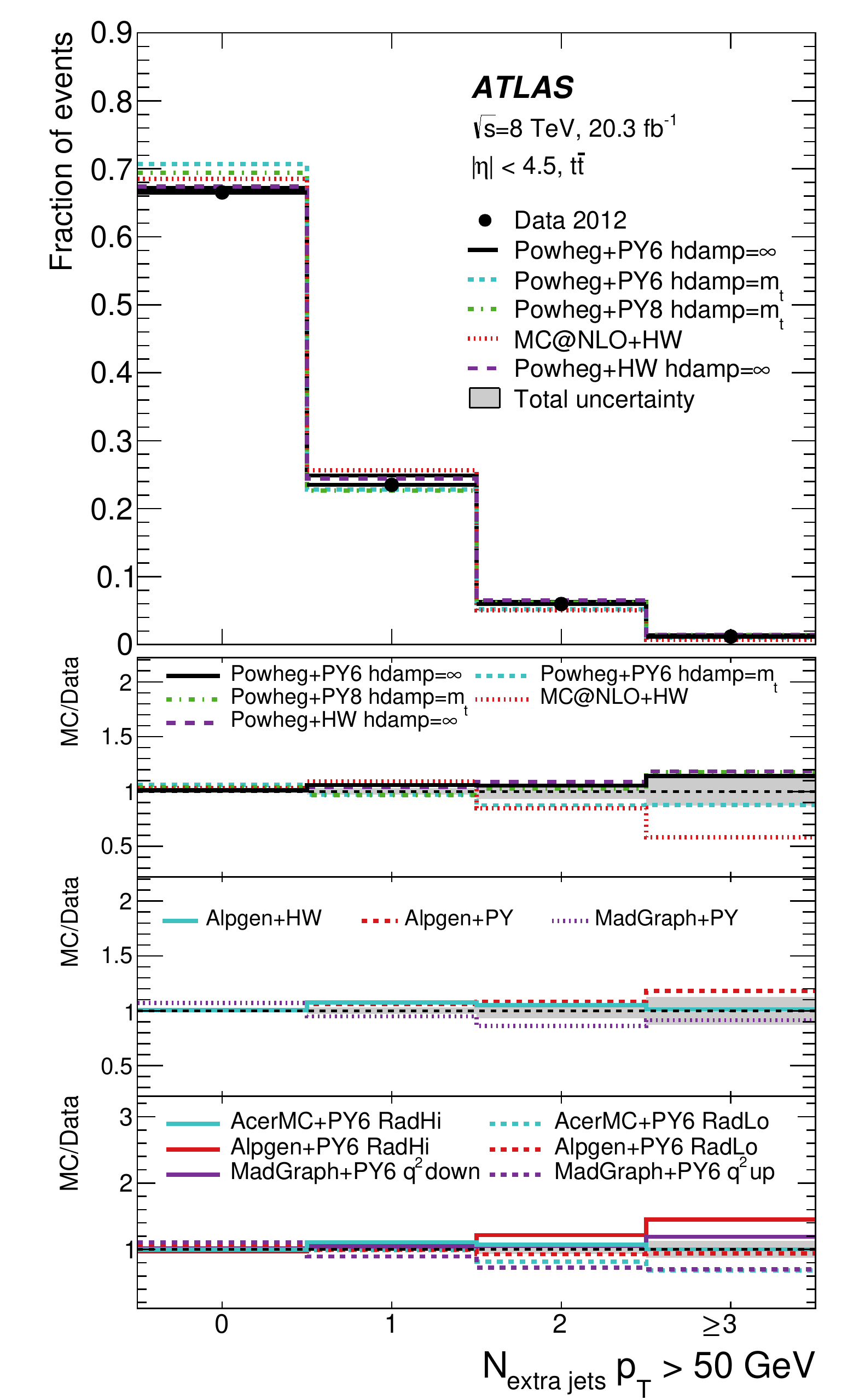}} \\
\caption{Unfolded normalised distributions of particle-level additional-jet 
multiplicity with \pt > (a) 40\,\GeV\ and (b) 50\,\GeV\ in selected
\emubb\ events. The data are 
shown as points with error bars indicating the statistical uncertainty,
and are compared to simulation from several NLO \ttbar\ generator 
configurations. The $Wt$ contribution taken from
{\sc Powheg\,+\,Pythia6} is subtracted from the data.
The lower plots show the ratios of the different simulation predictions to
data, with the shaded bands including both the statistical and systematic 
uncertainties of the data.}
\label{fig:unfmult2}
\end{figure}
\begin{figure}[htp]
\centering
\subfloat[][]{
\hspace{-8mm}
\includegraphics[width=0.53\textwidth]{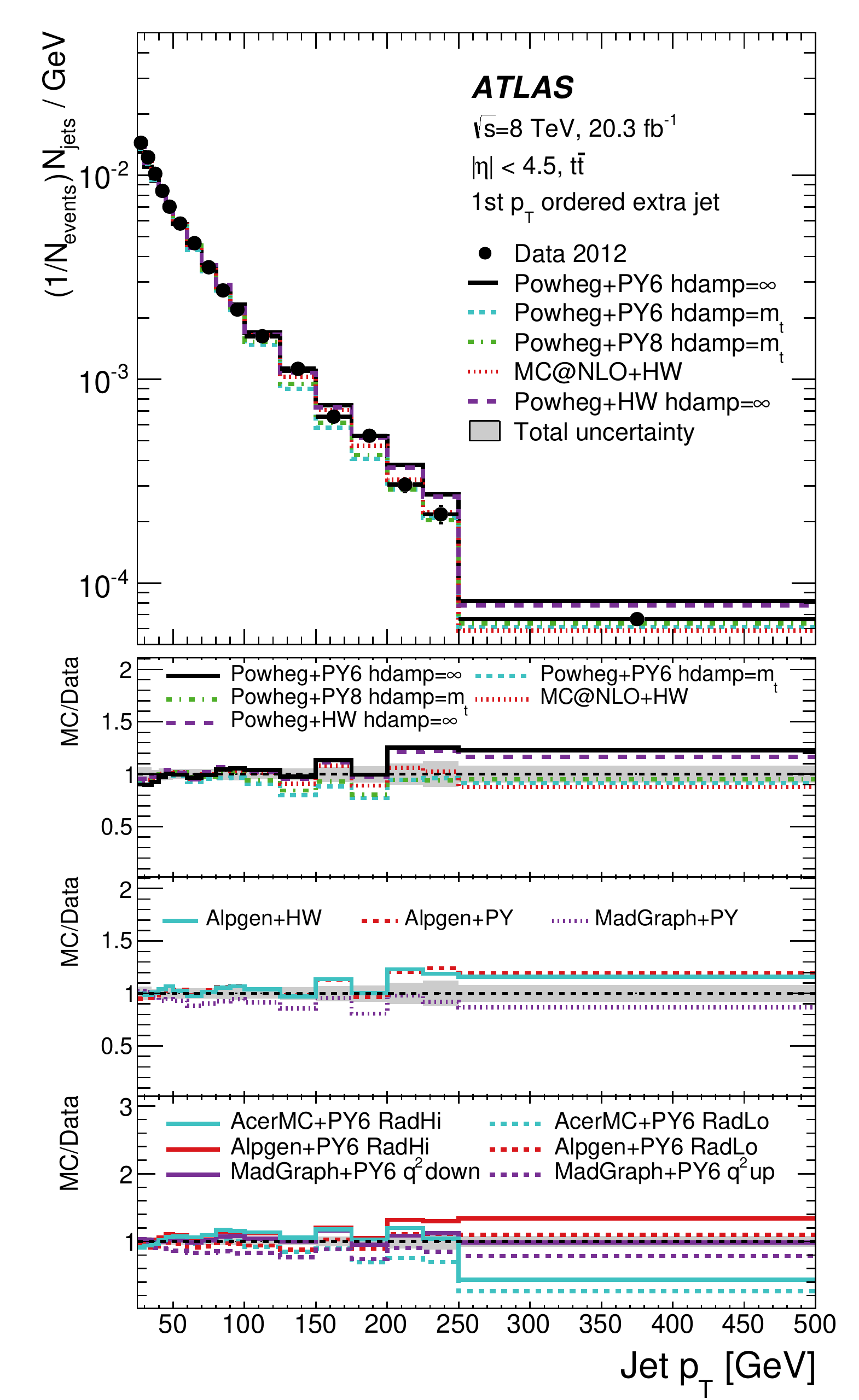}}
\subfloat[][]{
\includegraphics[width=0.53\textwidth]{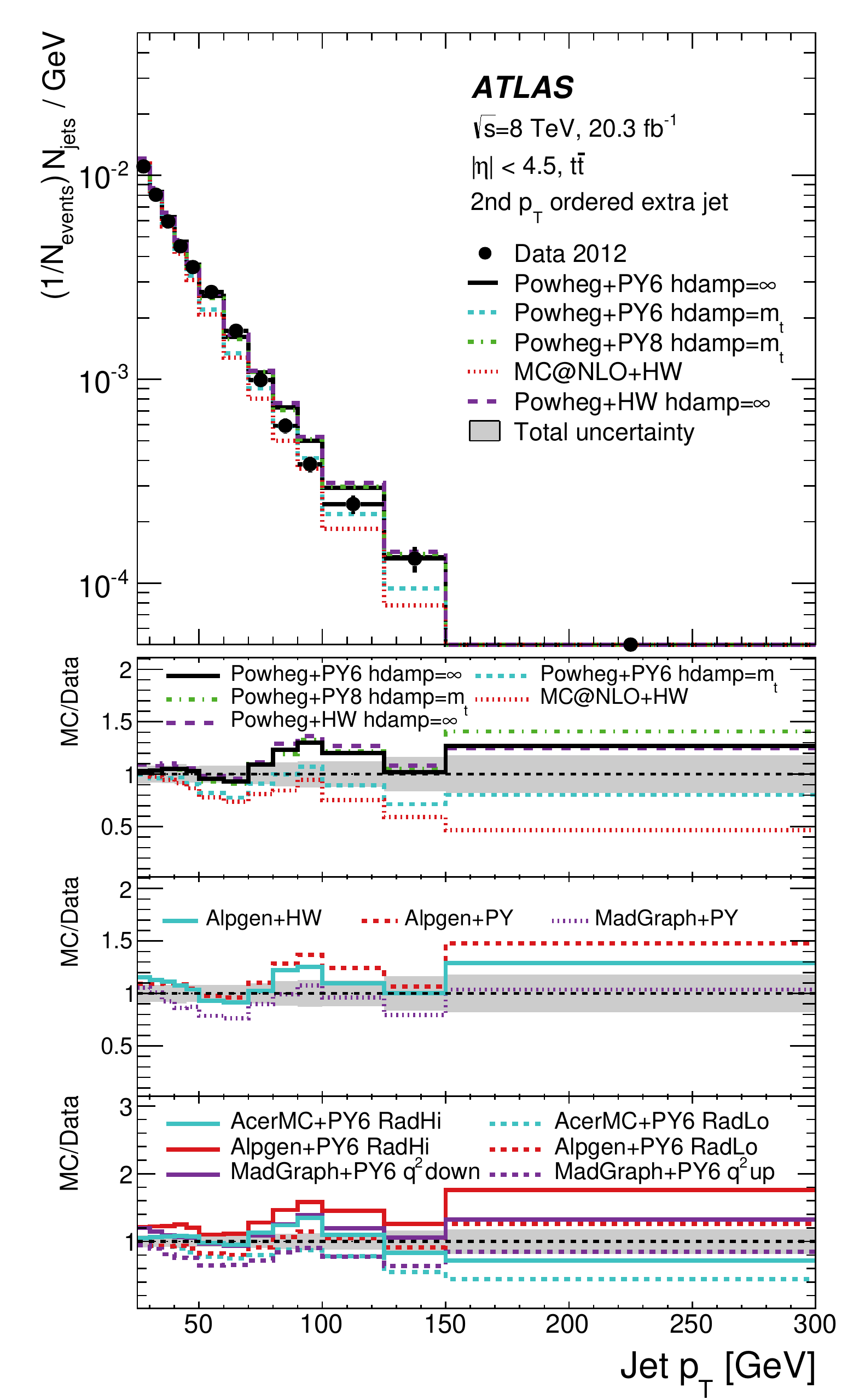}} \\
\caption{Unfolded normalised distributions of particle-level jet \pt\ for the 
first and second additional jet in selected \emubb\ events. The data are 
shown as points with error bars indicating the statistical uncertainty,
and are compared to simulation from several NLO \ttbar\ generator 
configurations. The $Wt$ contribution taken from
{\sc Powheg\,+\,Pythia6} is subtracted from the data.
The lower plots show the ratios of the different simulation predictions to
data, with the shaded bands including both the statistical and systematic 
uncertainties of the data.}
\label{fig:unfpt1}
\end{figure}
\begin{figure}[htp]
\centering
\subfloat[][]{
\hspace{-8mm}
\includegraphics[width=0.53\textwidth]{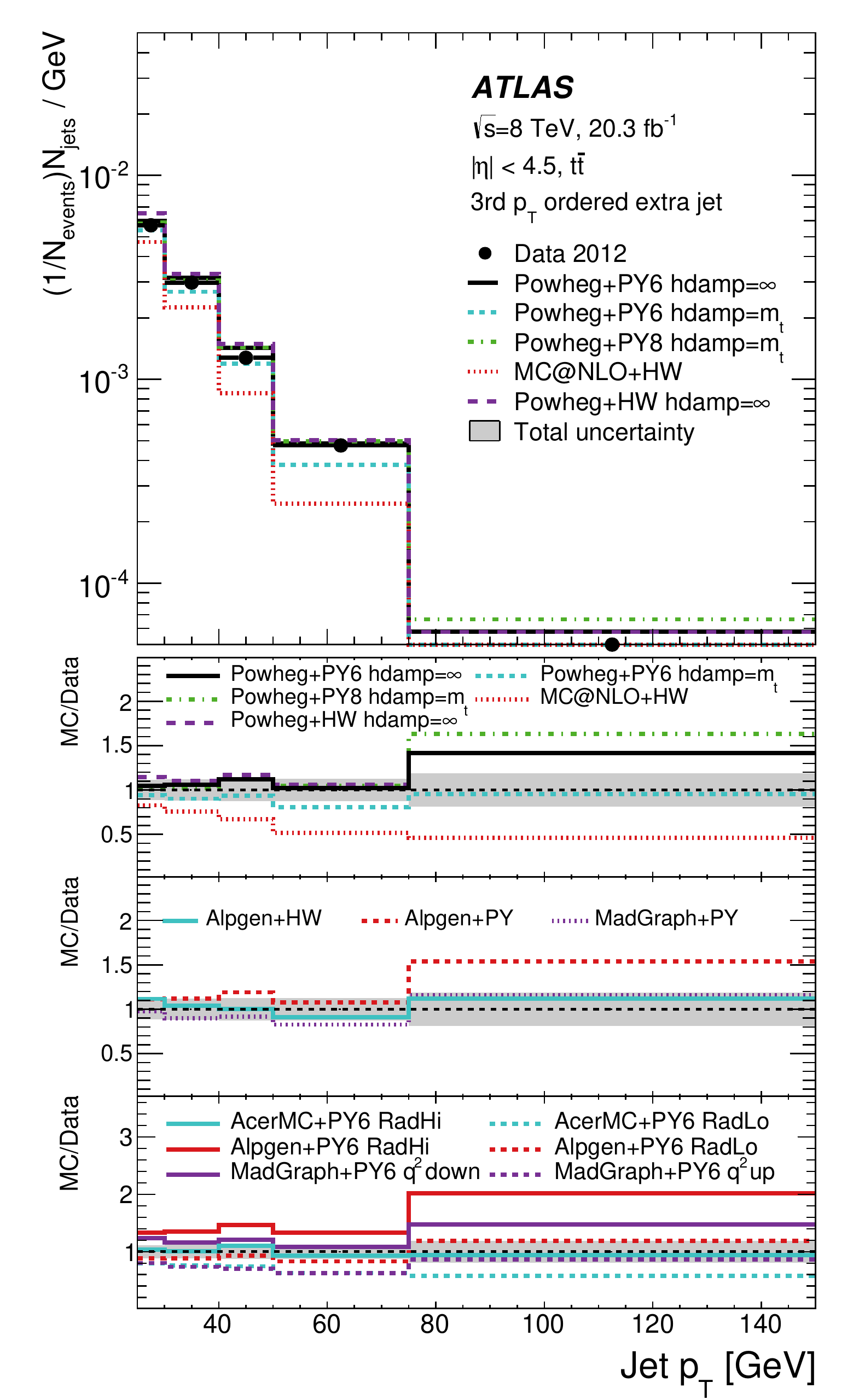}}
\subfloat[][]{
\includegraphics[width=0.53\textwidth]{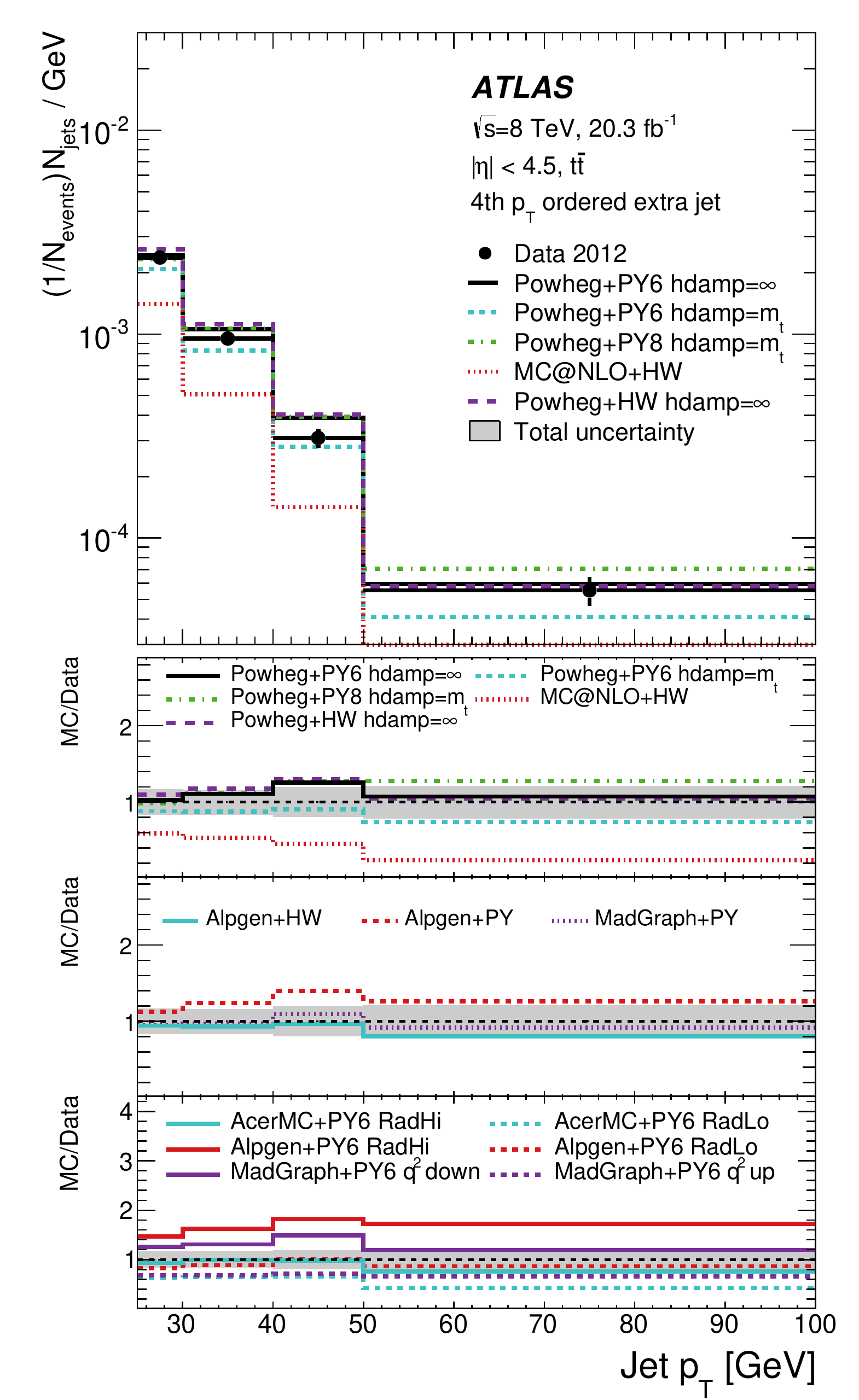}} 
\caption{Unfolded normalised distributions of particle-level jet \pt\ for the 
third and fourth additional jet in selected \emubb\ events. The data are 
shown as points with error bars indicating the statistical uncertainty,
and are compared to simulation from several NLO \ttbar\ generator 
configurations. The $Wt$ contribution taken from
{\sc Powheg\,+\,Pythia6} is subtracted from the data.
The lower plots show the ratios of the different simulation predictions to
data, with the shaded bands including both the statistical and systematic 
uncertainties of the data.}
\label{fig:unfpt2}
\end{figure}

All the NLO generators provide a reasonable description of the leading jet,
which might be expected since they include one additional jet in the 
matrix-element calculation of the \ttbar\ process. 
Differences among the generators become larger with increasing jet rank, where
the prediction from the NLO generators is determined mainly by the parton
shower. In this region, the 
generators predict significantly different rates of additional-jet production. 
They also predict some differences in the shapes of the jet \pt\ spectra. 
The {\sc MC@NLO\,+\,Herwig} sample predicts the lowest rate of additional-jet
 production and underestimates the number of events with at least four 
additional jets by 40\,\%.

The same fully corrected data are compared to the leading-order multi-leg
generators {\sc Alpgen\,+\,Pythia6}, {\sc Alpgen\,+\,Herwig}
and {\sc MadGraph\,+\,Pythia6}
in the second set of ratio plots in 
Figures~\ref{fig:unfmult1}--\ref{fig:unfpt2}.  
In all cases, the renormalisation and
factorisation scales are set to the defaults provided by the code authors.  
For
leading-order generators, the predicted cross-section can depend strongly on 
the choice of QCD scales and parton shower parameters; Figures~\ref{fig:unfmult1}--\ref{fig:unfpt2} also show the effects 
of the variations discussed in Section~\ref{sec:simulation} for
samples generated with {\sc AcerMC\,+\,Pythia6}, {\sc Alpgen\,+\,Pythia6}
and {\sc MadGraph\,+\,Pythia6}.
The measurement gives an uncertainty in the differential cross-sections that is
smaller than the range spanned by these variations in the leading-order generators.


The level of agreement between the generator predictions and the data was 
assessed quantitatively using a \chisq\ test taking into account all bins of 
the measured jet \pt\ distributions with rank one to five.
Since the systematic uncertainties and unfolding corrections induce large 
correlations between bins, the \chisq\ was calculated from the
full covariance matrix. 
Table~\ref{t:chi2} presents the resulting  $\chi^2$ values.
Among the NLO generators, \pow+\hw, and {\sc Powheg\,+\,Pythia6} with
$\hdmp=\infty$ or \mtop, agree reasonably well with the data. \peight\
is disfavoured and \mcnlohw\ gives a very poor description of the data.
The leading-order multi-leg generators {\sc Alpgen\,+\,Pythia6} and \madpy\ agree 
reasonably well with data, whilst \alpg+\hw\ is slightly disfavoured. 
Of the three variations of {\sc Alpgen\,+\,Pythia6}, the `RadLo' variation
with less radiation agrees best with data, suggesting that the scale used
in the baseline ATLAS tune predicts too much radiation in the fiducial region
of this measurement. For \madpy\, the opposite is true, and the `$q^2$ down' tune,
which corresponds to more radiation than the baseline tune, agrees best with
data. The \acermc\,+\,\py\ samples do not reproduce the data well, 
regardless of parameter choice.

\begin{table}
\begin{center}
\begin{tabular}{|l|rr|}
\hline
Generator & $\chi^2$ & \multicolumn{1}{c|}{$p$-value}\\
\hline
\powpy\  &  55.3 & 6.7\e{-2} \\ 
\hdamp &  57.4 & 4.6\e{-2} \\ 
\peight\ $h_{\rm damp}=m_{t}$ &  78.0 & 4.4\e{-4} \\ 
\mcnlohw &  108.2 & 5.8\e{-8} \\ 
\pow+\hw\ $\hdmp=\infty$ &  51.4 & 1.3\e{-1} \\ 
\hline
\alpg+\hw &  64.0 & 1.2\e{-2} \\ 
\alpg+\py &  55.5 & 6.4\e{-2} \\ 
\madpy &  54.7 & 7.4\e{-2} \\ 
\hline
\acermc+\py\ RadHi &  138.4 & 1.8\e{-12} \\ 
\acermc+\py\ RadLo &  148.1 & 4.9\e{-14} \\ 
\alpg+\py\ RadHi &  104.7 & 1.8\e{-7} \\ 
\alpg+\py\ RadLo &  47.9 & 2.1\e{-1} \\ 
\madpy\ $q^{2}$ down &  50.2 & 1.5\e{-1} \\ 
\madpy\ $q^{2}$ up &  78.7 & 3.6\e{-4} \\ 
\hline

\end{tabular}
\caption{Values of $\chi^2$ for the comparison of the full set of additional-jet
\pt\ spectra in data with the predictions from  various \ttbar\  generator configurations, including both the statistical and systematic uncertainties.
The additional jets correspond to the full pseudorapidity range ($|\eta|<4.5$).
The $\chi^2$ and $p$-values correspond to 41 degrees of freedom.} 
\label{t:chi2}
\end{center}
\end{table}

\FloatBarrier
\section{Gap fraction measurements}
\label{sec:gapfrac}

The gap fraction \fqzero\ as defined in Equation~(\ref{e:fgap}) was measured 
by using the analogous definition for reconstructed jets, counting the
number of selected \emubb\ events $N$ and the number $n(\qzero)$ of them
that have no additional jets with $\pt>\qzero$
within the rapidity interval \dely:
\begin{equation}
\frecoqz \equiv \frac{n(\qzero)}{N}
\end{equation}
and similarly for the gap fraction based on \qsum. The values of $N$ and 
$n$ were first corrected to remove the background contributions estimated 
from simulation, including the $Wt$ contribution, as this study focuses on 
the comparison of measured gap fractions with the predictions from the \ttbar\ 
generators discussed in Section~\ref{sec:simulation}.
The measured gap fraction \frecoqz\ was then multiplied by
a correction factor \corqz\ to obtain the particle-level gap fraction
\fpartqz\ defined as in Equation~(\ref{e:fgap}) using the particle-level
definitions given in Section~\ref{ss:partlevel}. The correction factor
was evaluated using the values of \frecoqz\ and \fpartqz\ obtained from the 
baseline {\sc Powheg\,+\,Pythia6} \ttbar\ simulation sample: 
\begin{equation}
\corqz \equiv \frac{\fpartqz}{\frecoqz} .
\end{equation}
Systematic uncertainties arise in this procedure from the uncertainties in
\corqz\ and the backgrounds subtracted before the calculation of $N$ and $n$.

The gap fractions \fqzero\ and \fqsum\ were measured for the same
rapidity regions as used in Ref.~\cite{TOPQ-2011-21}, namely
$|y|<0.8$, $0.8<|y|<1.5$, $1.5<|y|<2.1$ and the inclusive region
$|y|<2.1$. The sets of \qzero\ and \qsum\ threshold values chosen also
correspond to those in Ref.~\cite{TOPQ-2011-21}, and the steps correspond
approximately to one standard deviation of the jet energy resolution.
The values of the correction factor \corqz\ (and similarly for \qsum) 
deviate by at most 5\,\% from unity 
at low \qzero\ and \qsum, and approach unity at higher threshold values.
The small corrections reflect the high selection efficiency and high purity
of the event samples; at each threshold \qzero, the baseline simulation 
predicts that
around 80\,\% of the selected reconstructed events that do not have a jet
with $\pt>\qzero$ also have no particle-level jet with $\pt>\qzero$. Therefore,
a simple bin-by-bin correction method is adequate, 
rather than a full unfolding as used for the differential jet cross-section
measurement.  

The systematic uncertainties in the gap fraction measurements
were evaluated as discussed in Section~\ref{sec:systematics}, and the 
uncertainties from different sources added in quadrature. The results
are shown in Figure~\ref{fig:gfsyst} as relative uncertainties $\Delta f/f$
in the measured gap fraction for two illustrative rapidity intervals,
$|y|<0.8$ and $|y|<2.1$.

\begin{figure}[tp]
\centering
\subfloat[][]{
\includegraphics[width=0.5\textwidth]{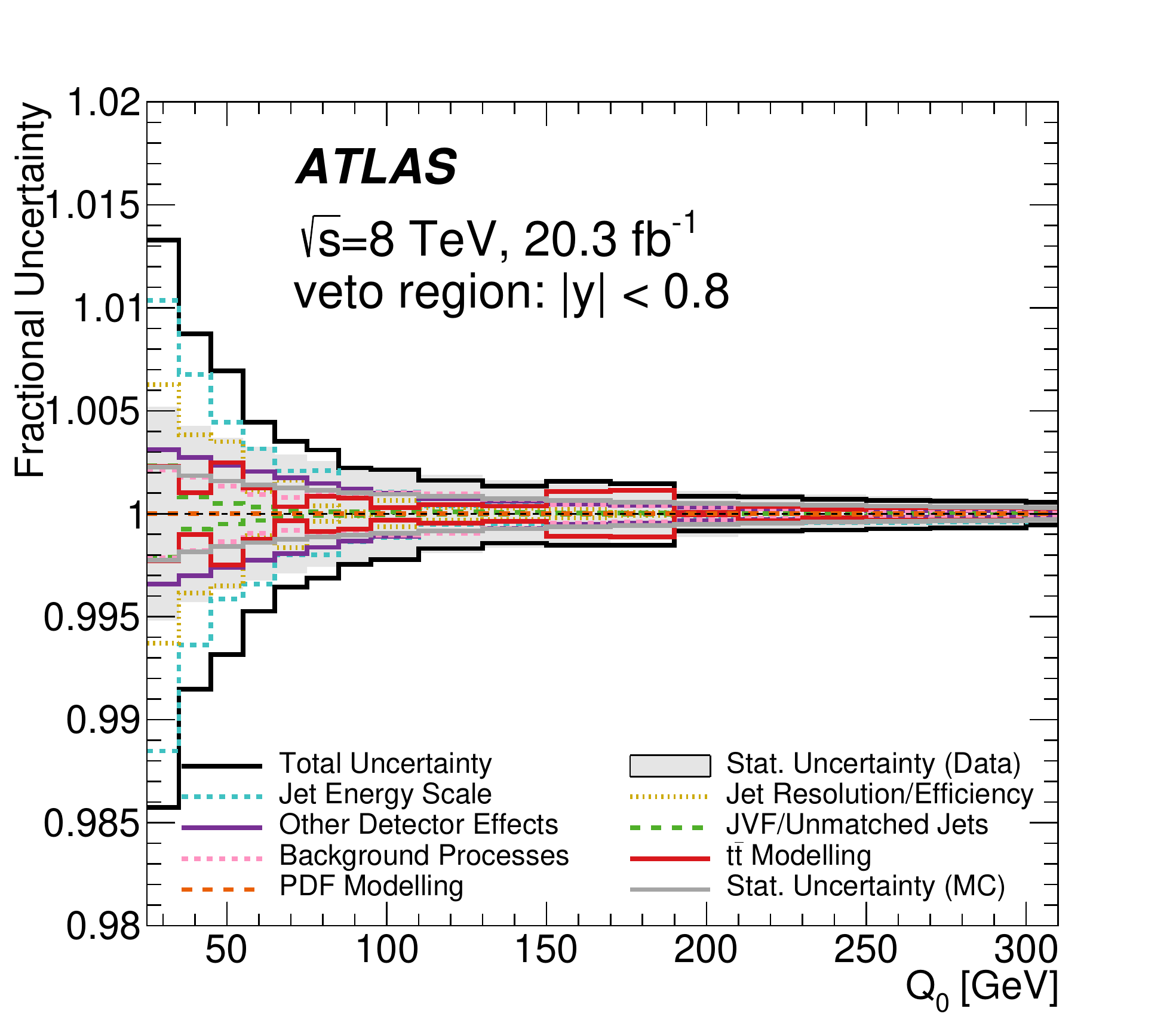}}
\subfloat[][]{
\includegraphics[width=0.5\textwidth]{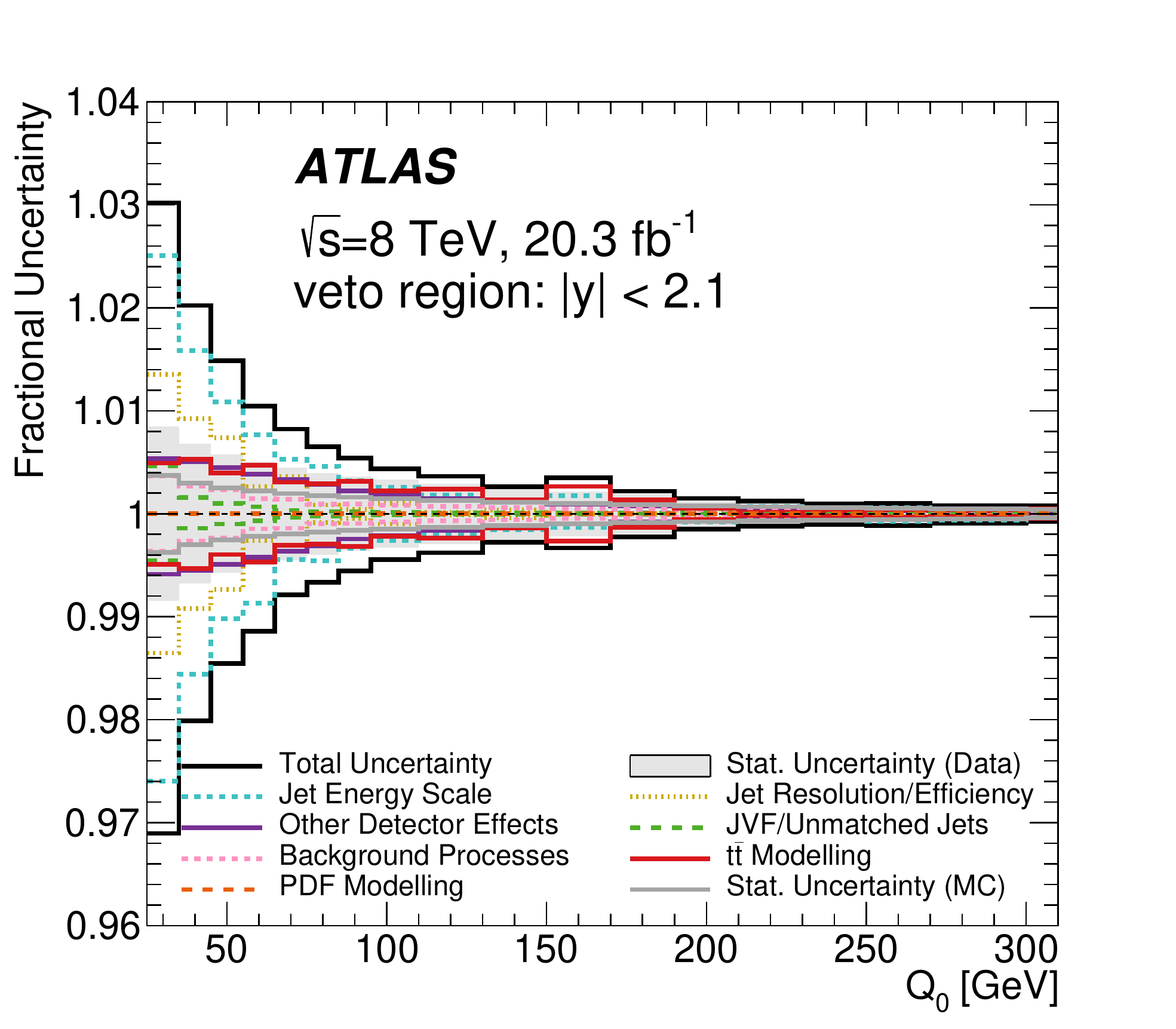}} \\
\caption{Envelope of fractional uncertainties $\Delta f/f$ in the gap fraction \fqzero\ for 
(a) $|y|<0.8$ and (b) $|y|<2.1$. The statistical uncertainty is shown by the
hatched area, and the total uncertainty by the solid black line. 
The systematic uncertainty is shown broken down into several groups,
each of which includes various individual components (see text).}
\label{fig:gfsyst}
\end{figure}

\subsection{Gap fraction results in rapidity regions}\label{ssec:gfracresrap}

Figures \ref{fig:gapfracparta} and~\ref{fig:gapfracpartb} show the resulting
measurements of the gap fraction \fqzero\ in data, corrected to the
particle level. Figure~\ref{fig:gapfracparttota} shows the analogous 
results for \fqsum, for the $|y|<0.8$ and $|y|<2.1$ regions only.
The gap fraction plots and the first sets of ratio plots 
compare the data to the same NLO generator configurations as studied in
Section~\ref{ssec:jetres}.
The middle ratio plots compare the data to the predictions of the leading-order
multi-leg generators {\sc Alpgen\,+\,Pythia6}, {\sc Alpgen\,+\,Herwig}
and {\sc MadGraph\,+\,Pythia6}. The lower ratio plots compare the data
to {\sc AcerMC\,+\,Pythia6}, {\sc Alpgen\,+\,Pythia6} and 
{\sc MadGraph\,+\,Pythia6} samples with increased and decreased levels
of parton shower radiation. The numerical values of the gap fraction 
measurements are presented as a function of 
\qzero\ in Table~\ref{table:q0rapresults} and as a function of 
\qsum\ in Table~\ref{table:qsumrapresults}, together with the values
predicted by the generators shown in the upper plots of
Figures~\ref{fig:gapfracparta}, \ref{fig:gapfracpartb} and 
\ref{fig:gapfracparttota}.
The matrix of statistical and systematic correlations is shown in Figure \ref{fig:corrcomb} for the gap fraction measurement at different values of $Q_0$ for the full central $|y|<2.1$ rapidity region. Nearby points in $Q_0$ are highly correlated, while well-separated $Q_0$ points are less correlated. The
full covariance matrix including correlations was used to calculate 
a $\chi^2$ value for the consistency of each of the NLO generator predictions 
with the data in each veto region. The results are shown in 
Tables~\ref{table:chirapQ0} and~\ref{table:chirapQsum}.

\begin{table}[]
\centering
\small
\begin{tabular}{llllllll}
\hline & \multicolumn{6}{c}{\fqzero\ [\%]}& \\ \cline{2-7}
\begin{tabular}[c]{@{}l@{}}$Q_0$\\{[\gev]}\end{tabular} &\begin{tabular}[c]{@{}l@{}}Data\\$\pm$(stat.)$\pm$(syst.)\end{tabular}& \begin{tabular}[c]{@{}l@{}}\sc{Powheg}\\ +\sc{Pythia6} \\ $\hdmp=\infty$ \end{tabular} & \begin{tabular}[c]{@{}l@{}}\sc{Powheg}\\ +\sc{Pythia6}\\ $h_{\mathrm{damp}}=m_t$\end{tabular} & \begin{tabular}[c]{@{}l@{}}\sc{Powheg}\\ +\sc{Pythia8}\\ $\hdmp=m_t$\end{tabular} & \begin{tabular}[c]{@{}l@{}}\sc{MC@NLO}\\ +\sc{Herwig}\end{tabular} & \begin{tabular}[c]{@{}l@{}}\sc{Powheg}\\ +\sc{Herwig}\\ $\hdmp=\infty$ \end{tabular} & \begin{tabular}[c]{@{}l@{}}$\rho_j^i$\\ (stat.+syst.)\end{tabular}\\ \hline
\multicolumn{3}{l}{veto region: |y| < 0.8}&&&&& \\[0.5pt]
25&76.5$\pm$0.4$\pm 1.1$&76.0$\pm$0.2&78.1$\pm$0.2&76.1$\pm$0.2&79.1$\pm$0.2&74.6$\pm$0.2&$\rho^{25}_{75}=$ 0.65\\[1pt]
75&93.2$\pm$0.2$\pm 0.3$&92.3$\pm$0.1&93.8$\pm$0.1&93.0$\pm$0.1&94.3$\pm$0.1&92.3$\pm$0.1&$\rho^{75}_{150}=$ 0.56\\[1pt]
150&97.8$\pm$0.1$\pm 0.2$&97.3$\pm$0.1&98.0$\pm$0.1&97.8$\pm$0.1&98.3$\pm$0.1&97.4$\pm$0.1&$\rho^{150}_{25}=$ 0.31\\[1pt]
\hline\multicolumn{3}{l}{veto region: 0.8 < |y| < 1.5}&&&&& \\[0.5pt]
25&79.8$\pm$0.4$\pm 1.1$&79.7$\pm$0.2&81.6$\pm$0.2&80.1$\pm$0.2&81.8$\pm$0.2&79.2$\pm$0.2&$\rho^{25}_{75}=$ 0.59\\[1pt]
75&94.5$\pm$0.2$\pm 0.3$&93.5$\pm$0.1&94.7$\pm$0.1&94.3$\pm$0.1&94.7$\pm$0.1&93.7$\pm$0.1&$\rho^{75}_{150}=$ 0.77\\[1pt]
150&98.2$\pm$0.1$\pm 0.2$&97.8$\pm$0.1&98.3$\pm$0.1&98.3$\pm$0.1&98.3$\pm$0.1&97.9$\pm$0.1&$\rho^{150}_{25}=$ 0.39\\[1pt]
\hline\multicolumn{3}{l}{veto region: 1.5 < |y| < 2.1}&&&&& \\[0.5pt]
25&85.3$\pm$0.3$\pm 0.9$&84.9$\pm$0.2&86.1$\pm$0.2&85.4$\pm$0.2&85.5$\pm$0.2&84.7$\pm$0.2&$\rho^{25}_{75}=$ 0.77\\[1pt]
75&96.0$\pm$0.2$\pm 0.4$&95.5$\pm$0.1&96.2$\pm$0.1&96.0$\pm$0.1&95.5$\pm$0.1&95.5$\pm$0.1&$\rho^{75}_{150}=$ 0.89\\[1pt]
150&98.7$\pm$0.1$\pm 0.2$&98.6$\pm$0.1&98.9$\pm$0.0&98.9$\pm$0.0&98.6$\pm$0.1&98.6$\pm$0.1&$\rho^{150}_{25}=$ 0.64\\[1pt]
\hline\multicolumn{3}{l}{veto region: |y| < 2.1}&&&&& \\[0.5pt]
25&53.9$\pm$0.5$\pm 1.7$&53.6$\pm$0.2&56.7$\pm$0.2&54.5$\pm$0.2&56.2$\pm$0.2&52.0$\pm$0.2&$\rho^{25}_{75}=$ 0.66\\[1pt]
75&84.8$\pm$0.3$\pm 0.6$&82.9$\pm$0.2&85.8$\pm$0.2&85.0$\pm$0.2&85.3$\pm$0.2&83.0$\pm$0.2&$\rho^{75}_{150}=$ 0.74\\[1pt]
150&94.9$\pm$0.2$\pm 0.3$&93.8$\pm$0.1&95.4$\pm$0.1&95.2$\pm$0.1&95.3$\pm$0.1&94.1$\pm$0.1&$\rho^{150}_{25}=$ 0.34\\[1pt]
\hline\end{tabular}
\caption{The measured gap fraction values \fqzero\ for different veto-region
rapidity intervals and \qzero\ values of 25, 75 and 150 \gev\ in data compared to the predictions from various \ttbar\ simulation samples. The combination of statistical and systematic correlations between measurements at $\qzero=i$ and $\qzero=j$ is given as $\rho_j^i$.}
\label{table:q0rapresults}
\end{table}
\begin{table}[]
\centering
\small
\begin{tabular}{llllllll}
\hline & \multicolumn{6}{c}{\fqsum\ [\%]}& \\ \cline{2-7}
\begin{tabular}[c]{@{}l@{}}$Q_{\mathrm{sum}}$\\{[\gev]}\end{tabular} &\begin{tabular}[c]{@{}l@{}}Data\\$\pm$(stat.)$\pm$(syst.)\end{tabular}& \begin{tabular}[c]{@{}l@{}}\sc{Powheg}\\ +\sc{Pythia6} \\ $\hdmp=\infty$ \end{tabular} & \begin{tabular}[c]{@{}l@{}}\sc{Powheg}\\ +\sc{Pythia6}\\ $h_{\mathrm{damp}}=m_t$\end{tabular} & \begin{tabular}[c]{@{}l@{}}\sc{Powheg}\\ +\sc{Pythia8}\\ $\hdmp=m_t$\end{tabular} & \begin{tabular}[c]{@{}l@{}}\sc{MC@NLO}\\ +\sc{Herwig}\end{tabular} & \begin{tabular}[c]{@{}l@{}}\sc{Powheg}\\ +\sc{Herwig}\\ $\hdmp=\infty$ \end{tabular} & \begin{tabular}[c]{@{}l@{}}$\rho_j^i$\\ (stat.+syst.)\end{tabular}\\ \hline
\multicolumn{3}{l}{veto region: |y| < 0.8}&&&&& \\[0.5pt]
55&88.1$\pm$0.3$\pm 0.5$&87.7$\pm$0.2&89.5$\pm$0.1&88.1$\pm$0.2&90.2$\pm$0.1&87.2$\pm$0.2&$\rho^{55}_{150}=$ 0.70\\[1pt]
150&97.0$\pm$0.2$\pm 0.2$&96.4$\pm$0.1&97.3$\pm$0.1&96.9$\pm$0.1&97.9$\pm$0.1&96.5$\pm$0.1&$\rho^{150}_{300}=$ 0.57\\[1pt]
300&99.4$\pm$0.1$\pm 0.1$&99.1$\pm$0.0&99.4$\pm$0.0&99.3$\pm$0.0&99.6$\pm$0.0&99.2$\pm$0.0&$\rho^{300}_{55}=$ 0.47\\[1pt]
\hline\multicolumn{3}{l}{veto region: 0.8 < |y| < 1.5}&&&&& \\[0.5pt]
55&90.5$\pm$0.3$\pm 0.6$&89.8$\pm$0.1&91.3$\pm$0.1&90.5$\pm$0.1&91.4$\pm$0.1&89.9$\pm$0.1&$\rho^{55}_{150}=$ 0.68\\[1pt]
150&97.8$\pm$0.1$\pm 0.2$&97.2$\pm$0.1&97.9$\pm$0.1&97.7$\pm$0.1&98.0$\pm$0.1&97.4$\pm$0.1&$\rho^{150}_{300}=$ 0.42\\[1pt]
300&99.6$\pm$0.1$\pm 0.1$&99.4$\pm$0.0&99.6$\pm$0.0&99.5$\pm$0.0&99.7$\pm$0.0&99.5$\pm$0.0&$\rho^{300}_{55}=$ 0.31\\[1pt]
\hline\multicolumn{3}{l}{veto region: 1.5 < |y| < 2.1}&&&&& \\[0.5pt]
55&93.1$\pm$0.2$\pm 0.8$&92.8$\pm$0.1&93.8$\pm$0.1&93.4$\pm$0.1&93.2$\pm$0.1&92.8$\pm$0.1&$\rho^{55}_{150}=$ 0.89\\[1pt]
150&98.5$\pm$0.1$\pm 0.3$&98.3$\pm$0.1&98.7$\pm$0.1&98.6$\pm$0.1&98.5$\pm$0.1&98.3$\pm$0.1&$\rho^{150}_{300}=$ 0.64\\[1pt]
300&99.8$\pm$0.0$\pm 0.1$&99.7$\pm$0.0&99.8$\pm$0.0&99.8$\pm$0.0&99.8$\pm$0.0&99.7$\pm$0.0&$\rho^{300}_{55}=$ 0.60\\[1pt]
\hline\multicolumn{3}{l}{veto region: |y| < 2.1}&&&&& \\[0.5pt]
55&72.7$\pm$0.4$\pm 1.3$&71.8$\pm$0.2&75.2$\pm$0.2&73.3$\pm$0.2&74.7$\pm$0.2&71.1$\pm$0.2&$\rho^{55}_{150}=$ 0.89\\[1pt]
150&91.3$\pm$0.3$\pm 0.5$&89.9$\pm$0.1&92.2$\pm$0.1&91.2$\pm$0.1&92.6$\pm$0.1&90.1$\pm$0.1&$\rho^{150}_{300}=$ 0.79\\[1pt]
300&98.0$\pm$0.1$\pm 0.2$&97.3$\pm$0.1&98.1$\pm$0.1&97.7$\pm$0.1&98.6$\pm$0.1&97.5$\pm$0.1&$\rho^{300}_{55}=$ 0.73\\[1pt]
\hline\end{tabular}
\caption{The measured gap fraction values \fqsum\ for different veto-region
rapidity intervals and \qsum\ values of 55, 150 and 300 \gev\ in data compared to the predictions from various \ttbar\ simulation samples. The combination of statistical and systematic correlations between measurements at $\qsum=i$ and $\qsum=j$ is given as $\rho_j^i$.}
\label{table:qsumrapresults}
\end{table}

\begin{figure}[tp]
\centering
\subfloat[][]{\includegraphics[width=0.52\linewidth]{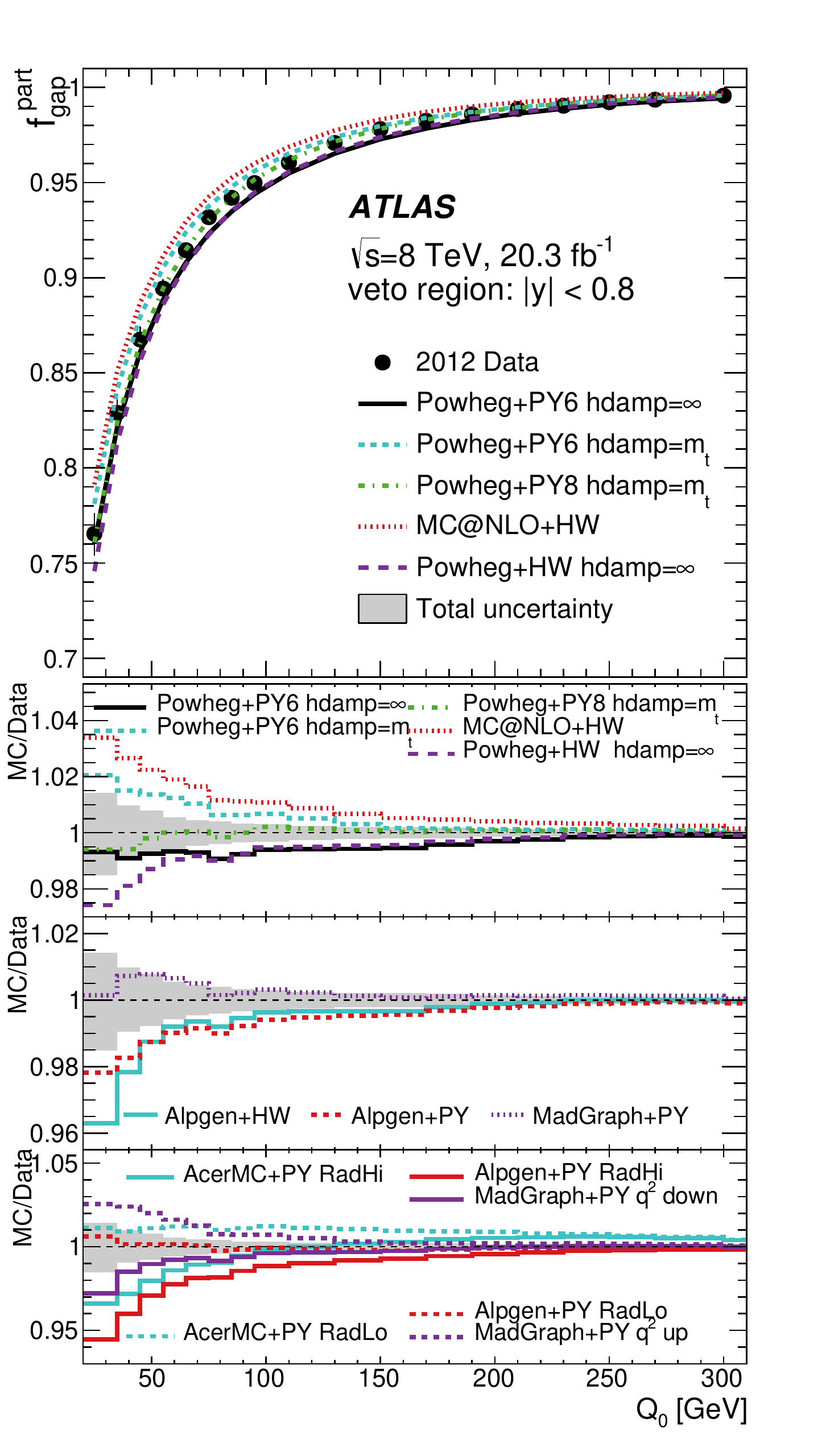}}
\subfloat[][]{\includegraphics[width=0.52\linewidth]{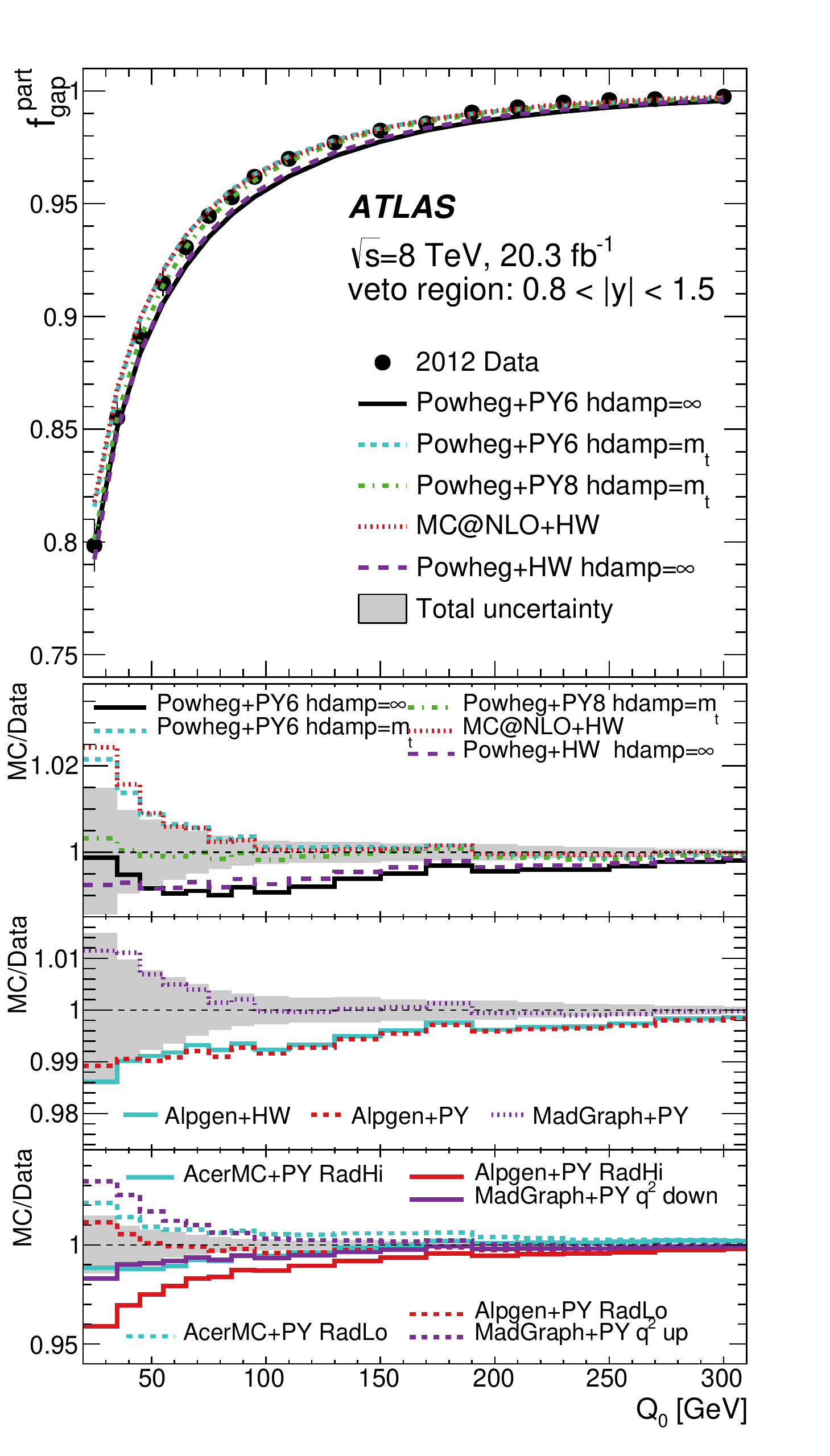}}
\caption{The measured gap fraction \fqzero\ as a function of \qzero\ in 
different veto-region rapidity intervals \dely, for (a) $|y|<0.8$ and (b) $0.8<|y|<1.5$. 
The data are shown by the points with
error bars indicating the total uncertainty, and compared to the predictions
from various \ttbar\ simulation samples (see text) shown as smooth curves.
The lower plots show the
ratio of predictions to data, with the data uncertainty being indicated by
the shaded band, and the \qzero\ thresholds corresponding to the left edges
of the histogram bins, except for the first bin.}
\label{fig:gapfracparta}
\end{figure}

\begin{figure}[tp]
\centering
\subfloat[][]{\includegraphics[width=0.52\linewidth]{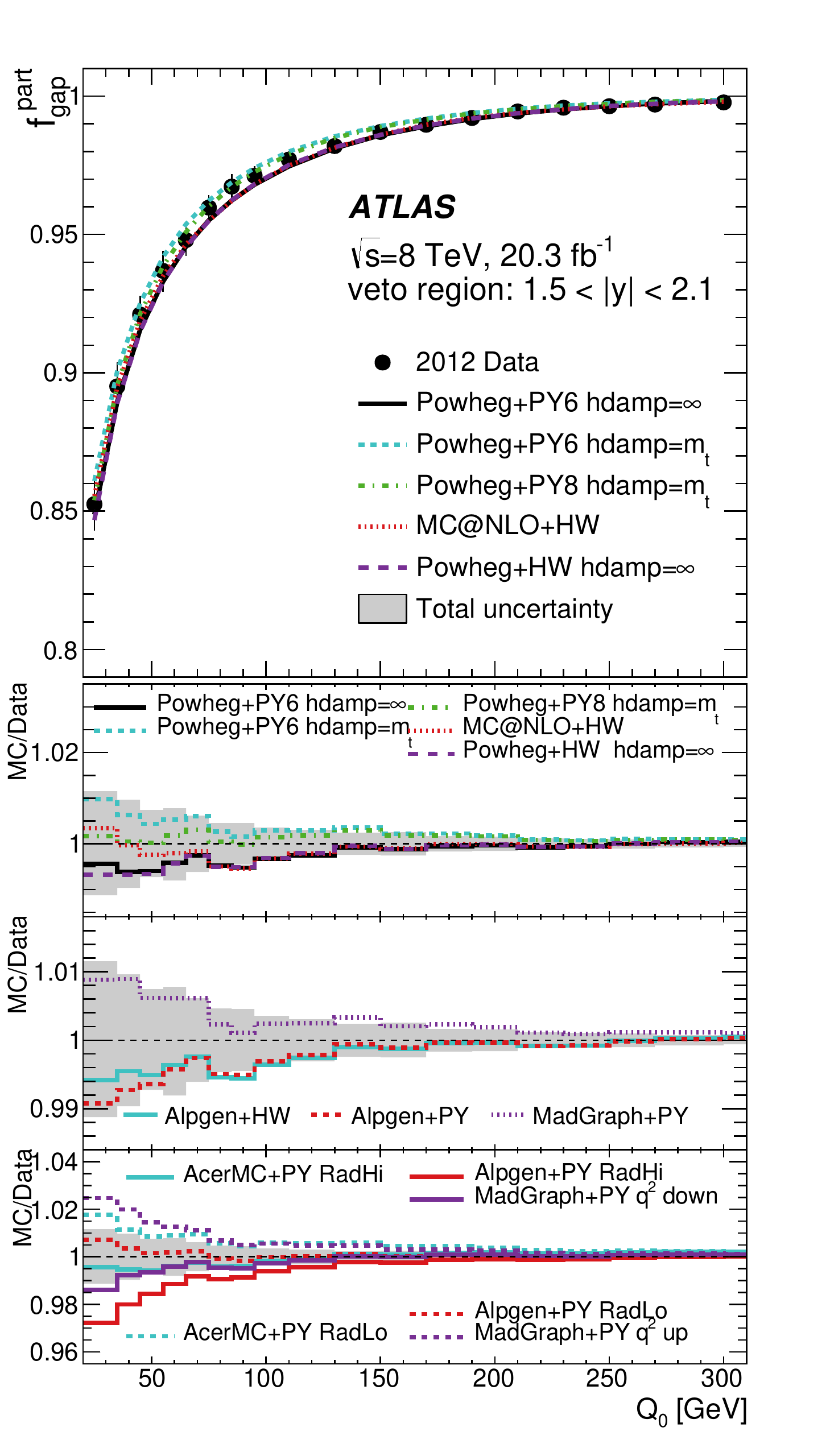}}
\subfloat[][]{\includegraphics[width=0.52\linewidth]{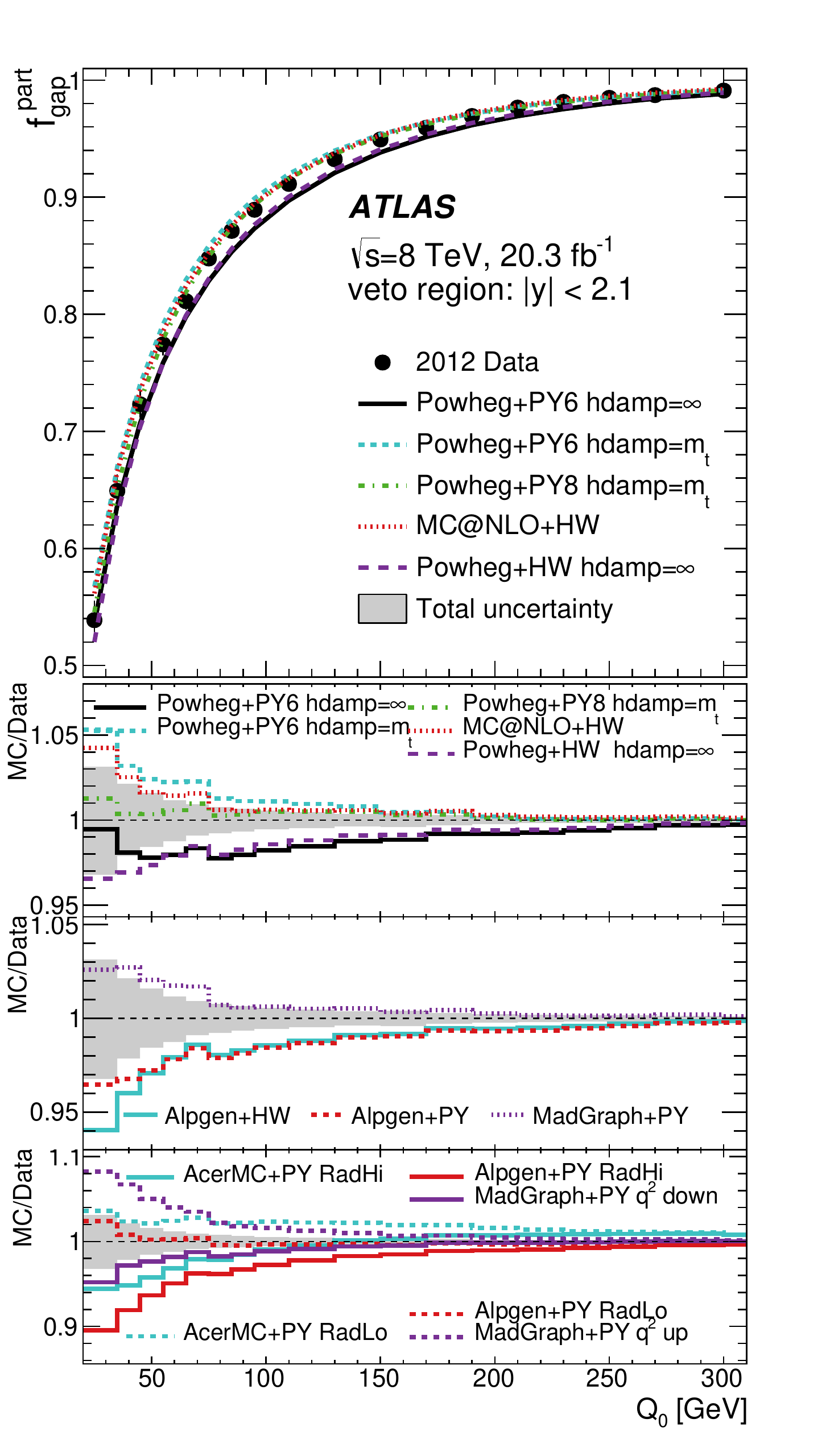}}
\caption{The measured gap fraction \fqzero\ as a function of \qzero\ in 
different veto-region rapidity intervals \dely, for (a) $1.5<|y|<2.1$ and (b) $|y|<2.1$. 
The data are shown by the points with
error bars indicating the total uncertainty, and compared to the predictions
from various \ttbar\ simulation samples (see text) shown as smooth curves.
The lower plots show the
ratio of predictions to data, with the data uncertainty being indicated by
the shaded band, and the \qzero\ thresholds corresponding to the left edges
of the histogram bins, except for the first bin.}
\label{fig:gapfracpartb}
\end{figure}

\begin{figure}[tp]
\centering
\subfloat[][]{\includegraphics[width=0.52\linewidth]{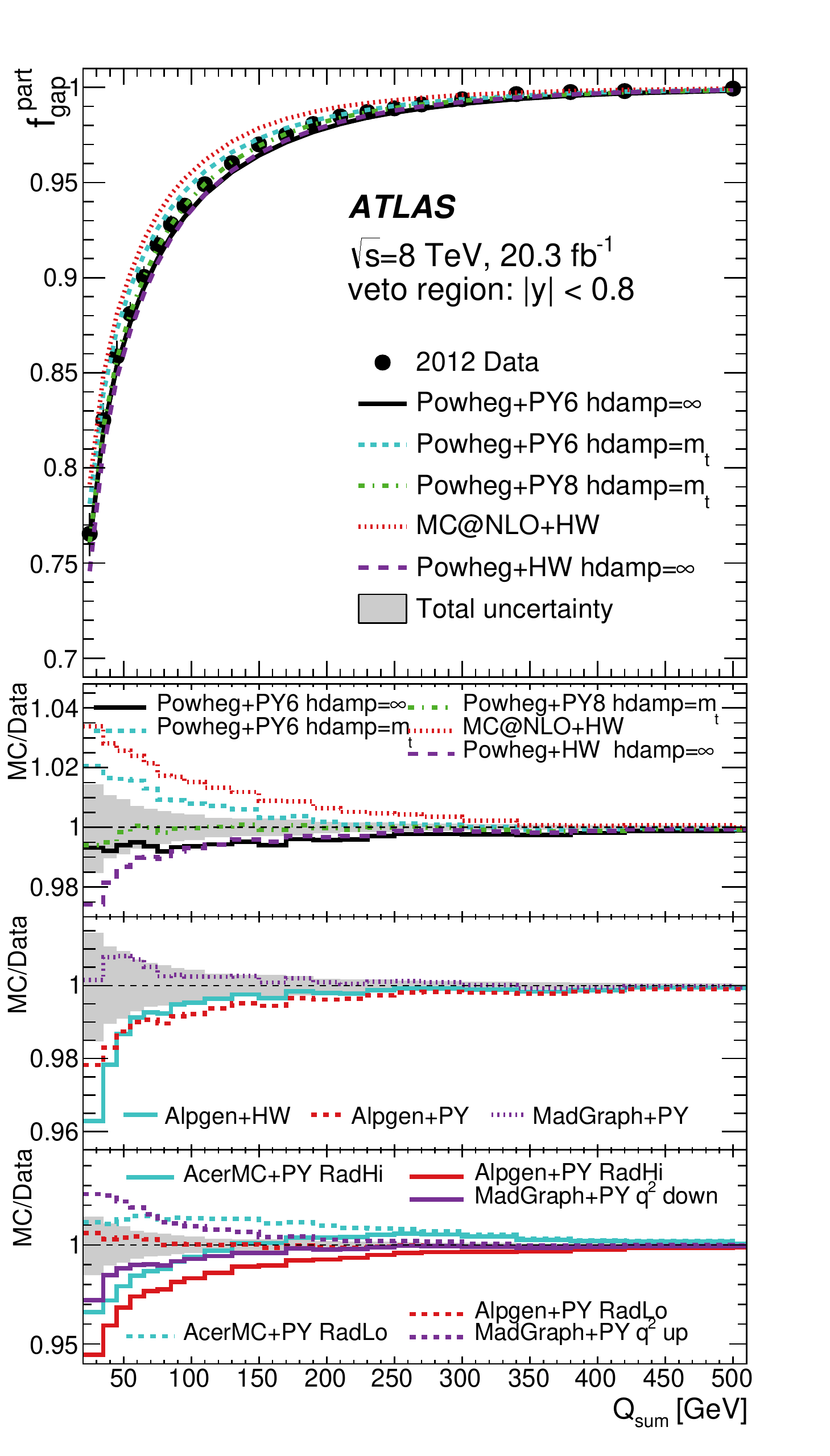}}
\subfloat[][]{\includegraphics[width=0.52\linewidth]{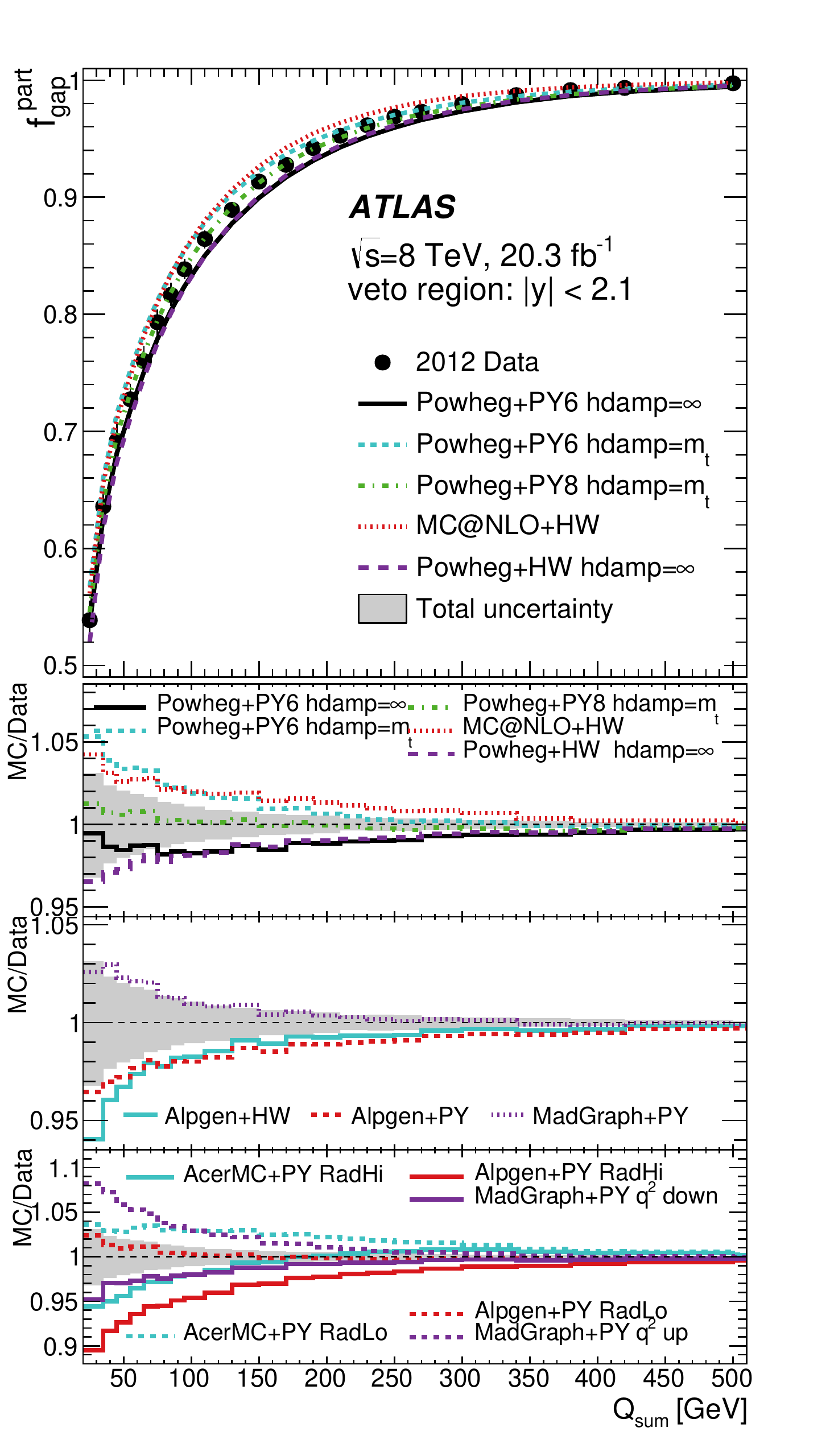}}
\caption{The measured gap fraction \fqsum\ as a function of \qsum\ in 
different veto-region rapidity intervals \dely, for (a) $|y|<0.8$ and 
(b) $|y|<2.1$. The data are shown by the points with
error bars indicating the total uncertainty, and compared to the predictions
from various \ttbar\ simulation samples (see text) shown as smooth curves. 
The lower plots show the
ratio of predictions to data, with the data uncertainty being indicated by
the shaded band, and the \qsum\ thresholds corresponding to the left edges
of the histogram bins, except for the first bin.}
\label{fig:gapfracparttota}
\end{figure}

\begin{figure}[tp]
\centering
\includegraphics[width=0.5\linewidth]{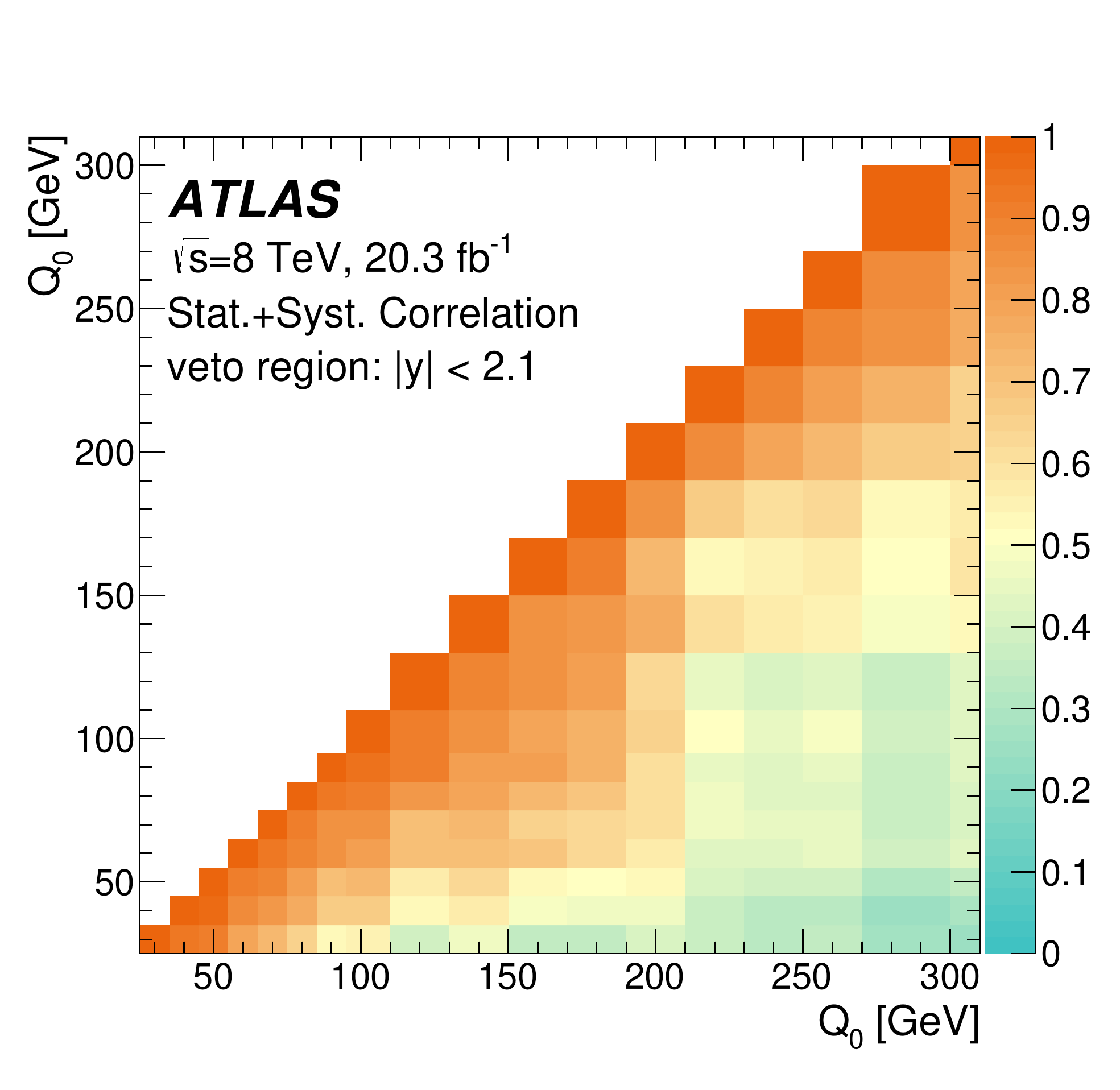}
\caption{The correlation matrix (including statistical and systematic correlations) for the gap fraction measurement at different values of $Q_0$ for the full central rapidity region $|y|<2.1$.}
\label{fig:corrcomb}
\end{figure}

\begin{table}[]
\centering
\footnotesize
\begin{tabular}{|l|ll|ll|ll|ll|}
\hline
$Q_0$&\multicolumn{2}{l|}{$|y|<0.8$}&\multicolumn{2}{l|}{$0.8<|y|<1.5$}&\multicolumn{2}{l|}{$1.5<|y|<2.1$}&\multicolumn{2}{l|}{$|y|<2.1$}\\
Generator&$\chi^2$ & $p$-value & $\chi^2$ & $p$-value & $\chi^2$    & $p$-value      & $\chi^2$  & $p$-value    \\ \hline
{\sc Powheg+Pythia}6 $h_{\rm{damp}}=\infty$ & 15.6 & 6.2$\times 10^{-1}$ & 29.8 & 3.9$\times 10^{-2}$ & 26.3 & 9.3$\times 10^{-2}$ & 31.1 & 2.8$\times 10^{-2}$ \\
{\sc Powheg+Pythia}6 $h_{\rm{damp}}=m_t$ & 17.3 & 5.0$\times 10^{-1}$ & 20.4 & 3.1$\times 10^{-1}$ & 28.6 & 5.4$\times 10^{-2}$ & 25.6 & 1.1$\times 10^{-1}$ \\
{\sc Powheg+Pythia}8 $h_{\rm{damp}}=m_t$ & 11.1 & 8.9$\times 10^{-1}$ & 16.8 & 5.4$\times 10^{-1}$ & 23.2 & 1.8$\times 10^{-1}$ & 16.6 & 5.5$\times 10^{-1}$ \\
{\sc MC@NLO+Herwig} & 22.9 & 2.0$\times 10^{-1}$ & 17.9 & 4.7$\times 10^{-1}$ & 29.9 & 3.9$\times 10^{-2}$ & 18.5 & 4.3$\times 10^{-1}$ \\
{\sc Powheg+Herwig} $h_{\rm{damp}}=\infty$ & 16.7 & 5.5$\times 10^{-1}$ & 24.1 & 1.5$\times 10^{-1}$ & 29.4 & 4.4$\times 10^{-2}$ & 21.5 & 2.5$\times 10^{-1}$ \\ \hline
{\sc Alpgen+Herwig} & 21.8 & 2.4$\times 10^{-1}$ & 27.0 & 8.0$\times 10^{-2}$ & 35.3 & 8.8$\times 10^{-3}$ & 21.9 & 2.4$\times 10^{-1}$ \\
{\sc Alpgen+Pythia}6 & 13.2 & 7.8$\times 10^{-1}$ & 27.4 & 7.2$\times 10^{-2}$ & 29.0 & 4.8$\times 10^{-2}$ & 24.8 & 1.3$\times 10^{-1}$ \\
{\sc MadGraph+Pythia}6 & 12.3 & 8.3$\times 10^{-1}$ & 19.7 & 3.5$\times 10^{-1}$ & 28.9 & 5.0$\times 10^{-2}$ & 16.3 & 5.7$\times 10^{-1}$ \\ \hline
{\sc AcerMC+Pythia}6 RadHi & 81.0 & 5.8$\times 10^{-10}$ & 44.6 & 4.7$\times 10^{-4}$ & 40.2 & 2.0$\times 10^{-3}$ & 112.5 & 1.1$\times 10^{-15}$ \\
{\sc AcerMC+Pythia}6 RadLo & 55.5 & 1.1$\times 10^{-5}$ & 38.4 & 3.4$\times 10^{-3}$ & 41.5 & 1.3$\times 10^{-3}$ & 93.9 & 2.9$\times 10^{-12}$ \\
{\sc Alpgen+Pythia}6 RadHi & 35.1 & 9.2$\times 10^{-3}$ & 47.0 & 2.1$\times 10^{-4}$ & 38.8 & 3.0$\times 10^{-3}$ & 40.7 & 1.7$\times 10^{-3}$ \\
{\sc Alpgen+Pythia}6 RadLo & 11.2 & 8.9$\times 10^{-1}$ & 19.0 & 4.0$\times 10^{-1}$ & 25.2 & 1.2$\times 10^{-1}$ & 18.8 & 4.1$\times 10^{-1}$ \\
{\sc MadGraph+Pythia}6 $q^2$ down & 17.8 & 4.7$\times 10^{-1}$ & 25.2 & 1.2$\times 10^{-1}$ & 33.7 & 1.4$\times 10^{-2}$ & 21.1 & 2.8$\times 10^{-1}$ \\
{\sc MadGraph+Pythia}6 $q^2$ up & 21.0 & 2.8$\times 10^{-1}$ & 25.3 & 1.2$\times 10^{-1}$ & 32.4 & 1.9$\times 10^{-2}$ & 28.0 & 6.2$\times 10^{-2}$ \\
\hline
\end{tabular}
\caption{Values of $\chi^2$ for the comparison of the measured
gap fraction distributions with the predictions from various \ttbar\
generator configurations, for the four rapidity regions as a function of \qzero.
The $\chi^2$ and $p$-values correspond to 18 degrees of freedom.}
\label{table:chirapQ0}
\end{table}
\begin{table}[]
\centering
\footnotesize
\begin{tabular}{|l|ll|ll|ll|ll|}
\hline
$Q_{\rm{sum}}$&\multicolumn{2}{l|}{$|y|<0.8$}&\multicolumn{2}{l|}{$0.8<|y|<1.5$}&\multicolumn{2}{l|}{$1.5<|y|<2.1$}&\multicolumn{2}{l|}{$|y|<2.1$}\\
Generator&$\chi^2$ & $p$-value & $\chi^2$ & $p$-value & $\chi^2$    & $p$-value      & $\chi^2$  & $p$-value    \\ \hline

{\sc Powheg+Pythia}6 $h_{\rm{damp}}=\infty$ 	& 22.3 & 4.4$\times 10^{-1}$ & 41.9 & 6.5$\times 10^{-3}$ & 25.8 & 2.6$\times 10^{-1}$ & 39.1 & 1.4$\times 10^{-2}$ \\
{\sc Powheg+Pythia}6 $h_{\rm{damp}}=m_t$ 	& 25.6 & 2.7$\times 10^{-1}$ & 33.8 & 5.2$\times 10^{-2}$ & 27.8 & 1.8$\times 10^{-1}$ & 47.0 & 1.5$\times 10^{-3}$ \\
{\sc Powheg+Pythia}8 $h_{\rm{damp}}=m_t$ 	& 18.3 & 6.9$\times 10^{-1}$ & 27.3 & 2.0$\times 10^{-1}$ & 21.9 & 4.6$\times 10^{-1}$ & 34.3 & 4.6$\times 10^{-2}$ \\
{\sc MC@NLO+Herwig} 					& 32.2 & 7.4$\times 10^{-2}$ & 28.5 & 1.6$\times 10^{-1}$ & 32.1 & 7.5$\times 10^{-2}$ & 44.6 & 3.0$\times 10^{-3}$ \\
{\sc Powheg+Herwig} $h_{\rm{damp}}=\infty$ 	& 21.4 & 5.0$\times 10^{-1}$ & 37.7 & 2.0$\times 10^{-2}$ & 29.6 & 1.3$\times 10^{-1}$ & 29.9 & 1.2$\times 10^{-1}$ \\ \hline

{\sc Alpgen+Herwig} 					& 33.1 & 6.1$\times 10^{-2}$ & 37.7 & 2.0$\times 10^{-2}$ & 32.4 & 7.1$\times 10^{-2}$ & 31.4 & 8.8$\times 10^{-2}$ \\
{\sc Alpgen+Pythia}6 					& 23.0 & 4.0$\times 10^{-1}$ & 39.7 & 1.2$\times 10^{-2}$ & 28.8 & 1.5$\times 10^{-1}$ & 37.0 & 2.4$\times 10^{-2}$ \\
{\sc MadGraph+Pythia}6 					& 20.1 & 5.7$\times 10^{-1}$ & 30.3 & 1.1$\times 10^{-1}$ & 28.1 & 1.7$\times 10^{-1}$ & 31.9 & 7.9$\times 10^{-2}$ \\ \hline

{\sc AcerMC+Pythia}6 RadHi 				& 75.0 & 1.1$\times 10^{-7}$ & 48.0 & 1.1$\times 10^{-3}$ & 35.5 & 3.4$\times 10^{-2}$ & 91.1 & 2.2$\times 10^{-10}$ \\
{\sc AcerMC+Pythia}6 RadLo 				& 62.4 & 9.9$\times 10^{-6}$ & 46.4 & 1.8$\times 10^{-3}$ & 35.6 & 3.4$\times 10^{-2}$ & 94.8 & 5.2$\times 10^{-11}$ \\
{\sc Alpgen+Pythia}6 RadHi 				& 44.3 & 3.3$\times 10^{-3}$ & 62.3 & 1.0$\times 10^{-5}$ & 42.2 & 5.9$\times 10^{-3}$ & 61.3 & 1.4$\times 10^{-5}$ \\
{\sc Alpgen+Pythia}6 RadLo 				& 17.4 & 7.4$\times 10^{-1}$ & 34.6 & 4.2$\times 10^{-2}$ & 26.8 & 2.2$\times 10^{-1}$ & 34.6 & 4.2$\times 10^{-2}$ \\
{\sc MadGraph+Pythia}6 $q^2$ down 		& 22.5 & 4.3$\times 10^{-1}$ & 35.3 & 3.6$\times 10^{-2}$ & 31.1 & 9.5$\times 10^{-2}$ & 29.3 & 1.4$\times 10^{-1}$ \\
{\sc MadGraph+Pythia}6 $q^2$ up 			& 25.0 & 3.0$\times 10^{-1}$ & 38.4 & 1.7$\times 10^{-2}$ & 34.3 & 4.6$\times 10^{-2}$ & 50.8 & 4.6$\times 10^{-4}$ \\ 
\hline\end{tabular}
\caption{Values of $\chi^2$ for the comparison of the measured
gap fraction distributions with the predictions from various \ttbar\
generator configurations, for the four rapidity regions as a function of \qsum.
The $\chi^2$ and $p$-values correspond to 22 degrees of freedom.}
\label{table:chirapQsum}
\end{table}

All the NLO generators provide a reasonable description of the \fqzero\ 
distribution in the regions  $|y|<0.8$ and $0.8<|y|<1.5$. All these 
generators are also consistent with the data in the most forward 
region ($1.5<|y|<2.1$), whereas at $\sqrt{s}=7$\,\TeV, they tended to lie
below the data \cite{TOPQ-2011-21}.
However, the current measurements are significantly more 
precise in this region, thanks in particular to improvements in the jet energy
scale calibration. Over the full rapidity range ($|y|<2.1$), 
{\sc Powheg\,+\,Pythia8} provides the best description of the data,
whilst {\sc Powheg\,+\,Pythia6} with $\hdmp=\mtop$ and {\sc MC@NLO\,+\,Herwig}
predict slightly less radiation, and {\sc Powheg\,+\,Pythia6} with 
$\hdmp=\infty$
and {\sc Powheg\,+\,Herwig} predict slightly more. {\sc Powheg\,+\,Pythia8}
also provides the best description across the individual $|y|$ regions. The
results for \fqsum, which are sensitive to all the additional jets within
the rapidity interval, show somewhat larger differences between the
generators than those for \fqzero. Over the rapidity
region $|y|<2.1$, {\sc Powheg\,+\,Pythia6} with $\hdmp=\mtop$ and
{\sc MC@NLO\,+\,Herwig} are disfavoured. The latter generator combination
also performs poorly
for the differential cross-section measurements discussed in Section~\ref{sec:jets}.

The leading-order generators {\sc Alpgen\,+\,Pythia6}, {\sc Alpgen\,+\,Herwig}
and {\sc MadGraph\,+Pythia6} also provide a reasonable description of the gap 
fraction as a function of \qzero\ and \qsum.  The pairs of samples with
increased/decreased radiation also bracket the data in all rapidity regions,
except for {\sc AcerMC\,+\,Pythia6}, which always predicts higher gap fractions
than observed at high \qzero\ and \qsum. As in the differential cross-section
measurements, the data show a clear preference for the `RadLo' variation
for {\sc Alpgen\,+\,Pythia6} and the `$q^2$ down' variation for 
\madpy, across all rapidity regions. These
data should therefore allow the uncertainties due to radiation modelling
in \ttbar\ events to be significantly reduced, once the models are tuned to
these more precise $\sqrt{s}=8$\,\TeV\ results rather than the 
$\sqrt{s}=7$\,\TeV\ results used previously \cite{TOPQ-2011-21}.

\subsection{Gap fraction results in \emubb\ mass regions}\label{ssec:gfracresmass}

The gap fraction was also measured over the full rapidity veto region $|y|<2.1$ 
after dividing the data sample into
four regions of \memubb. The distribution of reconstructed \memubb\ in the
selected \emubb\ events is shown in Figure~\ref{fig:massdist}, and is 
reasonably well-reproduced by the baseline \ttbar\ simulation sample. The
distribution was divided into four regions at both reconstruction and 
particle level: $\memubb<300$\,\GeV, 
$300<\memubb<425$\,\GeV, $425<\memubb<600$\,\GeV\ and $\memubb>600$\,GeV.
These boundaries were chosen to minimise migration between the regions; in
the baseline simulation, around 85\,\% of the reconstructed events in each
\memubb\ region come from the corresponding region at particle level. 
The corresponding correction factors \cormqz\ which translate the measured
gap fraction in the reconstruction-level \memubb\ region to the corresponding
particle-level gap fractions \fmqzero\ and \fmqsum, 
are of similar size to \corqz, with the exception 
of the highest \memubb\ region, where they reach about 1.1 at low \qzero.
The systematic uncertainties in the gap fraction measurement 
in two \memubb\ regions are shown in Figure~\ref{fig:gfsystmass}.
The magnitudes of the systematic uncertainties are comparable to those in the 
full \memubb\ range, except for the highest \memubb\ region where they
are significantly larger.

\begin{figure}[tp]
\centering
\includegraphics[width=0.6\textwidth]{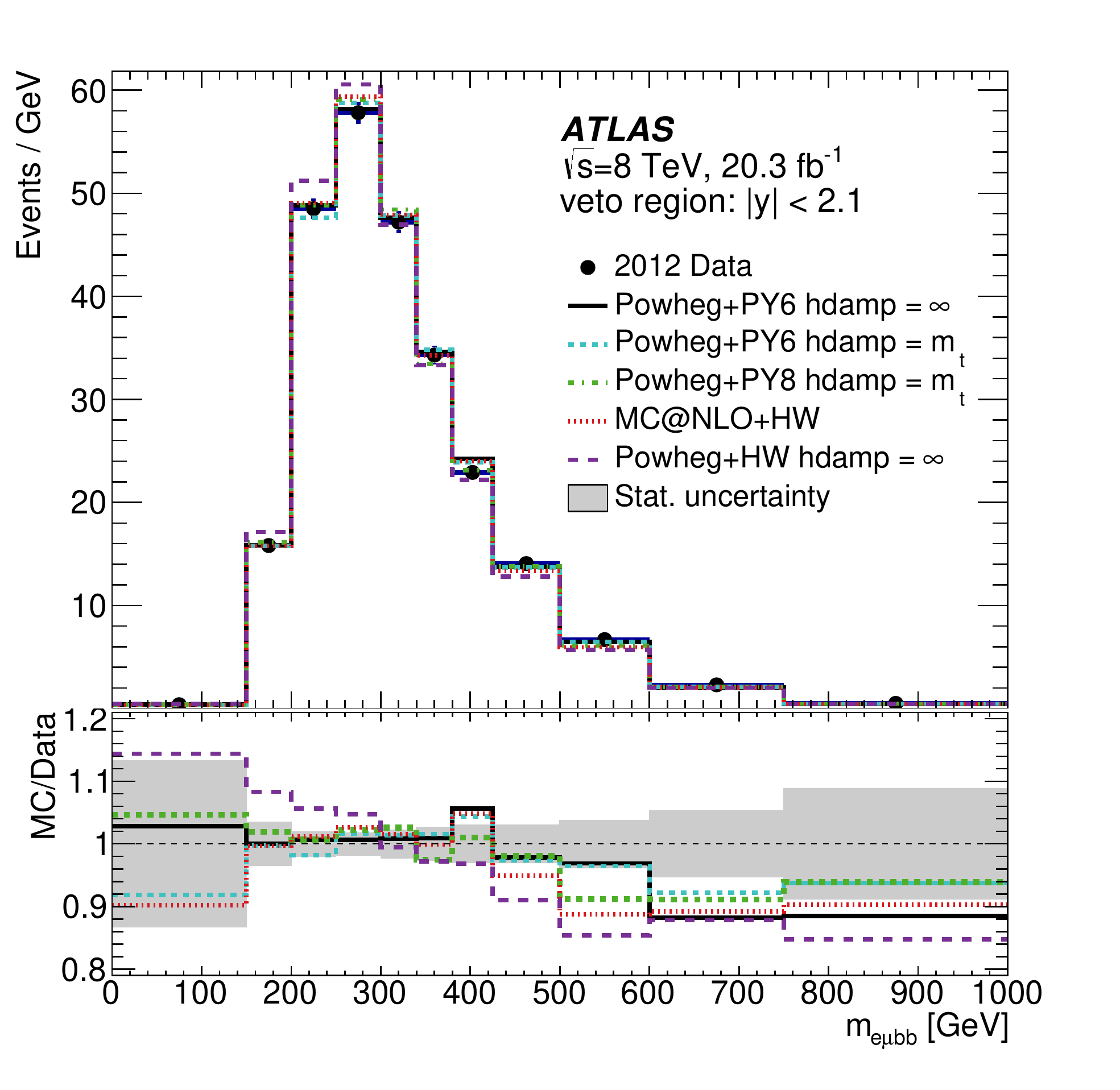}
\caption{Distribution of the reconstructed invariant mass of the \emubb\ system
\memubb\ in data, compared to simulation using various \ttbar\ generators. 
The shaded band represents the statistical uncertainty in data. The lower plot shows the ratio of the distribution of invariant mass in data to that in each of the simulation samples.}
\label{fig:massdist}
\end{figure}

\begin{table}[] 
\centering
\small
\begin{tabular}{llllllll}
\hline & \multicolumn{6}{c}{\fmqzero\ [\%]}& \\ \cline{2-7}
\begin{tabular}[c]{@{}l@{}}$Q_0$\\{[\gev]}\end{tabular} &\begin{tabular}[c]{@{}l@{}}Data\\$\pm$(stat.)$\pm$(syst.)\end{tabular}& \begin{tabular}[c]{@{}l@{}}\sc{Powheg}\\ +\sc{Pythia6} \\ $\hdmp=\infty$ \end{tabular} & \begin{tabular}[c]{@{}l@{}}\sc{Powheg}\\ +\sc{Pythia6}\\ $h_{\mathrm{damp}}=m_t$\end{tabular} & \begin{tabular}[c]{@{}l@{}}\sc{Powheg}\\ +\sc{Pythia8}\\ $\hdmp=m_t$\end{tabular} & \begin{tabular}[c]{@{}l@{}}\sc{MC@NLO}\\ +\sc{Herwig}\end{tabular} & \begin{tabular}[c]{@{}l@{}}\sc{Powheg}\\ +\sc{Herwig}\\ $\hdmp=\infty$ \end{tabular} & \begin{tabular}[c]{@{}l@{}}$\rho_j^i$\\ (stat.+syst.)\end{tabular}\\ \hline
\multicolumn{5}{l}{veto region: $|y|<2.1$, \memubb\ < 300 \gev}&&& \\[0.5pt]
25&56.0$\pm$0.6$\pm 2.0$&55.1$\pm$0.3&57.8$\pm$0.3&56.0$\pm$0.3&57.3$\pm$0.3&53.5$\pm$0.3&$\rho^{25}_{75}=$ 0.60\\[1pt]
75&86.7$\pm$0.5$\pm 0.7$&84.3$\pm$0.3&86.9$\pm$0.2&86.2$\pm$0.2&86.4$\pm$0.2&84.4$\pm$0.2&$\rho^{75}_{150}=$ 0.80\\[1pt]
150&95.7$\pm$0.3$\pm 0.3$&94.5$\pm$0.2&95.9$\pm$0.1&95.7$\pm$0.1&95.9$\pm$0.1&94.7$\pm$0.2&$\rho^{150}_{25}=$ 0.50\\[1pt]
\hline\multicolumn{5}{l}{veto region: $|y|<2.1$, 300 < \memubb\ < 425 \gev}&&& \\[0.5pt]
25&54.4$\pm$0.8$\pm 1.8$&53.5$\pm$0.4&57.0$\pm$0.4&54.6$\pm$0.4&56.2$\pm$0.4&52.2$\pm$0.4&$\rho^{25}_{75}=$ 0.63\\[1pt]
75&84.7$\pm$0.6$\pm 0.7$&82.7$\pm$0.3&85.8$\pm$0.3&84.9$\pm$0.3&85.4$\pm$0.3&83.0$\pm$0.3&$\rho^{75}_{150}=$ 0.48\\[1pt]
150&95.0$\pm$0.4$\pm 0.6$&93.8$\pm$0.2&95.4$\pm$0.2&95.2$\pm$0.2&95.4$\pm$0.2&94.1$\pm$0.2&$\rho^{150}_{25}=$ 0.52\\[1pt]
\hline\multicolumn{5}{l}{veto region: $|y|<2.1$, 425 < \memubb\ < 600 \gev}&&& \\[0.5pt]
25&47.6$\pm$1.3$\pm 1.7$&51.0$\pm$0.7&54.2$\pm$0.7&51.6$\pm$0.6&53.4$\pm$0.6&48.1$\pm$0.7&$\rho^{25}_{75}=$ 0.63\\[1pt]
75&79.0$\pm$1.0$\pm 1.0$&80.3$\pm$0.5&83.7$\pm$0.5&82.9$\pm$0.5&82.5$\pm$0.5&80.0$\pm$0.5&$\rho^{75}_{150}=$ 0.52\\[1pt]
150&92.7$\pm$0.7$\pm 0.8$&92.6$\pm$0.3&94.4$\pm$0.3&94.2$\pm$0.3&94.0$\pm$0.3&92.6$\pm$0.3&$\rho^{150}_{25}=$ 0.11\\[1pt]
\hline\multicolumn{5}{l}{veto region: $|y|<2.1$, \memubb\ > 600 \gev}&&& \\[0.5pt]
25&45.9$\pm$2.3$\pm 3.9$&45.2$\pm$1.2&49.8$\pm$1.2&46.8$\pm$1.2&51.0$\pm$1.2&43.9$\pm$1.2&$\rho^{25}_{75}=$ 0.82\\[1pt]
75&81.7$\pm$2.0$\pm 3.6$&75.7$\pm$1.0&80.3$\pm$1.0&78.8$\pm$1.0&79.8$\pm$1.0&75.9$\pm$1.1&$\rho^{75}_{150}=$ 0.85\\[1pt]
150&92.4$\pm$1.3$\pm 2.8$&89.7$\pm$0.7&92.6$\pm$0.6&92.3$\pm$0.6&91.5$\pm$0.7&90.5$\pm$0.7&$\rho^{150}_{25}=$ 0.72\\[1pt]
\hline\end{tabular}
\caption{The measured gap fraction values \fmqzero\ for the veto region $|y|<2.1$ and four invariant mass regions, for \qzero\ values of 25, 75 and 150 \gev\ in data compared to the predictions from various \ttbar\ simulation samples. The combination of statistical and systematic correlations between measurements at $\qzero=i$ and $\qzero=j$ is given as $\rho_j^i$.}
\label{table:q0massresults}
\end{table}
\begin{table}[] 
\centering
\small
\begin{tabular}{llllllll}
\hline & \multicolumn{6}{c}{\fmqsum\ [\%]}& \\ \cline{2-7}
\begin{tabular}[c]{@{}l@{}}$Q_{\mathrm{sum}}$\\{[\gev]}\end{tabular} &\begin{tabular}[c]{@{}l@{}}Data\\$\pm$(stat.)$\pm$(syst.)\end{tabular}& \begin{tabular}[c]{@{}l@{}}\sc{Powheg}\\ +\sc{Pythia6} \\ $\hdmp=\infty$ \end{tabular} & \begin{tabular}[c]{@{}l@{}}\sc{Powheg}\\ +\sc{Pythia6}\\ $h_{\mathrm{damp}}=m_t$\end{tabular} & \begin{tabular}[c]{@{}l@{}}\sc{Powheg}\\ +\sc{Pythia8}\\ $\hdmp=m_t$\end{tabular} & \begin{tabular}[c]{@{}l@{}}\sc{MC@NLO}\\ +\sc{Herwig}\end{tabular} & \begin{tabular}[c]{@{}l@{}}\sc{Powheg}\\ +\sc{Herwig}\\ $\hdmp=\infty$ \end{tabular} & \begin{tabular}[c]{@{}l@{}}$\rho_j^i$\\ (stat.+syst.)\end{tabular}\\ \hline
\multicolumn{5}{l}{veto region: $|y|<2.1$, \memubb\ < 300 \gev}&&& \\[0.5pt]
55&75.0$\pm$0.6$\pm 1.4$&73.5$\pm$0.3&76.7$\pm$0.3&74.9$\pm$0.3&76.1$\pm$0.3&72.8$\pm$0.3&$\rho^{55}_{150}=$ 0.80\\[1pt]
150&92.5$\pm$0.4$\pm 0.5$&91.0$\pm$0.2&93.0$\pm$0.2&92.2$\pm$0.2&93.5$\pm$0.2&91.1$\pm$0.2&$\rho^{150}_{300}=$ 0.71\\[1pt]
300&98.3$\pm$0.2$\pm 0.2$&97.8$\pm$0.1&98.4$\pm$0.1&98.1$\pm$0.1&98.9$\pm$0.1&97.9$\pm$0.1&$\rho^{300}_{55}=$ 0.71\\[1pt]
\hline\multicolumn{5}{l}{veto region: $|y|<2.1$, 300 < \memubb\ < 425 \gev}&&& \\[0.5pt]
55&72.8$\pm$0.7$\pm 1.3$&71.7$\pm$0.4&75.2$\pm$0.4&73.2$\pm$0.4&74.8$\pm$0.3&71.1$\pm$0.4&$\rho^{55}_{150}=$ 0.78\\[1pt]
150&91.4$\pm$0.5$\pm 0.8$&90.0$\pm$0.2&92.2$\pm$0.2&91.2$\pm$0.2&92.8$\pm$0.2&90.1$\pm$0.2&$\rho^{150}_{300}=$ 0.65\\[1pt]
300&98.1$\pm$0.2$\pm 0.2$&97.3$\pm$0.1&98.1$\pm$0.1&97.7$\pm$0.1&98.6$\pm$0.1&97.5$\pm$0.1&$\rho^{300}_{55}=$ 0.62\\[1pt]
\hline\multicolumn{5}{l}{veto region: $|y|<2.1$, 425 < \memubb\ < 600 \gev}&&& \\[0.5pt]
55&67.4$\pm$1.2$\pm 2.2$&68.7$\pm$0.6&72.5$\pm$0.6&70.3$\pm$0.6&71.5$\pm$0.6&67.3$\pm$0.6&$\rho^{55}_{150}=$ 0.61\\[1pt]
150&87.9$\pm$0.8$\pm 0.8$&87.9$\pm$0.4&90.6$\pm$0.4&89.5$\pm$0.4&90.6$\pm$0.4&87.7$\pm$0.4&$\rho^{150}_{300}=$ 0.61\\[1pt]
300&96.4$\pm$0.5$\pm 0.3$&96.4$\pm$0.2&97.4$\pm$0.2&97.1$\pm$0.2&98.0$\pm$0.2&96.6$\pm$0.2&$\rho^{300}_{55}=$ 0.31\\[1pt]
\hline\multicolumn{5}{l}{veto region: $|y|<2.1$, \memubb\ > 600 \gev}&&& \\[0.5pt]
55&63.2$\pm$2.3$\pm 4.2$&62.6$\pm$1.2&67.6$\pm$1.1&65.0$\pm$1.1&67.9$\pm$1.1&61.9$\pm$1.2&$\rho^{55}_{150}=$ 0.82\\[1pt]
150&87.3$\pm$1.7$\pm 3.0$&83.6$\pm$0.9&87.6$\pm$0.8&85.4$\pm$0.8&87.3$\pm$0.8&84.1$\pm$0.9&$\rho^{150}_{300}=$ 0.80\\[1pt]
300&97.3$\pm$0.8$\pm 2.3$&94.5$\pm$0.6&96.3$\pm$0.5&95.7$\pm$0.5&96.5$\pm$0.5&95.1$\pm$0.5&$\rho^{300}_{55}=$ 0.74\\[1pt]
\hline\end{tabular}
\caption{The measured gap fraction values \fmqsum\ for the veto region $|y|<2.1$ and four invariant mass regions, for \qsum\ values of 55, 150 and 300 \gev\ in data compared to the predictions from various \ttbar\ simulation samples. The combination of statistical and systematic correlations between measurements at $\qsum=i$ and $\qsum=j$ is given as $\rho_j^i$.}
\label{table:qsummassresults}
\end{table}

\begin{figure}[htp]
\centering
\subfloat[][]{
\includegraphics[width=0.5\textwidth]{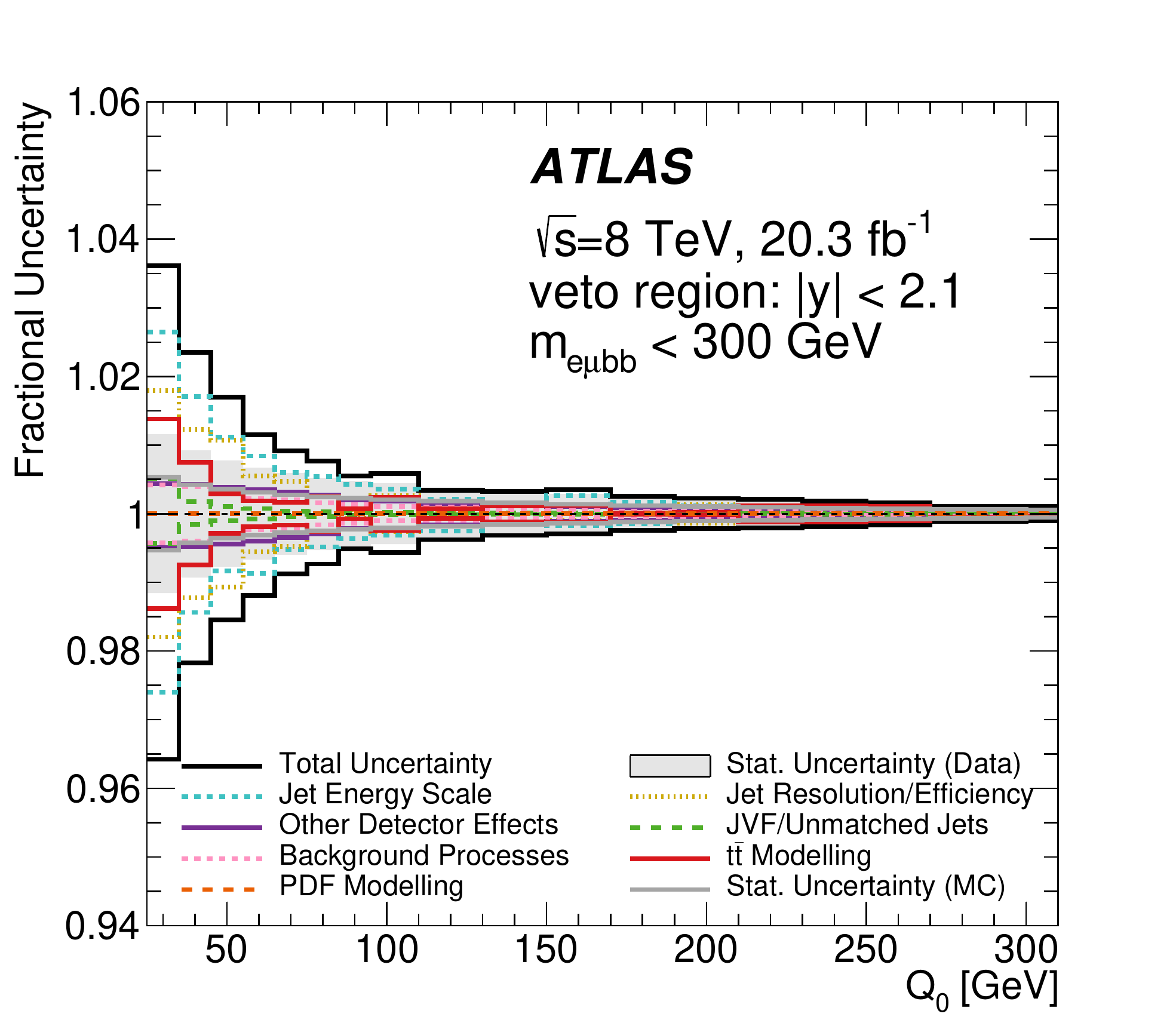}}
\subfloat[][]{
\includegraphics[width=0.5\textwidth]{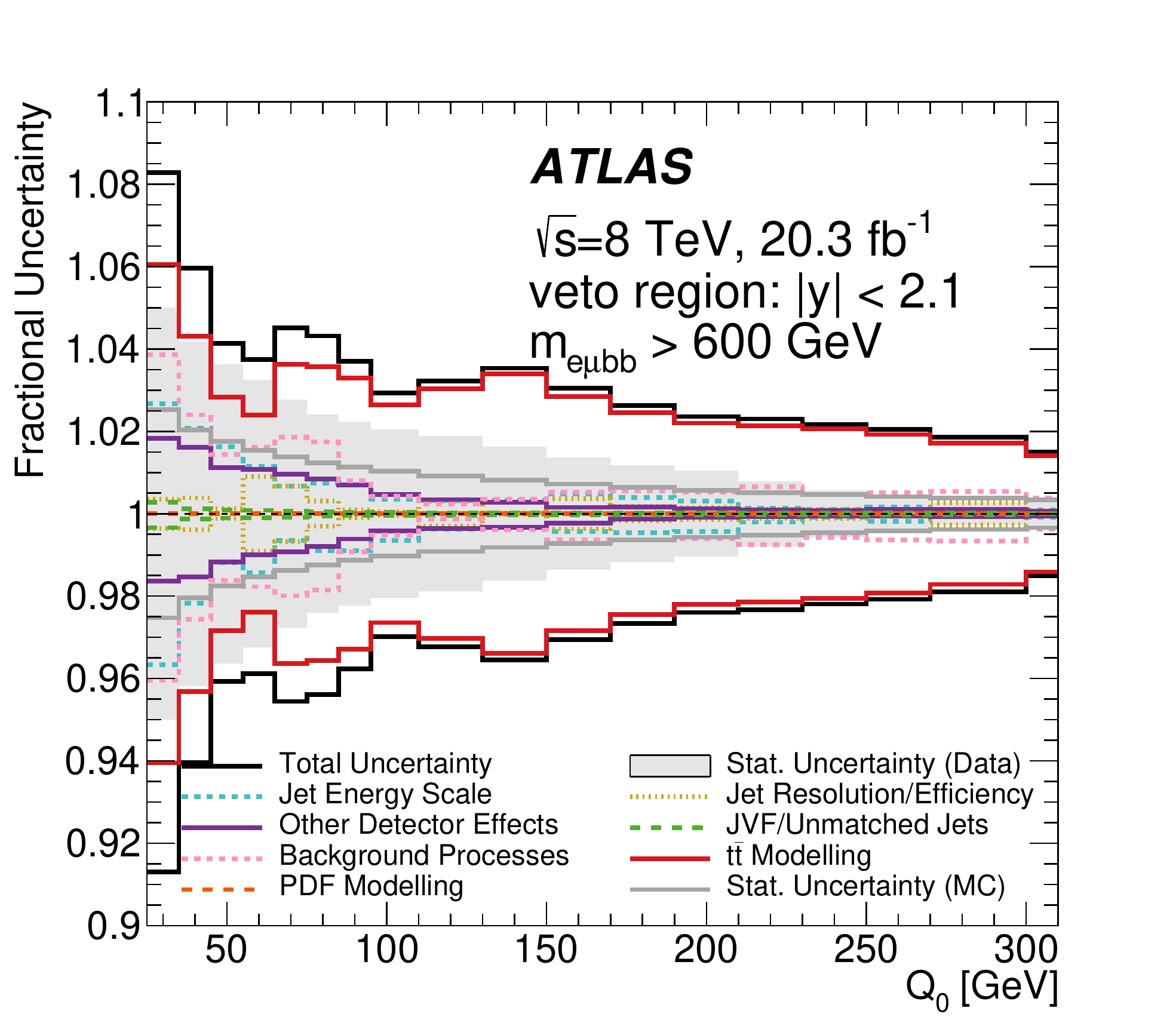}} \\
\caption{Envelope of fractional uncertainties $\Delta f/f$ in the gap fraction \fmqzero\ for 
(a) $\memubb<300$\,\GeV\ and (b) $\memubb>600$\,GeV. The statistical 
uncertainty is shown by the
hatched area, and the total systematic uncertainty by the solid black line. 
The systematic uncertainty is also shown broken down into several groups,
each of which includes various individual components (see text).}
\label{fig:gfsystmass}
\end{figure}

Figures \ref{fig:gapfracpartmassa} and~\ref{fig:gapfracpartmassb} show 
the resulting measurements of the gap fractions as a function of \qzero\ in
the four \memubb\ regions in
data, compared to the same set of predictions as shown in 
Figures~\ref{fig:gapfracparta}, \ref{fig:gapfracpartb} 
and~\ref{fig:gapfracparttota}.
Tables~\ref{table:q0massresults} and~\ref{table:qsummassresults} show the
gap fractions at selected \qzero\ and \qsum\ values in each invariant
mass region, again compared to predictions from the first set of generators.
Figure \ref{fig:altmass} gives an alternative presentation of the gap fraction 
\fmqzero\ as a function of \memubb\ for four different \qzero\ values. The 
$\chi^2$ values for the consistency of the prediction from each NLO 
generator with data in the four mass regions are
given in Tables~\ref{table:chimassQ0} and~\ref{table:chimassQsum}.

\begin{figure}[htp]
\centering
\subfloat[][]{\includegraphics[width=0.52\linewidth]{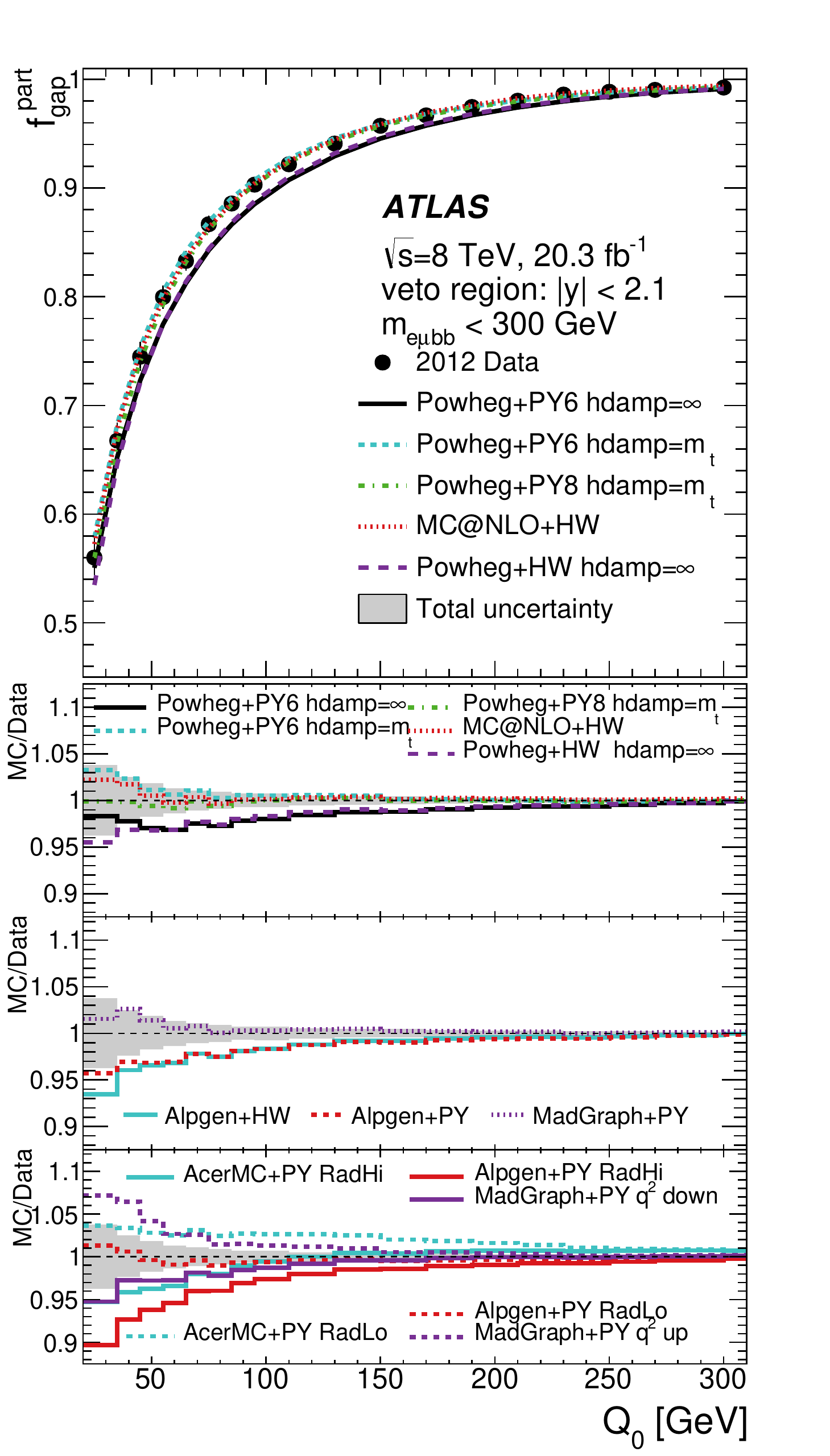}}
\subfloat[][]{\includegraphics[width=0.52\linewidth]{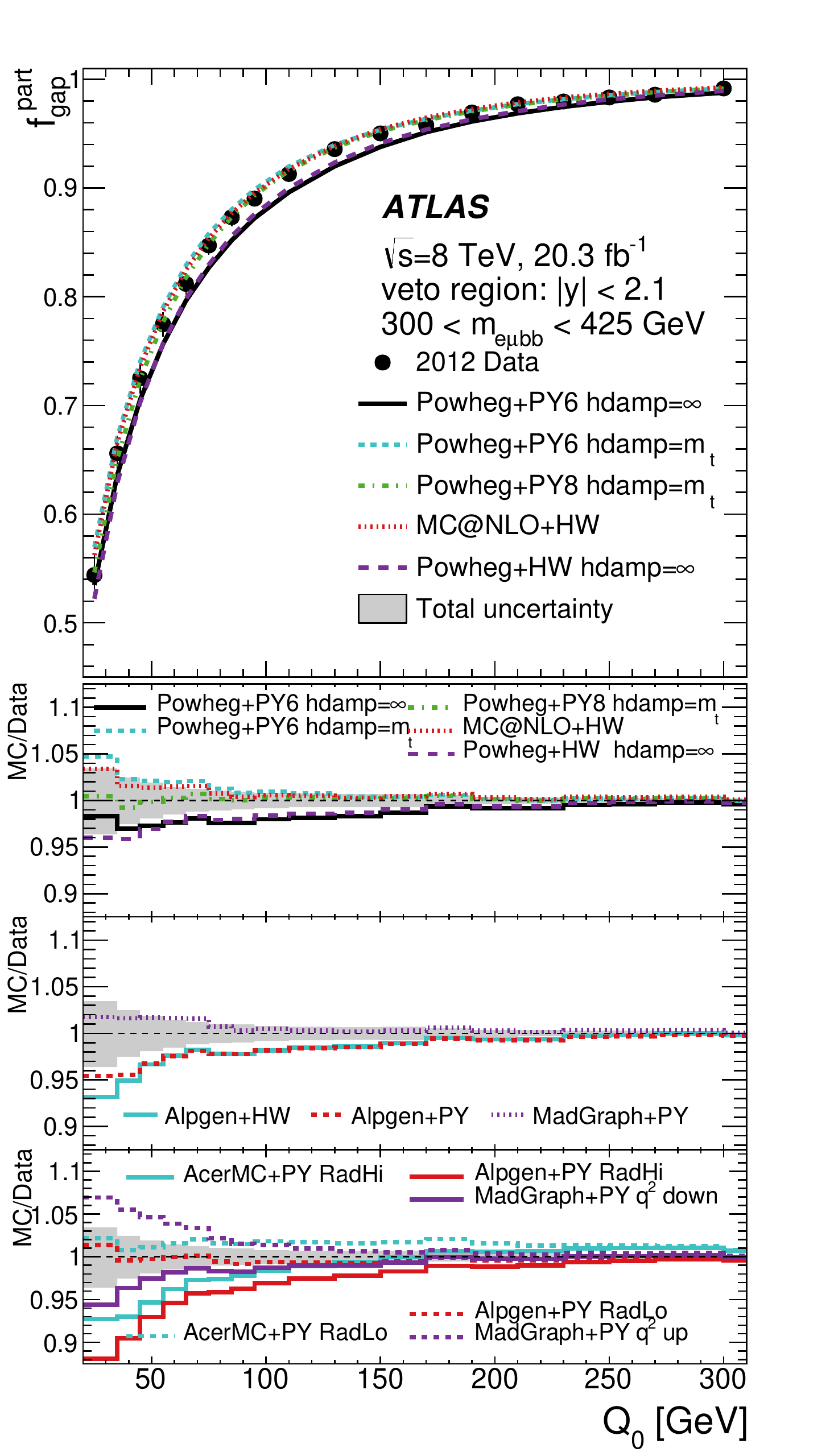}}
\caption{The measured gap fraction \fmqzero\ as a function of \qzero\ in the
veto region $|y|<2.1$ for the invariant mass regions (a) $\memubb<300$\,\GeV\ and (b) $300<\memubb<425$\,\GeV.
The data are shown by the points with
error bars indicating the total uncertainty, and compared to the predictions
from various \ttbar\ simulation samples (see text) shown as smooth curves. 
The lower plots show the
ratio of predictions to data, with the data uncertainty being indicated by
the shaded band, and the \qzero\ thresholds corresponding to the left edges
of the histogram bins, except for the first bin.}
\label{fig:gapfracpartmassa}
\end{figure}

\begin{figure}[htp]
\centering
\subfloat[][]{\includegraphics[width=0.52\linewidth]{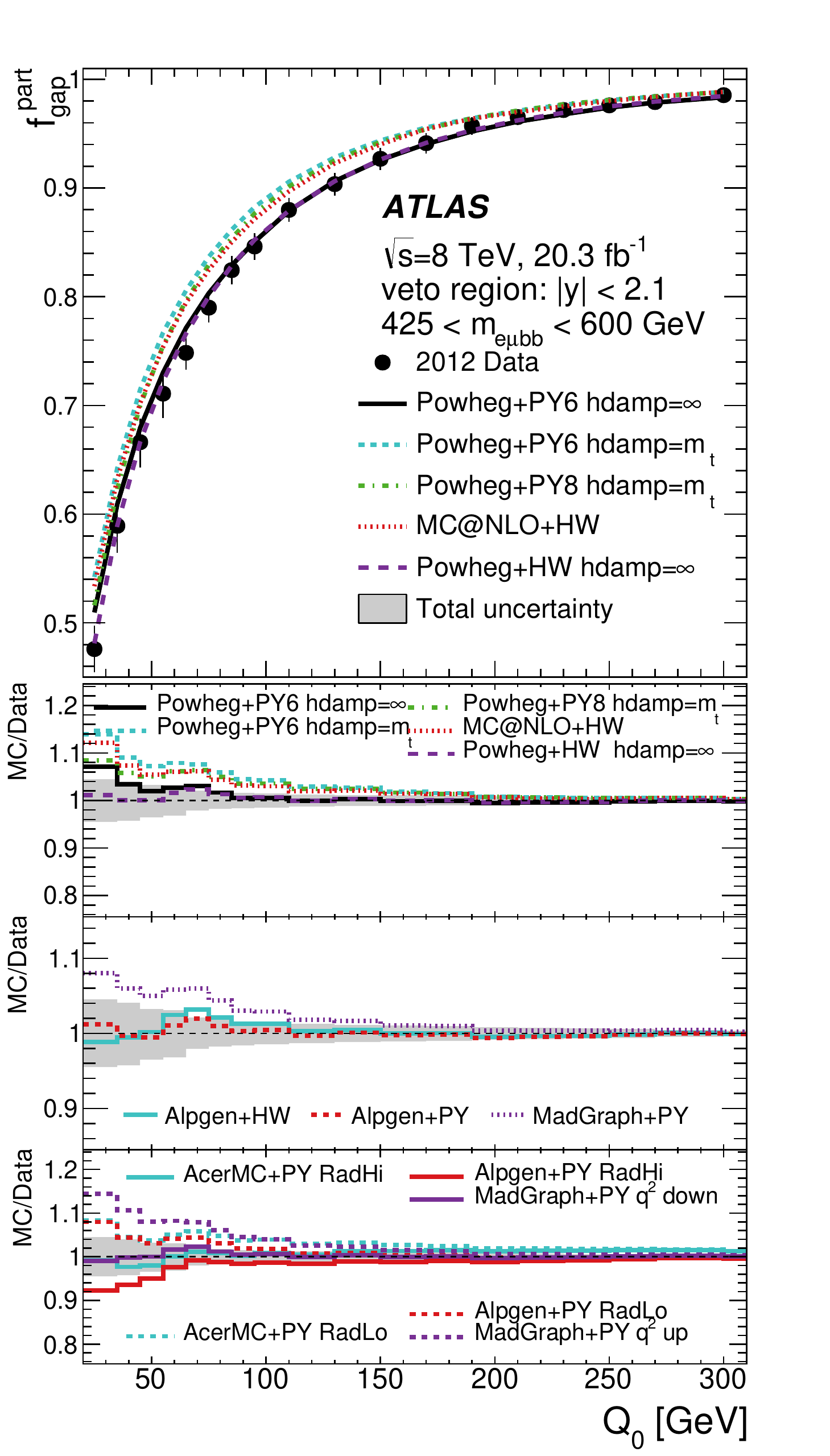}}
\subfloat[][]{\includegraphics[width=0.52\linewidth]{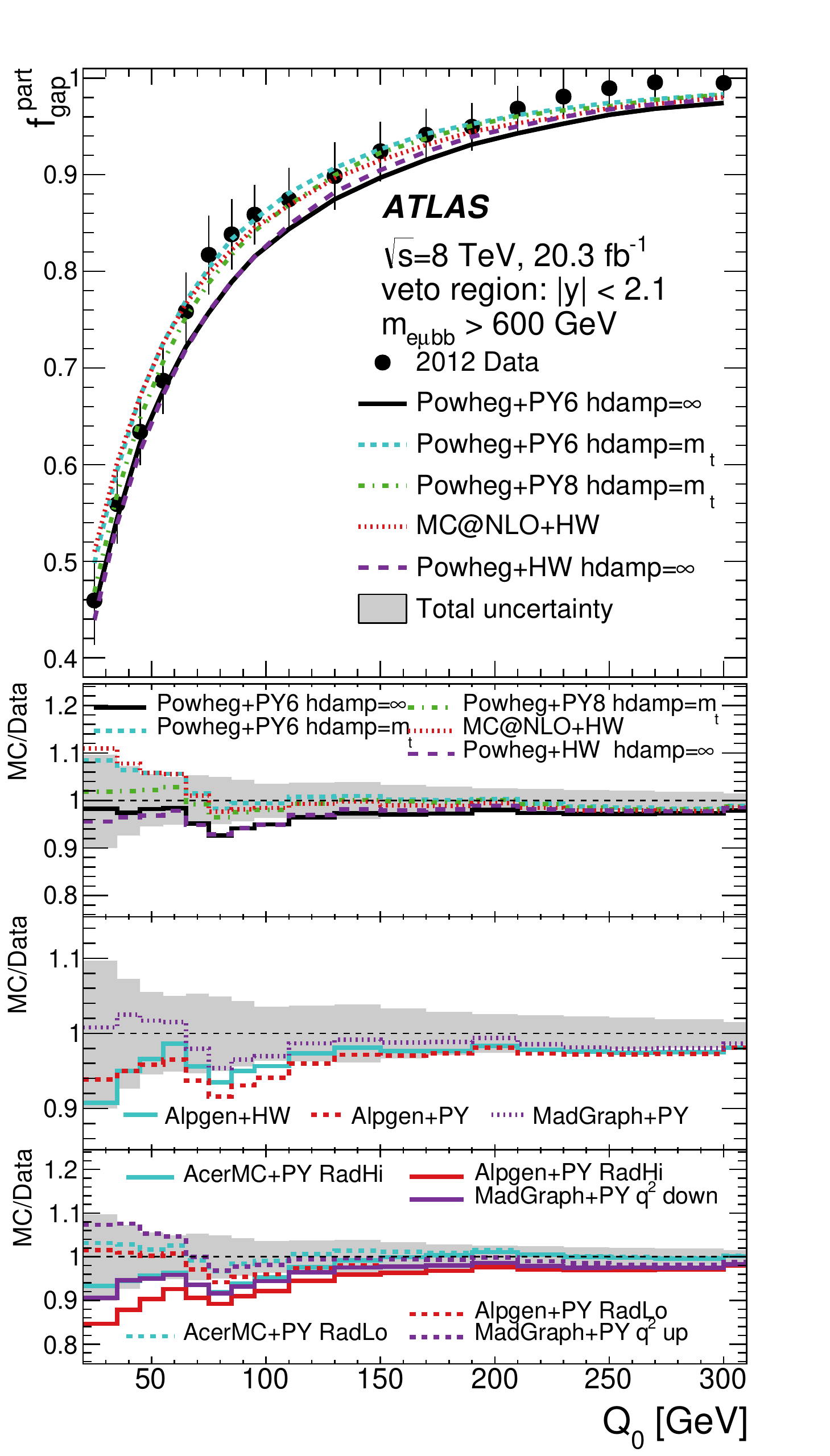}}
\caption{The measured gap fraction \fmqzero\ as a function of \qzero\ in the
veto region $|y|<2.1$ for the invariant mass regions (a) $425<\memubb<600$\,\GeV\ and (b) $\memubb>600$\,\GeV.
The data are shown by the points with
error bars indicating the total uncertainty, and compared to the predictions
from various \ttbar\ simulation samples (see text) shown as smooth curves. 
The lower plots show the
ratio of predictions to data, with the data uncertainty being indicated by
the shaded band, and the \qzero\ thresholds corresponding to the left edges
of the histogram bins, except for the first bin.}
\label{fig:gapfracpartmassb}
\end{figure}

\begin{figure}[htp]
\centering
\includegraphics[width=0.6\linewidth]{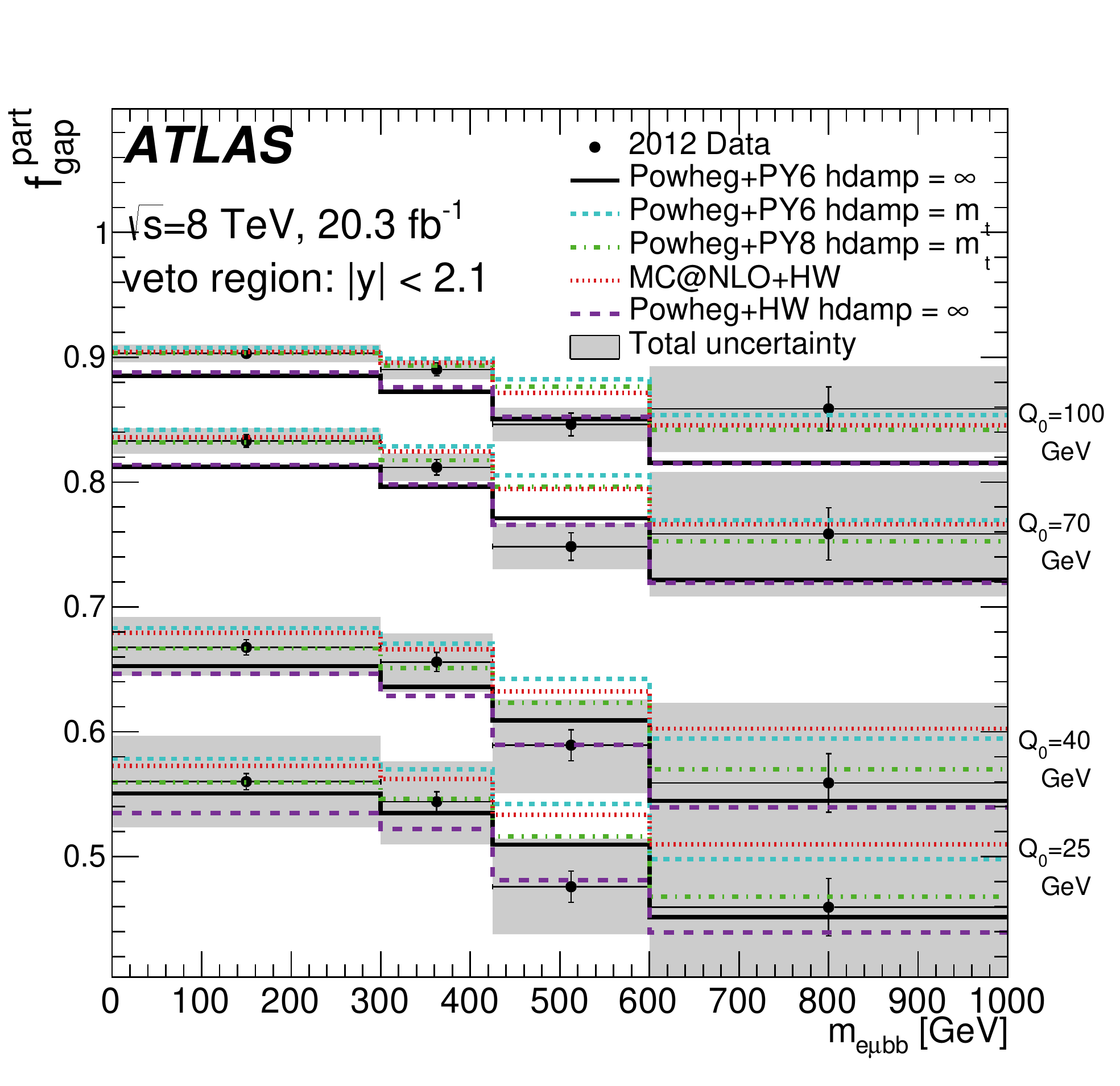}
\caption{The gap fraction measurement \fmqzero\ as a function of the
invariant mass \memubb, for
several different values of \qzero. The data are shown as points with error bars
indicating the statistical uncertainties and shaded boxes 
the total uncertainties. The
data are compared to the predictions from various \ttbar\ simulation samples.}
\label{fig:altmass}
\end{figure}
\begin{table}[]
\centering
\footnotesize
\begin{tabular}{|l|ll|ll|ll|ll|}
\hline
$Q_0$&\multicolumn{2}{l|}{$m<300$~\gev}&\multicolumn{2}{l|}{$300<m<425$~\gev}&\multicolumn{2}{l|}{$425<m<600$~\gev}&\multicolumn{2}{l|}{$m>600$~\gev}\\
Generator&$\chi^2$ & $p$-value & $\chi^2$ & $p$-value & $\chi^2$    & $p$-value      & $\chi^2$  & $p$-value    \\ \hline
{\sc Powheg+Pythia}6 $h_{\rm{damp}}=\infty$ 	& 25.5 & 1.1$\times 10^{-1}$ & 27.8 & 6.5$\times 10^{-2}$ & 17.4 & 5.0$\times 10^{-1}$ & 14.0 & 7.3$\times 10^{-1}$ \\
{\sc Powheg+Pythia}6 $h_{\rm{damp}}=m_t$ 	& 18.3 & 4.4$\times 10^{-1}$ & 22.2 & 2.2$\times 10^{-1}$ & 34.8 & 1.0$\times 10^{-2}$ & 20.0 & 3.3$\times 10^{-1}$ \\
{\sc Powheg+Pythia}8 $h_{\rm{damp}}=m_t$ 	& 14.1 & 7.2$\times 10^{-1}$ & 18.9 & 4.0$\times 10^{-1}$ & 22.6 & 2.1$\times 10^{-1}$ & 16.0 & 5.9$\times 10^{-1}$ \\
{\sc MC@NLO+Herwig} 					& 13.9 & 7.4$\times 10^{-1}$ & 18.6 & 4.2$\times 10^{-1}$ & 25.7 & 1.1$\times 10^{-1}$ & 21.9 & 2.4$\times 10^{-1}$ \\
{\sc Powheg+Herwig} $h_{\rm{damp}}=\infty$ 	& 22.5 & 2.1$\times 10^{-1}$ & 23.5 & 1.7$\times 10^{-1}$ & 13.3 & 7.7$\times 10^{-1}$ & 14.9 & 6.7$\times 10^{-1}$ \\ \hline

{\sc Alpgen+Herwig} 					& 24.6 & 1.4$\times 10^{-1}$ & 28.0 & 6.2$\times 10^{-2}$ & 19.2 & 3.8$\times 10^{-1}$ & 13.7 & 7.5$\times 10^{-1}$ \\
{\sc Alpgen+Pythia}6				 	& 21.6 & 2.5$\times 10^{-1}$ & 24.6 & 1.4$\times 10^{-1}$ & 13.5 & 7.6$\times 10^{-1}$ & 16.7 & 5.4$\times 10^{-1}$ \\
{\sc MadGraph+Pythia}6 					& 20.2 & 3.2$\times 10^{-1}$ & 19.1 & 3.9$\times 10^{-1}$ & 20.1 & 3.3$\times 10^{-1}$ & 14.6 & 6.9$\times 10^{-1}$ \\ \hline

{\sc AcerMC+Pythia}6 RadHi 				& 58.8 & 3.2$\times 10^{-6}$ & 68.7 & 7.4$\times 10^{-8}$ & 23.4 & 1.8$\times 10^{-1}$ & 21.1 & 2.7$\times 10^{-1}$ \\
{\sc AcerMC+Pythia}6 RadLo 				& 52.6 & 3.0$\times 10^{-5}$ & 49.7 & 8.3$\times 10^{-5}$ & 30.3 & 3.5$\times 10^{-2}$ & 16.4 & 5.6$\times 10^{-1}$ \\
{\sc Alpgen+Pythia}6 RadHi 				& 32.7 & 1.8$\times 10^{-2}$ & 39.2 & 2.7$\times 10^{-3}$ & 14.1 & 7.2$\times 10^{-1}$ & 17.8 & 4.7$\times 10^{-1}$ \\
{\sc Alpgen+Pythia}6 RadLo 				& 18.7 & 4.1$\times 10^{-1}$ & 23.1 & 1.9$\times 10^{-1}$ & 20.5 & 3.1$\times 10^{-1}$ & 18.8 & 4.1$\times 10^{-1}$ \\
{\sc MadGraph+Pythia}6 $q^2$ down 		& 24.1 & 1.5$\times 10^{-1}$ & 24.5 & 1.4$\times 10^{-1}$ & 12.6 & 8.1$\times 10^{-1}$ & 14.2 & 7.2$\times 10^{-1}$ \\
{\sc MadGraph+Pythia}6 $q^2$ up 			& 27.6 & 6.9$\times 10^{-2}$ & 22.5 & 2.1$\times 10^{-1}$ & 30.7 & 3.1$\times 10^{-2}$ & 20.5 & 3.1$\times 10^{-1}$ \\
\hline\end{tabular}
\caption{Values of $\chi^2$ for the comparison of the measured
gap fraction distributions with the predictions from various \ttbar\
generator configurations, for the four invariant mass \memubb\ regions as a 
function of \qzero. The $\chi^2$ and $p$-values correspond to 18 degrees of freedom.}
\label{table:chimassQ0}
\end{table}
\begin{table}[]
\centering
\footnotesize
\begin{tabular}{|l|ll|ll|ll|ll|}
\hline
$Q_{\rm{sum}}$&\multicolumn{2}{l|}{$m<300$~\gev}&\multicolumn{2}{l|}{$300<m<425$~\gev}&\multicolumn{2}{l|}{$425<m<600$~\gev}&\multicolumn{2}{l|}{$m>600$~\gev}\\
Generator&$\chi^2$ & $p$-value & $\chi^2$ & $p$-value & $\chi^2$    & $p$-value      & $\chi^2$  & $p$-value    \\ \hline
{\sc Powheg+Pythia}6 $h_{\rm{damp}}=\infty$ 	& 35.8 & 3.2$\times 10^{-2}$ & 26.4 & 2.3$\times 10^{-1}$ & 13.1 & 9.3$\times 10^{-1}$ & 31.1 & 9.4$\times 10^{-2}$ \\
{\sc Powheg+Pythia}6 $h_{\rm{damp}}=m_t$ 	& 34.7 & 4.1$\times 10^{-2}$ & 24.7 & 3.1$\times 10^{-1}$ & 26.8 & 2.2$\times 10^{-1}$ & 31.1 & 9.5$\times 10^{-2}$ \\
{\sc Powheg+Pythia}8 $h_{\rm{damp}}=m_t$ 	& 31.2 & 9.3$\times 10^{-2}$ & 21.7 & 4.8$\times 10^{-1}$ & 13.6 & 9.1$\times 10^{-1}$ & 29.7 & 1.3$\times 10^{-1}$ \\
{\sc MC@NLO+Herwig} 					& 33.5 & 5.5$\times 10^{-2}$ & 20.8 & 5.3$\times 10^{-1}$ & 24.0 & 3.5$\times 10^{-1}$ & 20.5 & 5.5$\times 10^{-1}$ \\
{\sc Powheg+Herwig} $h_{\rm{damp}}=\infty$ 	& 35.4 & 3.5$\times 10^{-2}$ & 23.6 & 3.7$\times 10^{-1}$ & 9.8 & 9.9$\times 10^{-1}$ & 30.7 & 1.0$\times 10^{-1}$ \\ \hline

{\sc Alpgen+Herwig} 					& 42.6 & 5.3$\times 10^{-3}$ & 25.3 & 2.8$\times 10^{-1}$ & 12.8 & 9.4$\times 10^{-1}$ & 30.8 & 9.9$\times 10^{-2}$ \\
{\sc Alpgen+Pythia}6 					& 39.0 & 1.4$\times 10^{-2}$ & 25.9 & 2.6$\times 10^{-1}$ & 12.1 & 9.6$\times 10^{-1}$ & 31.8 & 8.1$\times 10^{-2}$ \\
{\sc MadGraph+Pythia}6 					& 32.0 & 7.8$\times 10^{-2}$ & 16.7 & 7.8$\times 10^{-1}$ & 15.7 & 8.3$\times 10^{-1}$ & 29.3 & 1.4$\times 10^{-1}$ \\ \hline

{\sc AcerMC+Pythia}6 RadHi 				& 64.6 & 4.6$\times 10^{-6}$ & 44.9 & 2.8$\times 10^{-3}$ & 29.4 & 1.4$\times 10^{-1}$ & 27.8 & 1.8$\times 10^{-1}$ \\
{\sc AcerMC+Pythia}6 RadLo 				& 64.5 & 4.7$\times 10^{-6}$ & 30.6 & 1.0$\times 10^{-1}$ & 31.2 & 9.2$\times 10^{-2}$ & 23.3 & 3.8$\times 10^{-1}$ \\
{\sc Alpgen+Pythia}6 RadHi 				& 60.5 & 1.9$\times 10^{-5}$ & 44.8 & 2.8$\times 10^{-3}$ & 15.5 & 8.4$\times 10^{-1}$ & 38.9 & 1.5$\times 10^{-2}$ \\
{\sc Alpgen+Pythia}6 RadLo 				& 31.5 & 8.7$\times 10^{-2}$ & 22.6 & 4.3$\times 10^{-1}$ & 14.9 & 8.7$\times 10^{-1}$ & 25.8 & 2.6$\times 10^{-1}$ \\
{\sc MadGraph+Pythia}6 $q^2$ down 		& 37.2 & 2.3$\times 10^{-2}$ & 18.0 & 7.1$\times 10^{-1}$ & 11.7 & 9.6$\times 10^{-1}$ & 29.3 & 1.4$\times 10^{-1}$ \\
{\sc MadGraph+Pythia}6 $q^2$ up 			& 40.2 & 1.0$\times 10^{-2}$ & 22.3 & 4.4$\times 10^{-1}$ & 26.6 & 2.3$\times 10^{-1}$ & 27.8 & 1.8$\times 10^{-1}$ \\ 
\hline\end{tabular}
\caption{Values of $\chi^2$ for the comparison of the measured
gap fraction distributions with the predictions from various \ttbar\
generator configurations, for the four invariant mass \memubb\ regions as a 
function of \qsum. The $\chi^2$ and $p$-values correspond to 22 degrees of freedom.}
\label{table:chimassQsum}
\end{table}

In general, the different generator configurations provide a good model of the
evolution of the gap fraction distributions with \memubb, and similar trends 
in the predictions of individual generators are seen as for the inclusive 
$|y|<2.1$ results discussed in Section~\ref{ssec:gfracresrap}. However,
it can be seen from Figures~\ref{fig:gapfracpartmassb} and~\ref{fig:altmass} 
that in the $425<\memubb<600$\,\GeV\ region, 
the NLO generator predictions split 
into two groups, with {\sc Powheg\,+\,Herwig} and {\sc Powheg\,+\,Pythia6} with 
$\hdmp=\infty$ being consistent with the data, and {\sc Powheg\,+\,Pythia6} 
with $\hdmp=\mtop$, {\sc Powheg\,+\,Pythia8} and {\sc MC@NLO\,+\,Herwig} 
predicting a slightly
larger gap  fraction (and hence less radiation). In the region with 
$\memubb>600$\,\GeV, the measurement uncertainties are too large to 
discriminate between the predictions.



\FloatBarrier
\section{Conclusions}
\label{sec:conclusion}

Studies of the additional jet activity in dileptonic \ttbar\ events with an 
opposite-sign $e\mu$ pair and two $b$-tagged jets have been presented, using
20.3\,\ifb\ of $\sqrt{s}=8$\,\TeV\ $pp$ collision data collected by the
ATLAS detector at the LHC. The measurements were corrected to the 
particle level and defined in a fiducial region corresponding closely to the
experimental acceptance, facilitating comparisons with the predictions of
different Monte Carlo \ttbar\ event generators. The additional-jet multiplicity
for various jet \pt\ thresholds has been measured in the pseudorapidity
region $|\eta|<4.5$, together with the normalised differential cross-sections
as a function of the first to the fourth jet \pt. The gap fraction, the
fraction of events with no additional jet above a certain \pt\ 
threshold, has also been measured in the central rapidity region $|y|<2.1$,
for subsets of this $y$ region, and as a function of the invariant mass
of the \emubb\ system. Taken together, these measurements can help to
characterise the production of additional jets in \ttbar\
events, an important test of QCD and a significant source of systematic 
uncertainty in many measurements and searches for new physics at the LHC.
The results will be made available in the {\sc HepData} repository
and through the {\sc Rivet} analysis framework.

The measurements are generally well-described by the predictions of the 
next-to-leading-order generators used in ATLAS physics analyses. Both
{\sc Powheg} (interfaced to {\sc Pythia6}, {\sc Pythia8} or {\sc Herwig})
and {\sc MC@NLO\,+\,Herwig} give good descriptions of the \pt\ spectrum
of the first additional jet,
although {\sc MC@NLO\,+\,Herwig} does not describe higher jet 
multiplicities, or the gap fraction as a function of a threshold on the
sum of the \pt\ of all additional jets.
The leading-order multi-leg generators {\sc Alpgen}, 
interfaced to {\sc Pythia6} or {\sc Herwig}, and 
{\sc MadGraph} interfaced to {\sc Pythia6}, are also generally compatible
with the data.
The predictions of these generators are sensitive to the choice of QCD scale 
and parton shower parameters, and tuning to the precise
measurements presented here offers considerable scope for reducing the
range of parameter variations which need to be considered when evaluating
\ttbar\ modelling uncertainties, compared to the ranges  derived from previous 
analyses based on smaller $\sqrt{s}=7$\,\TeV\ ATLAS data samples.

\section*{Acknowledgements}


We thank CERN for the very successful operation of the LHC, as well as the
support staff from our institutions without whom ATLAS could not be
operated efficiently.

We acknowledge the support of ANPCyT, Argentina; YerPhI, Armenia; ARC, Australia; BMWFW and FWF, Austria; ANAS, Azerbaijan; SSTC, Belarus; CNPq and FAPESP, Brazil; NSERC, NRC and CFI, Canada; CERN; CONICYT, Chile; CAS, MOST and NSFC, China; COLCIENCIAS, Colombia; MSMT CR, MPO CR and VSC CR, Czech Republic; DNRF and DNSRC, Denmark; IN2P3-CNRS, CEA-DSM/IRFU, France; GNSF, Georgia; BMBF, HGF, and MPG, Germany; GSRT, Greece; RGC, Hong Kong SAR, China; ISF, I-CORE and Benoziyo Center, Israel; INFN, Italy; MEXT and JSPS, Japan; CNRST, Morocco; FOM and NWO, Netherlands; RCN, Norway; MNiSW and NCN, Poland; FCT, Portugal; MNE/IFA, Romania; MES of Russia and NRC KI, Russian Federation; JINR; MESTD, Serbia; MSSR, Slovakia; ARRS and MIZ\v{S}, Slovenia; DST/NRF, South Africa; MINECO, Spain; SRC and Wallenberg Foundation, Sweden; SERI, SNSF and Cantons of Bern and Geneva, Switzerland; MOST, Taiwan; TAEK, Turkey; STFC, United Kingdom; DOE and NSF, United States of America. In addition, individual groups and members have received support from BCKDF, the Canada Council, CANARIE, CRC, Compute Canada, FQRNT, and the Ontario Innovation Trust, Canada; EPLANET, ERC, FP7, Horizon 2020 and Marie Sk{\l}odowska-Curie Actions, European Union; Investissements d'Avenir Labex and Idex, ANR, R{\'e}gion Auvergne and Fondation Partager le Savoir, France; DFG and AvH Foundation, Germany; Herakleitos, Thales and Aristeia programmes co-financed by EU-ESF and the Greek NSRF; BSF, GIF and Minerva, Israel; BRF, Norway; Generalitat de Catalunya, Generalitat Valenciana, Spain; the Royal Society and Leverhulme Trust, United Kingdom.

The crucial computing support from all WLCG partners is acknowledged gratefully, in particular from CERN, the ATLAS Tier-1 facilities at TRIUMF (Canada), NDGF (Denmark, Norway, Sweden), CC-IN2P3 (France), KIT/GridKA (Germany), INFN-CNAF (Italy), NL-T1 (Netherlands), PIC (Spain), ASGC (Taiwan), RAL (UK) and BNL (USA), the Tier-2 facilities worldwide and large non-WLCG resource providers. Major contributors of computing resources are listed in Ref.~\cite{ATL-GEN-PUB-2016-002}.



\printbibliography

\newpage 
\begin{flushleft}
{\Large The ATLAS Collaboration}

\bigskip

M.~Aaboud$^\textrm{\scriptsize 135d}$,
G.~Aad$^\textrm{\scriptsize 86}$,
B.~Abbott$^\textrm{\scriptsize 113}$,
J.~Abdallah$^\textrm{\scriptsize 64}$,
O.~Abdinov$^\textrm{\scriptsize 12}$,
B.~Abeloos$^\textrm{\scriptsize 117}$,
R.~Aben$^\textrm{\scriptsize 107}$,
O.S.~AbouZeid$^\textrm{\scriptsize 137}$,
N.L.~Abraham$^\textrm{\scriptsize 149}$,
H.~Abramowicz$^\textrm{\scriptsize 153}$
% ...
\bigskip
\\
% ...
$^{12}$ Institute of Physics, Azerbaijan Academy of Sciences, Baku, Azerbaijan\\
% ...
$^{64}$ University of Iowa, Iowa City IA, United States of America\\
% ...
$^{86}$ CPPM, Aix-Marseille Universit{\'e} and CNRS/IN2P3, Marseille, France\\
% ...
$^{107}$ Nikhef National Institute for Subatomic Physics and University of Amsterdam, Amsterdam, Netherlands\\
% ...
$^{113}$ Homer L. Dodge Department of Physics and Astronomy, University of Oklahoma, Norman OK, United States of America\\
% ...
$^{117}$ LAL, Univ. Paris-Sud, CNRS/IN2P3, Universit{\'e} Paris-Saclay, Orsay, France\\
% ...
$^{135}$ $^{(a)}$ Facult{\'e} des Sciences Ain Chock, R{\'e}seau Universitaire de Physique des Hautes Energies - Universit{\'e} Hassan II, Casablanca; $^{(b)}$ Centre National de l'Energie des Sciences Techniques Nucleaires, Rabat; $^{(c)}$ Facult{\'e} des Sciences Semlalia, Universit{\'e} Cadi Ayyad, LPHEA-Marrakech; $^{(d)}$ Facult{\'e} des Sciences, Universit{\'e} Mohamed Premier and LPTPM, Oujda; $^{(e)}$ Facult{\'e} des sciences, Universit{\'e} Mohammed V, Rabat, Morocco\\
% ...
$^{137}$ Santa Cruz Institute for Particle Physics, University of California Santa Cruz, Santa Cruz CA, United States of America\\
% ...
$^{149}$ Department of Physics and Astronomy, University of Sussex, Brighton, United Kingdom\\
% ...
$^{153}$ Raymond and Beverly Sackler School of Physics and Astronomy, Tel Aviv University, Tel Aviv, Israel\\
% ...
$^{a}$ Also at Department of Physics, King's College London, London, United Kingdom\\
$^{b}$ Also at Institute of Physics, Azerbaijan Academy of Sciences, Baku, Azerbaijan\\
$^{c}$ Also at Novosibirsk State University, Novosibirsk, Russia\\
$^{d}$ Also at TRIUMF, Vancouver BC, Canada\\
$^{e}$ Also at Department of Physics {\&} Astronomy, University of Louisville, Louisville, KY, United States of America\\
% ...
\end{flushleft}


\end{document}